\documentclass[12pt,aps,prd,preprint,tightenlines,superscriptaddress,amsmath,amssymb,nofootinbib]{revtex4-1} 

\RequirePackage[colorlinks=true
,urlcolor=blue
,anchorcolor=blue
,citecolor=blue
,filecolor=blue
,linkcolor=blue
,menucolor=blue
,linktocpage=true
,pdfproducer=medialab
,pdfa=true
]{hyperref}

\allowdisplaybreaks

\usepackage{amsmath,amssymb,amsthm,amsfonts}
\usepackage{graphicx}
\usepackage{subfigure}
\usepackage{dcolumn}	
\usepackage{bm}	     	
\usepackage{epsfig}
\usepackage{epstopdf}
\usepackage{setspace}
\usepackage{slashed}
\usepackage{comment}
\usepackage{enumitem}
\usepackage{siunitx}
\usepackage{multirow}
\usepackage{datetime}
\usepackage{fancyvrb}
\usepackage{textgreek}
\usepackage{xcolor}
\usepackage{xspace}

\setlist[description]{leftmargin=0.4cm}

\DeclareMathOperator{\diag}{diag}

\newcommand{\PRE}[1]{{#1}} 

\newcommand{\be}{\begin{equation}\begin{aligned}}
\newcommand{\ee}{\end{aligned}\end{equation}}
\newcommand{\beq}{\begin{equation}}
\newcommand{\eeq}{\end{equation}}
\newcommand{\beqa}{\begin{eqnarray}}
\newcommand{\eeqa}{\end{eqnarray}}

\newcommand{\ifb}{\text{fb}^{-1}}

\newcommand{\K}{\text{K}}

\newcommand{\mev}{\text{MeV}}
\newcommand{\gev}{\text{GeV}}
\newcommand{\tev}{\text{TeV}}
\newcommand{\fb}{\text{fb}}

\newcommand{\mm}{\text{mm}}
\newcommand{\cm}{\text{cm}}
\newcommand{\m}{\text{m}}

\newcommand{\mrad}{\text{mrad}}
\newcommand{\murad}{\mu\text{rad}}
\newcommand{\g}{\text{g}}

\newcommand{\s}{\text{s}}

\renewcommand{\eqref}[1]{Eq.~(\ref{eq:#1})}

\newcommand{\secref}[1]{Sec.~\ref{sec:#1}}
\newcommand{\secsref}[2]{Secs.~\ref{sec:#1} and \ref{sec:#2}}

\newcommand{\figref}[1]{Fig.~\ref{fig:#1}}

\newcommand{\Figref}[1]{Figure~\ref{fig:#1}}
\newcommand{\tableref}[1]{Table~\ref{table:#1}}

\RequirePackage[normalem]{ulem}

\newcommand{\FASERnu}{FASER$\nu$\xspace}

\begin{document}

\preprint{CERN-EP-2019-160, KYUSHU-RCAPP-2019-003, SLAC-PUB-17460, UCI-TR-2019-19}

\title{
{\Large Detecting and Studying High-Energy Collider Neutrinos with FASER at the LHC}
\PRE{\vspace*{0.5in} \\
FASER Collaboration}
}

\begin{figure*}[h]
\centering
\includegraphics[width=0.6\textwidth]{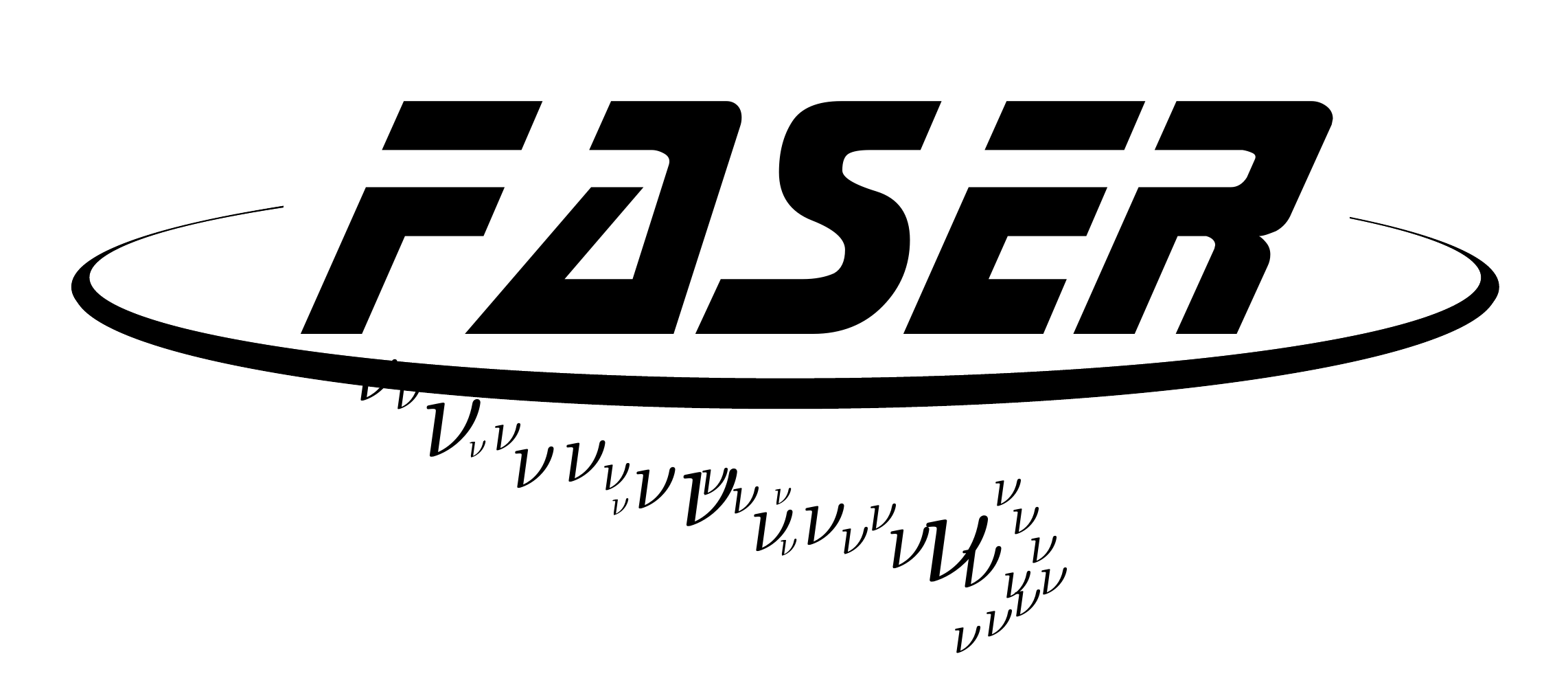}
\end{figure*}

\author{Henso Abreu}
\affiliation{Department of Physics and Astronomy, Technion---Israel Institute of Technology, Haifa 32000, Israel}

\author{Claire Antel}
\affiliation{D\'epartement de Physique Nucl\'eaire et Corpusculaire, 
University of Geneva, CH-1211 Geneva 4, Switzerland}

\author{Akitaka Ariga}
\email[Contact emails: ]{akitaka.ariga@lhep.unibe.ch, tomoko.ariga@cern.ch, felixk@slac.stanford.edu}
\affiliation{Universit\"at Bern, Sidlerstrasse 5, CH-3012 Bern, Switzerland}

\author{Tomoko Ariga}
\email[Contact emails: ]{akitaka.ariga@lhep.unibe.ch, tomoko.ariga@cern.ch, felixk@slac.stanford.edu}
\affiliation{Universit\"at Bern, Sidlerstrasse 5, CH-3012 Bern, Switzerland}
\affiliation{Kyushu University, Nishi-ku, 819-0395 Fukuoka, Japan}

\author{Jamie Boyd}
\affiliation{CERN, CH-1211 Geneva 23, Switzerland}

\author{Franck Cadoux}
\affiliation{D\'epartement de Physique Nucl\'eaire et Corpusculaire, 
University of Geneva, CH-1211 Geneva 4, Switzerland}

\author{David~W.~Casper}
\affiliation{Department of Physics and Astronomy, 
University of California, Irvine, CA 92697-4575, USA}

\author{Xin Chen}
\affiliation{Physics Department,Tsinghua University, Beijing, China}
 
\author{Andrea Coccaro}
\affiliation{INFN Sezione di Genova, Via Dodecaneso, 33--16146, Genova, Italy}
 
\author{Candan Dozen}
\affiliation{Physics Department,Tsinghua University, Beijing, China}

\author{Peter B.~Denton}
\email[FASER Associate.]{\strut}
\affiliation{Department of Physics, Brookhaven National Laboratory, Upton, NY 11973, USA}

\author{Yannick Favre}
\affiliation{D\'epartement de Physique Nucl\'eaire et Corpusculaire, 
University of Geneva, CH-1211 Geneva 4, Switzerland}

\author{Jonathan~L.~Feng}
\affiliation{Department of Physics and Astronomy, 
University of California, Irvine, CA 92697-4575, USA}

\author{Didier Ferrere}
\affiliation{D\'epartement de Physique Nucl\'eaire et Corpusculaire, 
University of Geneva, CH-1211 Geneva 4, Switzerland}

\author{Iftah Galon}
\affiliation{New High Energy Theory Center, Rutgers, The State University of New Jersey, Piscataway, New Jersey 08854-8019, USA}

\author{Stephen Gibson}
\affiliation{Royal Holloway, University of London, Egham, TW20 0EX, UK}

\author{Sergio Gonzalez-Sevilla}
\affiliation{D\'epartement de Physique Nucl\'eaire et Corpusculaire, 
University of Geneva, CH-1211 Geneva 4, Switzerland}

\author{Shih-Chieh Hsu}
\affiliation{Department of Physics, University of Washington, PO Box 351560, Seattle, WA 98195-1560, USA}

\author{Zhen Hu}
\affiliation{Physics Department,Tsinghua University, Beijing, China}

\author{Giuseppe Iacobucci}
\affiliation{D\'epartement de Physique Nucl\'eaire et Corpusculaire, 
University of Geneva, CH-1211 Geneva 4, Switzerland}

\author{Sune Jakobsen}
\affiliation{CERN, CH-1211 Geneva 23, Switzerland}

\author{Roland Jansky}
\affiliation{D\'epartement de Physique Nucl\'eaire et Corpusculaire, 
University of Geneva, CH-1211 Geneva 4, Switzerland}

\author{Enrique Kajomovitz}
\affiliation{Department of Physics and Astronomy, 
Technion---Israel Institute of Technology, Haifa 32000, Israel}

\author{Felix Kling}
\email[Contact emails: ]{akitaka.ariga@lhep.unibe.ch, tomoko.ariga@cern.ch, felixk@slac.stanford.edu}
\affiliation{Department of Physics and Astronomy, 
University of California, Irvine, CA 92697-4575, USA}
\affiliation{SLAC National Accelerator Laboratory, 2575 Sand Hill Road, Menlo Park, CA 94025, USA}

\author{Susanne Kuehn}
\affiliation{CERN, CH-1211 Geneva 23, Switzerland}

\author{Lorne Levinson}
\affiliation{Department of Particle Physics and Astrophysics, Weizmann Institute of Science, Rehovot 76100, Israel}

\author{Congqiao Li}
\affiliation{Department of Physics, University of Washington, PO Box 351560, Seattle, WA 98195-1560, USA}

\author{Josh McFayden}
\affiliation{CERN, CH-1211 Geneva 23, Switzerland}

\author{Sam Meehan}
\affiliation{CERN, CH-1211 Geneva 23, Switzerland}

\author{Friedemann Neuhaus}
\affiliation{Institut f\"ur Physik, Universität Mainz, Mainz, Germany}

\author{Hidetoshi Otono}
\affiliation{Kyushu University, Nishi-ku, 819-0395 Fukuoka, Japan}

\author{Brian Petersen}
\affiliation{CERN, CH-1211 Geneva 23, Switzerland}

\author{Helena Pikhartova}
\affiliation{Royal Holloway, University of London, Egham, TW20 0EX, UK}

\author{Michaela Queitsch-Maitland}
\affiliation{CERN, CH-1211 Geneva 23, Switzerland}

\author{Osamu Sato}
\affiliation{Nagoya University, Furo-cho, Chikusa-ku, Nagoya 464-8602, Japan}

\author{Kristof Schmieden}
\affiliation{CERN, CH-1211 Geneva 23, Switzerland}

\author{Matthias Schott}
\affiliation{Institut f\"ur Physik, Universität Mainz, Mainz, Germany}

\author{Anna Sfyrla}
\affiliation{D\'epartement de Physique Nucl\'eaire et Corpusculaire, 
University of Geneva, CH-1211 Geneva 4, Switzerland}

\author{Savannah Shively}
\affiliation{Department of Physics and Astronomy, 
University of California, Irvine, CA 92697-4575, USA}

\author{Jordan Smolinsky}
\affiliation{Department of Physics and Astronomy, 
University of California, Irvine, CA 92697-4575, USA}

\author{Aaron~M.~Soffa}
\affiliation{Department of Physics and Astronomy, 
University of California, Irvine, CA 92697-4575, USA}

\author{Yosuke Takubo}
\affiliation{Institute of Particle and Nuclear Study, 
KEK, Oho 1-1, Tsukuba, Ibaraki 305-0801, Japan}

\author{Eric Torrence}
\affiliation{University of Oregon, Eugene, OR 97403, USA}

\author{Sebastian Trojanowski}
\affiliation{Consortium for Fundamental Physics, School of Mathematics and  Statistics, University of Sheffield, Hounsfield Road, Sheffield, S3 7RH, UK
\PRE{\vspace*{.4in}}
}

\author{Callum Wilkinson}
\email[FASER Associate.]{\strut}
\affiliation{Universit\"at Bern, Sidlerstrasse 5, CH-3012 Bern, Switzerland}

\author{Dengfeng Zhang}
\affiliation{Physics Department,Tsinghua University, Beijing, China}

\author{Gang Zhang\PRE{\vspace*{.2in}}}
\affiliation{Physics Department,Tsinghua University, Beijing, China}

\begin{abstract}
\PRE{\vspace*{.2in}}
Neutrinos are copiously produced at particle colliders, but no collider neutrino has ever been detected.  Colliders produce both neutrinos and anti-neutrinos of all flavors at very high energies, and they are therefore highly complementary to those from other sources.  FASER, the Forward Search Experiment at the LHC, is ideally located to provide the first detection and study of collider neutrinos. We investigate the prospects for neutrino studies with FASER$\nu$, a proposed component of FASER, consisting of emulsion films interleaved with tungsten plates with a total target mass of 1.2 tons, to be placed on-axis at the front of FASER.  We estimate the neutrino fluxes and interaction rates, describe the FASER$\nu$ detector, and analyze the characteristics of the signals and primary backgrounds. For an integrated luminosity of $150~\ifb$ to be collected during Run 3 of the 14 TeV LHC in 2021-23, approximately 1300 electron neutrinos, 20,000 muon neutrinos, and 20 tau neutrinos will interact in FASER$\nu$, with mean energies of 600 GeV to 1 TeV.  With such rates and energies, FASER will measure neutrino cross sections at energies where they are currently unconstrained, will bound models of forward particle production, and could open a new window on physics beyond the standard model.
\end{abstract}

\maketitle

\renewcommand{\baselinestretch}{0.95}\normalsize
\tableofcontents
\renewcommand{\baselinestretch}{1.0}\normalsize


\clearpage
\section{Introduction}
\label{sec:introduction}

Since their discovery at a nuclear reactor in 1956~\cite{Cowan:1992xc}, neutrinos have been detected from an variety of sources: beam dump experiments~\cite{Danby:1962nd}, cosmic ray interactions in the atmosphere~\cite{Achar:1965ova,Reines:1965qk,Fukuda:1998mi}, the Sun~\cite{Davis:1968cp,Cleveland:1998nv}, the Earth~\cite{Araki:2005qa}, 
supernovae~\cite{Hirata:1987hu,Bionta:1987qt}, and other astrophysical bodies outside our galaxy~\cite{Aartsen:2014gkd}. The detection of neutrinos from these many sources has led to profound insights across the fields of particle physics, nuclear physics, and astrophysics.

At present, no neutrino produced at a particle collider has ever been detected.  Colliders copiously produce both neutrinos and anti-neutrinos of all flavors and, as we will discuss, most are produced at very high energies where neutrino interactions are not well studied.  Collider neutrinos are therefore highly complementary to those from other sources, and there has been a longstanding interest in detecting them; see, e.g., Refs.~\cite{DeRujula:1984pg, Vannucci:253670,DeRujula:1992sn,Park:2011gh,Buontempo:2018gta}.  Nevertheless, collider neutrinos have not yet been detected for at least two reasons.  First and most obviously, neutrinos interact very weakly.  The probability for a neutrino to interact in a meter of water is roughly $P \sim 4 \times 10^{-13} \, (E_{\nu}/\gev)$ for the GeV to TeV energies of interest.  Given the fluxes and energies of neutrinos produced at colliders and the typical sizes and coverage of collider detectors, neutrino interactions are very rare and difficult to detect above background.  Second, the highest energy collider neutrinos, and therefore the ones with the largest interaction cross sections, are produced along the beamline. Collider detectors have holes along the beamline to let the beams in, and so are blind in this region and miss the enormous flux of high-energy neutrinos streaming down the beam pipe.

FASER, the Forward Search Experiment at the Large Hadron Collider (LHC)~\cite{Feng:2017uoz}, covers this blind spot. FASER's main goal is long-lived particle searches, and it will be located 480 m downstream of the ATLAS interaction point (IP) along the beam collision axis in the existing side tunnel TI12.  The FASER location is currently being prepared with lighting and power, and a passarelle (stairs) and support structures are already in place to safely transport detector components around the LHC. In this location, starting in Run 3 from 2021-23, FASER will provide sensitive searches for many proposed light and very weakly-interacting particles, such as dark photons, axion-like particles, and light gauge bosons~\cite{Feng:2017uoz, Feng:2017vli, Batell:2017kty, Kling:2018wct, Helo:2018qej, Bauer:2018onh, Cheng:2018vaj, Feng:2018noy, Hochberg:2018rjs, Berlin:2018jbm, Dercks:2018eua, Ariga:2018uku, Beacham:2019nyx, deNiverville:2019xsx, Harigaya:2019shz, Deppisch:2019kvs, Cheng:2019yai}.

For the reasons noted above, the FASER location is also ideal to provide the first detection and studies of high-energy neutrinos produced at the LHC.  Collider neutrinos are predominantly produced in hadron decays, and hadrons are predominantly produced along the beamline.  Such neutrinos will easily pass through the 100 m of concrete and rock between the ATLAS IP and FASER, while almost all other standard model (SM) particles will be either deflected or absorbed before reaching FASER.  Preliminary measurements were made by small, pilot emulsion detectors that were installed and removed in Technical Stops in 2018. These detectors were placed in TI12 and TI18, a symmetric location on the other side of ATLAS that was considered for FASER.  As discussed below in \secsref{bg}{measurements}, these detectors have provided useful and encouraging data concerning background and signal rates in these locations~\cite{FASERmuondata}. FASER's neutrino detection capability was briefly discussed in the experiment's Letter of Intent~\cite{Ariga:2018zuc} and Technical Proposal~\cite{Ariga:2018pin}. More detailed studies on the detector design and physics sensitivities are reported here for the first time. A recent study of Beni et al.~\cite{Beni:2019gxv} has analyzed various locations around the LHC for neutrino detection and has found that, incorporating the background results from FASER, TI12 and TI18 are the best locations among those they considered.   

\begin{figure}[tp]
\centering
\includegraphics[width=0.98\textwidth]{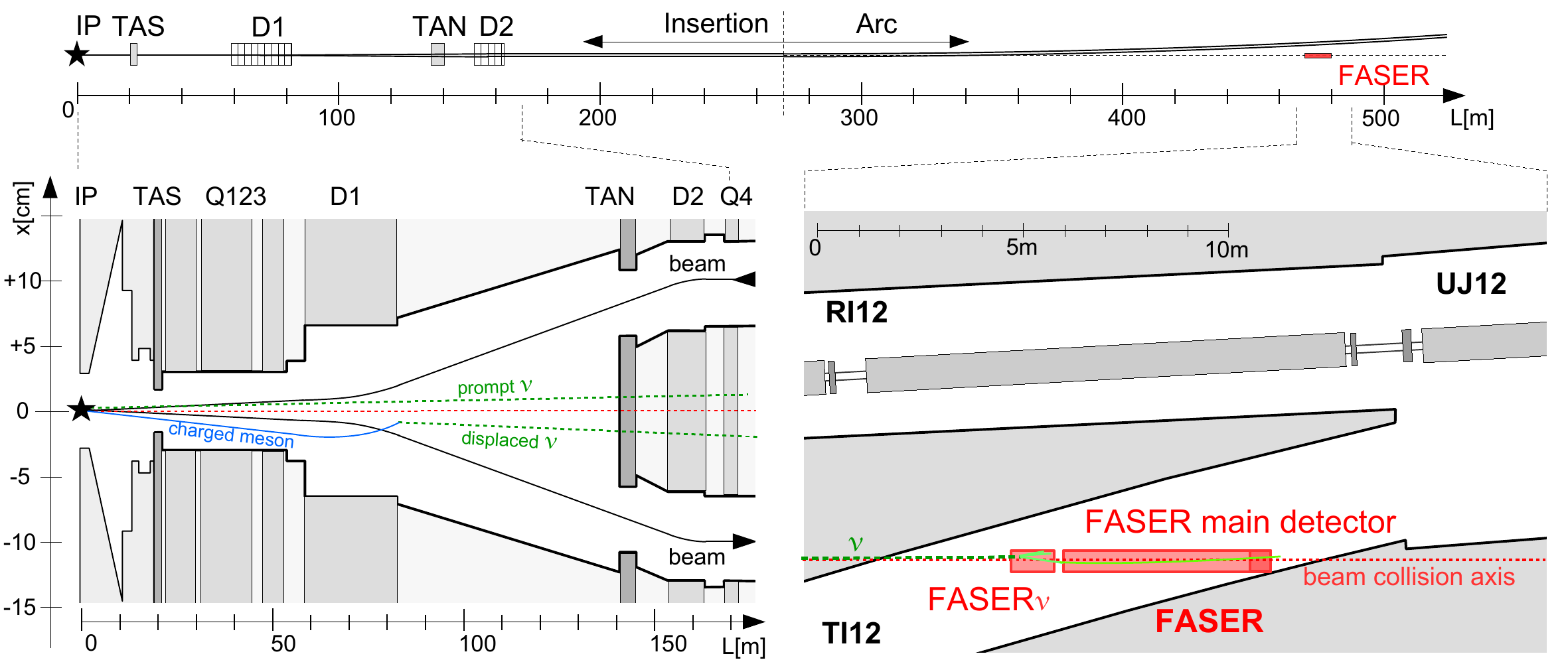} 
\caption{Schematic view of the far-forward region downstream of ATLAS. {\bf Upper panel}: FASER is located $480~\m$ downstream of ATLAS along the beam collision axis (dotted line) after the main LHC tunnel curves away.  {\bf Lower left panel}: High-energy particles produced at the IP in the far-forward direction.  Charged particles (solid lines) are deflected by LHC quadrupole (Q) and dipole (D) magnets. Neutral hadrons are absorbed by either the TAS front quadrupole absorber or by the TAN neutral particle absorber. Neutrinos (dashed lines) are produced either promptly or displaced and pass through the LHC infrastructure without interacting. Note the extreme difference in horizontal and vertical scales.  {\bf Lower right panel}: Neutrinos may then travel $\sim 480~\m$ further downstream into tunnel TI12 and interact in FASER$\nu$, which is located at the front of the FASER main detector.
} 
\label{fig:infrastructure}
\end{figure}

\begin{figure}[hbtp]
\centering
\includegraphics[width=0.55\textwidth]{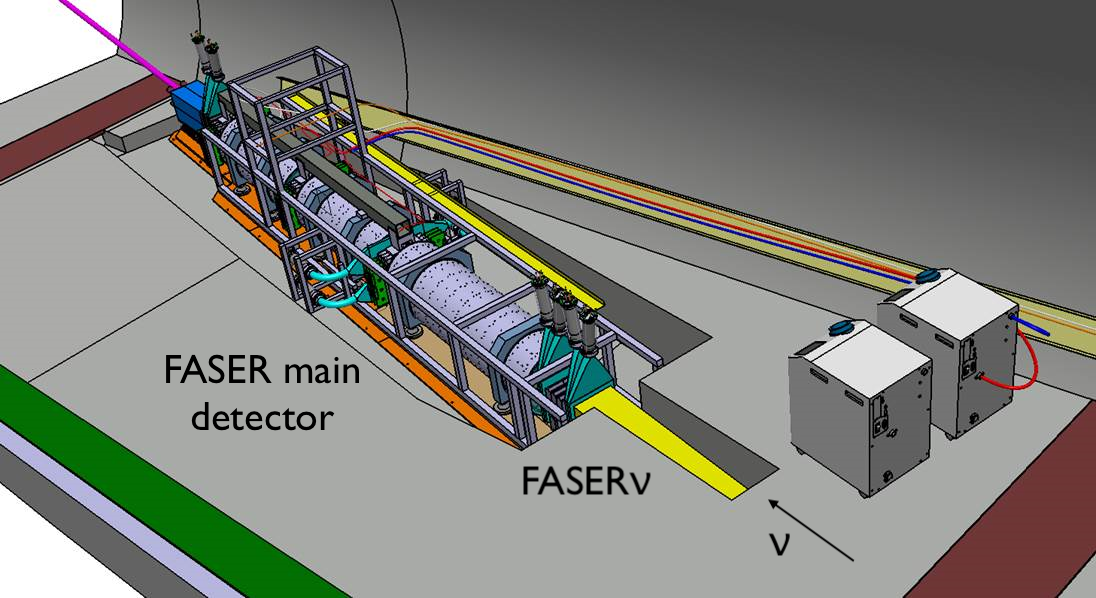} \quad
\includegraphics[width=0.40\textwidth]{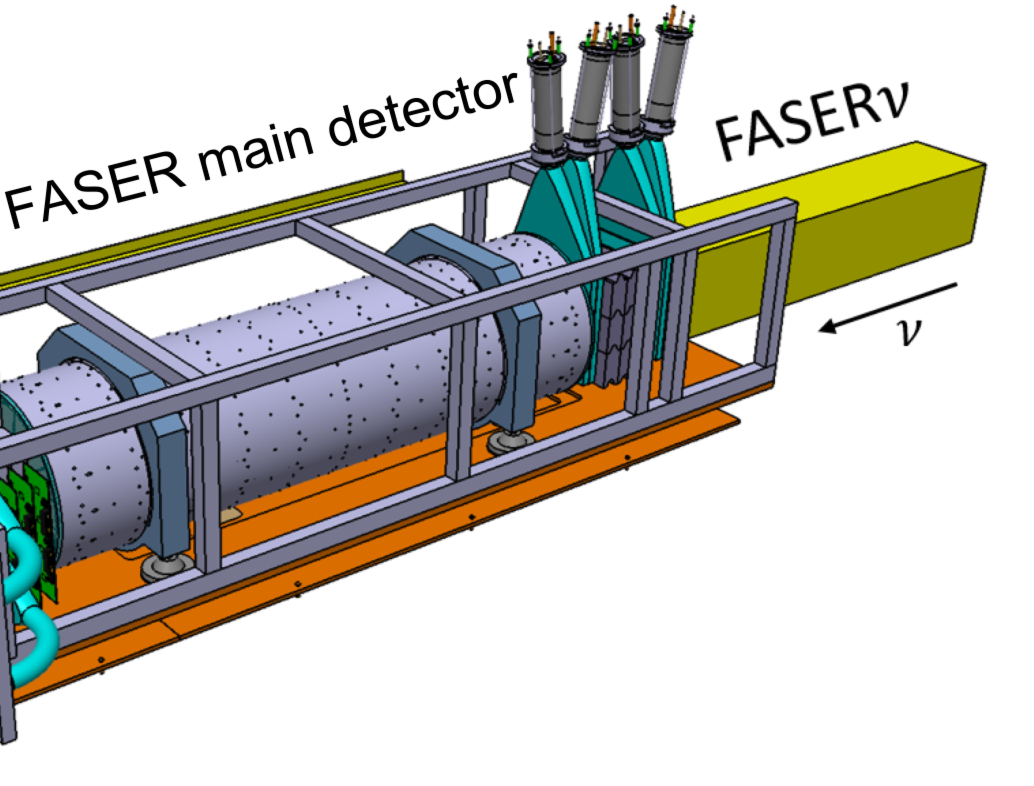}
\caption{View of FASER, including the FASER$\nu$ detector, in tunnel TI12, 480 m downstream from the ATLAS IP along the beam collision axis.  The FASER$\nu$ detector is a $25\,\cm \times 25\,\cm \times 1.35\,\m$ emulsion detector, consisting of 1000 layers of emulsion films interleaved with 1-mm-thick tungsten plates, with a total tungsten target mass of 1.2 tons.  It is located at the front of the FASER main detector in a narrow trench being excavated specifically to house it. 
}
\label{fig:diagram}
\end{figure}

The primary component of FASER for neutrino detection is FASER$\nu$, a proposed $25\,\cm \times 25\,\cm \times 1.35\,\m$ emulsion detector, consisting of 1000 layers of emulsion films interleaved with 1-mm-thick tungsten plates. This results in a total tungsten target mass of 1.2 tons.  FASER$\nu$ is planned to be located at the front of the FASER main detector, and the floor of TI12 is being lowered there to accommodate it.  A schematic view of the ATLAS far-forward region, neutrino trajectories, and TI12 is given in \figref{infrastructure}, and a view of the FASER main detector and FASER$\nu$ as they are proposed to be installed in TI12 is shown in \figref{diagram}.

This study explores FASER's ability to detect collider neutrinos for the first time and study their properties. For an integrated luminosity of $150~\ifb$ collected during Run 3 of the 14 TeV LHC from 2021-23, $2 \times 10^{11}$ electron neutrinos, $6 \times 10^{12}$ muon neutrinos, and $4 \times 10^{9}$ tau neutrinos, along with a comparable number of anti-neutrinos of each flavor, will stream through FASER$\nu$.  Assuming SM cross sections, this implies that 850 $\nu_e$, 450 $\bar{\nu}_{e}$, 14,000 $\nu_{\mu}$, 6000 $\bar{\nu}_\mu$, 14 $\nu_{\tau}$, and 7 $\bar{\nu}_{\tau}$ will interact in FASER$\nu$. Notably, the mean energy of neutrinos that interact in FASER$\nu$ is between 600 GeV and 1 TeV, depending on the flavor.  These energies are above the energies of previously detected man-made neutrinos, but also below the energies of the neutrinos detected by IceCube.  FASER$\nu$ will therefore not only detect the first collider neutrinos, but will also probe neutrino interactions in an energy range where neutrino cross sections are unconstrained, with a host of interesting implications.  As examples, FASER$\nu$ will be able to bound $\nu_{\mu}$ interactions in the TeV energy range, add significantly to the handful of $\nu_{\tau}$ events that have been directly detected so far, and detect the highest energy man-made electron and tau neutrinos, and may also be able to constrain the charm content of the nucleon and probe non-standard neutrino interactions, possibly opening up a new window on physics beyond the standard model (BSM).

In \secref{existing} we briefly summarize existing constraints on neutrino cross sections from non-collider sources.  In \secref{fluxes_rates} we then estimate the neutrino fluxes passing through FASER$\nu$ and the number of neutrino interactions expected in FASER$\nu$.   In \secref{detector} we present the detector design for FASER$\nu$, the signal characteristics in FASER$\nu$, and estimate the precision with which the neutrino energy can be measured.  Background rates and the prospects for differentiating signal from background are discussed in \secref{bg}.  In \secsref{measurements}{studies}, we discuss the detection of collider neutrinos, cross section measurements, and additional physics studies that can be performed at FASER$\nu$.  Our conclusions are collected in \secref{conclusions}.

\section{Existing Neutrino Cross Section Measurements}
\label{sec:existing}

Neutrino scattering provides a window on the interactions of leptons with matter, particularly in the deep inelastic scattering (DIS) regime with neutrino energies $E_\nu \agt 10~\gev$, where one can access fundamental neutrino-parton interactions.  The main sources of high-energy neutrino measurements have been the neutrino beams from the CERN SPS (400 GeV proton) and Fermilab Tevatron (800 GeV proton) accelerators. The maximum attainable neutrino energy from these accelerator sources is 360 GeV, and there is not much data available at such high energies.  The existing neutrino-nucleon charged current (CC) scattering cross section measurements for $\nu_e$, $\nu_{\mu}$, and $\nu_{\tau}$ are shown in \figref{nucross}. We now discuss each flavor in turn:

\begin{itemize}

\item[\boldmath{$\nu_e$}:] There are several measurements of electron neutrino cross sections, however, most are at low energy. The most cited result on the DIS cross section is one from Gargamelle~\cite{Blietschau:1977mu}, which reported $\nu_e$-nucleon cross sections up to 12 GeV. At higher energies, E53~\cite{Baltay:1988au} reported $\nu_e$-$\nu_\mu$ universality in their neutrino cross sections. DONuT~\cite{Kodama:2007aa} also reported consistency with lepton universality. There is no direct data available for $\nu_e$ energies above 250 GeV.  An indirect bound has been reported by HERA~\cite{Ahmed:1994fa}, which studied the reaction $e^- p \to \nu_e X$ and showed that their results are consistent with the SM and constrain the inverse reaction of $\nu_e p$ interactions, where the equivalent fixed target neutrino energy is 50 TeV. 

\item[\boldmath{$\nu_\mu$}:] Muon neutrinos have been intensively studied thanks to the ease with which they are produced and detected. Data from accelerator-based experiments exist up to neutrino energies of 360 GeV~\cite{Tanabashi:2018oca}. Using atmospheric neutrinos, IceCube has constrained cross sections for neutrino energies above 6.3 TeV, albeit with relatively large uncertainties~\cite{Aartsen:2017kpd, Bustamante:2017xuy}. The neutrino energy range between 360 GeV and 6.3 TeV is unexplored.

\item[\boldmath{$\nu_\tau$}:] The tau neutrino is the least studied neutrino. DONuT~\cite{Kodama:2007aa} is the only experiment that has reported the DIS cross section with accelerator neutrinos. The DONuT Collaboration has parameterized the $\nu_{\tau}$-nucleon CC interaction cross section as $\sigma = C E_{\nu} K(E_{\nu})$, where $C$ is a constant with units of $\cm^2/\gev$, the expected linear dependence on neutrino energy $E_{\nu}$ is explicitly identified, and $K(E_{\nu})$ encodes the suppression of the cross section due to the non-negligible $\tau$ mass. They then report constraints on $C$, the ``energy-independent part of the cross section.'' At lower energies, there are cross section measurements with oscillated $\nu_\tau$ from OPERA~\cite{Agafonova:2018auq} and SuperKamiokande~\cite{Li:2017dbe}.  Most recently, IceCube~\cite{Aartsen:2019tjl} also reported evidence of oscillated $\nu_\tau$. The results from OPERA, SuperKamiokande, and IceCube are not easily compared with DONuT, as they have not provided constraints on the energy-independent part of the cross section, but their energy ranges are indicated in the right panel of \figref{nucross}. Due to the size of the earth, the energy range of oscillated neutrinos from sources on earth is limited to several 10 GeV. There are no measurements of $\nu_{\tau}$ cross sections for energies $E_{\nu_\tau}>250~\gev$.

\end{itemize}

\begin{figure}[t]
\centering
\includegraphics[width=\textwidth]{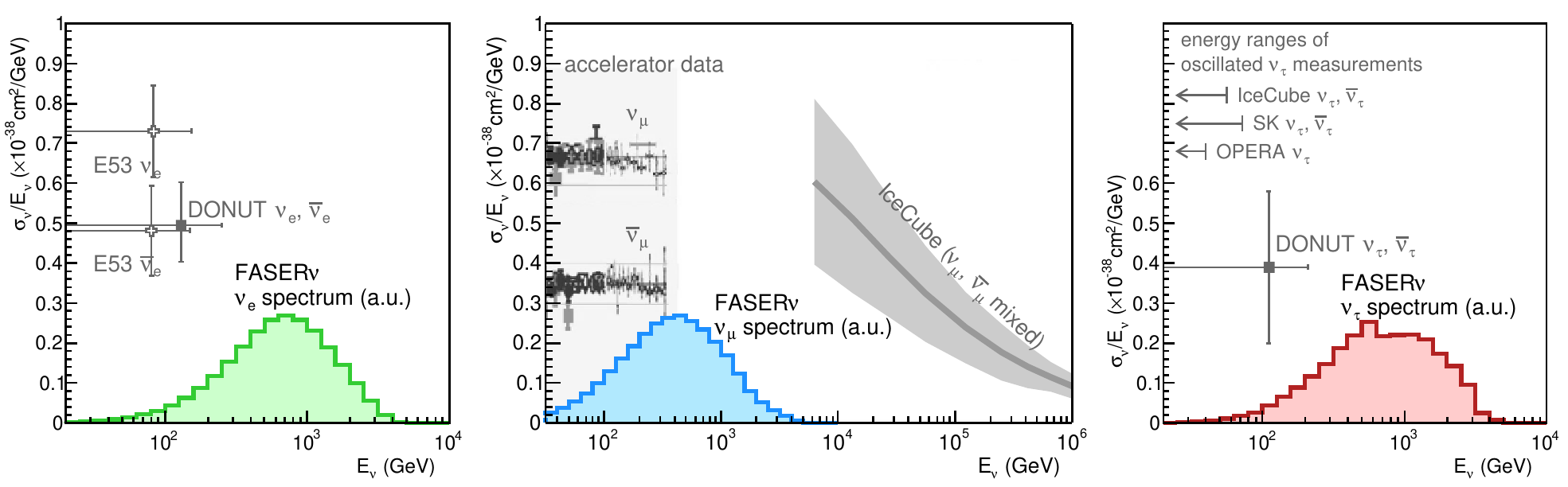}
\caption{Existing measurements of $\nu N$ CC scattering cross sections, where $N$ refers to an isoscalar nucleon in the target, and the expected energy spectra of neutrinos that interact in FASER$\nu$.  For all three flavors, the FASER$\nu$ energy spectra, shown as colored histograms, are peaked at energies that are currently unconstrained.  {\bf Left, \boldmath{$\nu_{e}$} constraints}: Bounds from E53~\cite{Baltay:1988au} and DONuT~\cite{Kodama:2007aa}. The bounds from E53 on $\sigma_{\nu_e} / \sigma_{\nu_\mu}$ and $\sigma_{\bar{\nu}_e} / \sigma_{\bar{\nu}_\mu}$ are multiplied by the current values of $\sigma_{\nu_\mu}$ and $\sigma_{\bar{\nu}_\mu}$, respectively. {\bf Center, \boldmath{$\nu_{\mu}$} constraints}: Bounds from accelerator neutrinos at energies below 360 GeV~\cite{Tanabashi:2018oca} and from IceCube at energies above 6.3 TeV~\cite{Aartsen:2017kpd, Bustamante:2017xuy}. From Ref.~\cite{Aartsen:2017kpd}. {\bf Right, \boldmath{$\nu_{\tau}$} constraints}: The constraint on the energy-independent part of the cross section $C$ from DONuT~\cite{Kodama:2007aa} is shown (see text). DONuT's main systematic uncertainty from the $D_s$ differential production cross section is not included. For OPERA~\cite{Agafonova:2018auq}, SuperKamiokande~\cite{Li:2017dbe}, and IceCube~\cite{Aartsen:2019tjl}, we indicate the energy ranges of $\nu_\tau$ cross section results, but not the measured cross sections themselves (see text). 
}
\label{fig:nucross}
\end{figure}

The energy spectra of neutrinos that interact in FASER$\nu$ are also shown as colored histograms for the three flavors in \figref{nucross}. (For the derivation of these spectra, see \secref{rates}.) As one can see, FASER$\nu$ has the potential to provide neutrino cross section measurements in currently uncharted energy ranges. This is especially notable for $\nu_e$ and $\nu_\tau$, where the cross sections will be measured at the highest energy ever.

The comparison of cross sections for three flavors will provide a test of lepton universality in neutrino scattering. In recent years, especially since the shutdown of the Tevatron in 2011, the focus of precision neutrino studies has been on low energies below 10 GeV~\cite{Acciarri:2018ahy, Acciarri:2016smi, Abe:2015biq}. The goal of these studies is primarily to constrain the $\nu_e$ and $\nu_\mu$ cross sections to minimize systematic uncertainties for  neutrino oscillation experiments~\cite{Acciarri:2016crz, Abe:2011ks, Abe:2015zbg}. At these energies, nuclear effects play an important role, and it is desirable to measure the cross sections experimentally.

On the other hand, recent results from collider experiments may suggest lepton universality violation in the decays $B \to D^{*} \ell \nu$~\cite{Lees:2013uzd, Hirose:2016wfn, Aaij:2017uff}, $B \to K^{*} \ell\ell$~\cite{Aaij:2017vbb}, and $B^+ \to K^+ \ell\ell$~\cite{Aaij:2019wad}. The results may be explained by new, weak-scale physics.  These flavor anomalies therefore motivate testing lepton universality in high-energy neutrino scattering, which may favor or exclude some of the proposed explanations.  

More generally, it is also true that for many new physics scenarios, the BSM effects are most pronounced for heavy and 3rd generation fermions.  This fact motivates probes of $\nu_{\tau}$ properties, which are almost completely unconstrained, as well as $\nu_e$ and $\nu_{\tau}$ production mechanisms, which receive significant or dominant contributions from charm and beauty hadron decays.

\section{Neutrino Rates}
\label{sec:fluxes_rates}

\subsection{Neutrino Flux at the FASER Location}
\label{sec:fluxes}

Neutrinos in the forward direction are predominantly produced in the decays of hadrons. These decays can either occur promptly at the IP or further down the beam pipe, depending on the lifetime of the hadron. Reliable estimates of the neutrino flux in FASER therefore require accurate modeling of the SM hadron spectra and also the LHC infrastructure in the far-forward region.

Hadronic interaction models, designed to describe inelastic collisions at both particle colliders and cosmic ray experiments, have improved greatly in recent years. We exploit this progress and use the Monte Carlo (MC) generators \textsc{Epos-Lhc}~\cite{Pierog:2013ria}, \textsc{Qgsjet-ii-04}~\cite{Ostapchenko:2010vb}, and \textsc{Sibyll 2.3c}~\cite{Ahn:2009wx, Riehn:2015oba, Riehn:2017mfm, Fedynitch:2018cbl}, as implemented in the \textsc{Crmc} simulation package~\cite{CRMC}. Additionally, we obtain the spectra of heavy mesons by simulating inelastic processes in \textsc{Pythia~8}~\cite{Sjostrand:2006za, Sjostrand:2007gs} using the \textsc{Monash}-tune~\cite{Skands:2014pea} and the minimum bias \textsc{A2}-tune~\cite{ATLAS:2012uec}.  

Charm and beauty hadrons decay approximately promptly, which allows us to simulate their decays with the MC generators. In contrast, the light hadrons are long-lived and decay downstream from the IP, which requires us to model their propagation and absorption in the LHC beam pipe. The forward LHC infrastructure~\cite{Aad:2013zwa} is shown in the bottom left panel of \figref{infrastructure}. Located about 20 m downstream from the IP is the TAS front quadrupole absorber, a $1.8~\m$-long copper block with an inner radius of 17 mm that essentially absorbs all hadrons traveling at angles relative to the beam axis of $\theta > 0.85~\mrad$. The TAS shields the inner triplet of quadrupole magnets (Q1-3) and the separator dipole magnet (D1).\footnote{The beam optics will be modified for the HL-LHC, which will impact the forward neutrino flux. Note, however, that the muon rate at the FASER location was estimated with FLUKA for both LHC and HL-LHC optics, and the results were compatible~\cite{FLUKAstudy}. This suggests that beam optics changes between the LHC and HL-LHC will not have a big effect on neutrino production.} The inner triplet consists of two MQXA quadrupoles (Q1 and Q3), oriented in the same direction as each other and orthogonal to the two MQXB magnets (Q2a and Q2b). Both the MQXA and MQXB magnets have field gradients of magnitude 205 T/m and apertures of diameter $70~\mm$~\cite{Siegel:2000fm}. The D1 magnet has a field strength of 3.5 T and an aperture of diameter 60 mm, and it is located $59-83~\m$ downstream. In particular, the dipole magnet will deflect charged particles so that they are not moving towards FASER and will eventually collide with the beam pipe. Located $\sim 140~\m$ downstream is the TAN neutral particle absorber~\cite{Adriani:2015iwv}, which absorbs photons and neutral hadrons produced in the forward direction. For the purpose of this study, we ignore the variable beam crossing angle.  A beam half crossing angle of $150~\murad$ shifts the position of the beam collision axis by $7.2~\cm$ at FASER and leads to a small (3 - 10\%) reduction in the number of neutrinos interacting in FASER$\nu$, where the exact reduction depends on the neutrino flavor.

In modeling the propagation of charged mesons through the forward infrastructure, we treat the inner triplet as thin lens quadrupoles~\cite{Willeke:1988zu}, while the separator dipole field is treated classically. These magnets tend to deflect lighter, softer mesons away from FASER, and so the neutrino spectrum below $\sim 100~\gev$ is dominated by neutrinos produced in the vacant first 20 m of forward infrastructure. Higher energy mesons are focused towards the beamline by the inner triplet, then deflected away by the dipole. As a result, most of the neutrinos above $1~\tev$ interacting in FASER are expected to be produced in the inner triplet region, while a subleading fraction comes from the dipole or unmagnetized regions.

Neutrinos are produced in the weak decay of the lightest mesons and baryons of a given flavor. A list of the considered particles, the MC generators used to simulate them, and their main decay channels into neutrinos can be found in \tableref{processes}. Note that except for pions and charged kaons, we simulated the meson decays using \textsc{Pythia~8} and therefore also considered additional subleading decay channels not shown in \tableref{processes}.  Additional production modes, for example, from the decay of $D$ mesons produced at the TAN or in collisions of the beam halo with the beam pipe, require a more careful study but are expected to be subleading. 

\begin{table}[t]
  \centering
  \begin{tabular}{|c|c|c|c|c|c|c|}
  \hline \hline
	{\bf Type} & \, {\bf Particles} \ & \, {\bf Main Decays} \ & \, {\sc E} \ & \, 
	{\sc  Q} \ & \, {\sc  S} \ & \, {\sc  P} \ \\ \hline \hline
    Pions   &  $ \pi^+$ &
    $\pi^+\to \mu \nu$ &
    $\surd$ &$\surd$ &$\surd$ & --- \\
    \hline
    Kaons   &  $ \K^+$, $K_S$, $K_L$ &
    $K^+ \to  \mu\nu$, $K \to \pi \ell \nu$ &
    $\surd$ &$\surd$ &$\surd$ & --- \\
    \hline
    Hyperons & $\Lambda$, $\Sigma^+$, $\Sigma^-$, $\Xi^0$, $\Xi^-$, $\Omega^-$ &
    $\Lambda\to  p \ell \nu$ &
    $\surd$ &$\surd$ &$\surd$ & --- \\
    \hline
    Charm   &  $D^+$, $D^0$, $D_s$, $\Lambda_c$, $\Xi_c^0$, $\Xi_c^+$ &
    $D \to  K\ell\nu$, $D_s \to  \tau \nu$,  $\Lambda_c \to \Lambda \ell \nu$ &
    ---  &---  & $\surd$& $\surd$ \\
    \hline
    Bottom  &  $B^+$, $B^0$, $B_s$, $\Lambda_b$, $\dots$ &
    $B \to D\ell\nu$,  $\Lambda_b \to  \Lambda_c \ell \nu$ &
    ---  &---  & ---& $\surd$ \\
    \hline \hline
  \end{tabular}
  \caption{Decays considered for the estimate of forward neutrino production.  For each type in the first column, we list the considered particles in the second column and the main decay modes contributing to neutrino production in the third column. In the last four columns we show which generators were used to obtain the meson spectra: \textsc{Epos-Lhc} (E)~\cite{Pierog:2013ria}, \textsc{Qgsjet-ii-04} (Q)~\cite{Ostapchenko:2010vb},  \textsc{Sibyll 2.3c} (S)~\cite{Ahn:2009wx, Riehn:2015oba, Riehn:2017mfm, Fedynitch:2018cbl}, and \textsc{Pythia~8} (P)~\cite{Sjostrand:2006za, Sjostrand:2007gs}, using both the \textsc{Monash}-tune~\cite{Skands:2014pea} and the minimum bias \textsc{A2}-tune~\cite{ATLAS:2012uec}.} 
\label{table:processes}
\end{table}

We use our simulation to obtain the neutrino flux as a function of energy and angle with respect to the beam collision axis for all flavours.\footnote{This implies that the neutrino flux is treated as being symmetric around the beam collision axis. This symmetry will be broken by the dipole magnets, although this effect is expected to be small since the dipole magnets quickly deflect charged particles away from the beam collision axis.} The energy spectra of neutrinos going through FASER$\nu$, with a cross sectional area of $25~\cm \times 25~\cm$ centered around the beam collision axis, are shown in \figref{spectra}. The shaded band in \figref{spectra} indicates the range of predictions obtained from different MC simulators, while the solid line corresponds to their average. We separate the different processes contributing to neutrino production into pion (blue), kaon (red), hyperon (orange), charm (dark green), and bottom (light green) decays. 
The 2-body decays of charged pions $\pi^\pm \to \mu \nu_{\mu}$ and kaons $K^\pm \to \mu \nu_{\mu}$ are the dominant sources of $\nu_{\mu}$ production. Note that kaon decays provide a larger contribution at higher energies due to the larger fraction of the parent meson energy obtained by the neutrino in kaon decays. Electron neutrinos are predominantly produced in 3-body kaon decays $K \to \pi e \nu_e$, with $K^\pm$, $K_L$, and $K_S$ providing similar contributions. 

Hyperon decays only provide a subleading contribution, with the notable exception of $\bar{\nu}_e$ production through the decay $\Lambda \to p e \bar{\nu}_e$, due to the enhanced forward $\Lambda$ production rate.  Charm decays can provide a sizable contribution to both $\nu_e$ and $\nu_\mu$ production, both through $D$ meson decays, as well as $\Lambda_c^+ \to \Lambda \ell^+ \nu$ decays. Tau neutrino production mainly proceeds through both the decay $D_s \to \tau \nu_\tau$ and subsequent $\tau$ decay. The decay of beauty hadrons does not constitute a sizable source of neutrino production in the forward direction, due to both the lower production rate of $B$ particles and the fact that the resulting neutrinos have a broader distribution of transverse momentum with $p_T \sim m_b$.

\begin{figure}[tbp]
\centering
\includegraphics[width=0.99\textwidth]{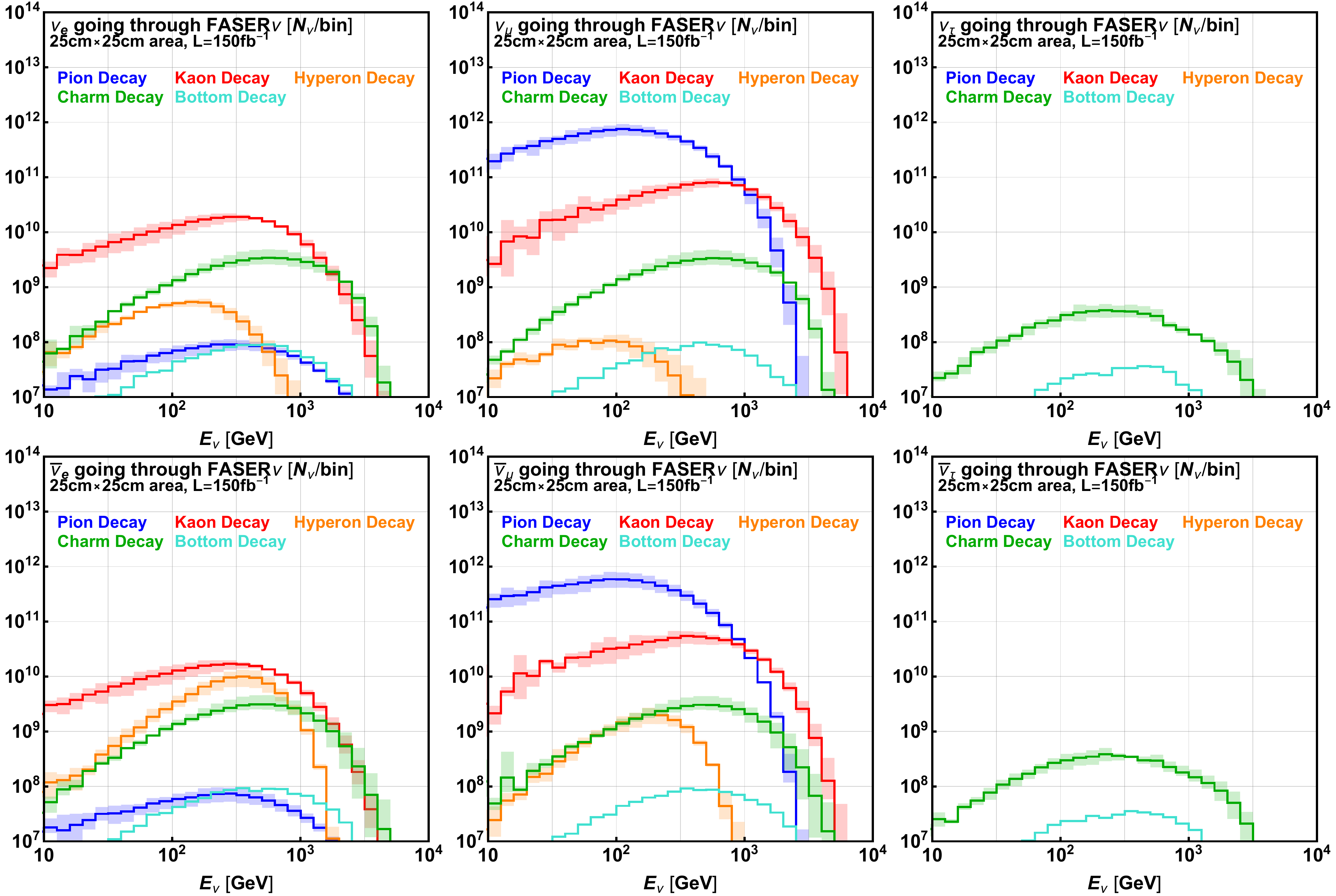} 
\caption{The estimated number of neutrinos that pass through the $25~\cm \times 25~\cm $ transverse area of FASER$\nu$, assuming an integrated luminosity of $150~\ifb$ for Run 3 at the 14 TeV LHC.  The event rates are for electron (left), muon (center), and tau (right) neutrinos (upper) and anti-neutrinos (lower). The shaded bands indicate the range of predictions from the different MC generators listed in \tableref{processes}, and the solid contours are the average results of these MC generators.}
\label{fig:spectra}
\end{figure}

\subsection{Neutrino Interactions with the Detector}
\label{sec:rates}

At the energies relevant for this study, $E_\nu > 100~\gev$, neutrino interactions can be described by DIS~\cite{Formaggio:2013kya,Gandhi:1995tf}. The corresponding differential cross section for neutrino scattering on protons has the form
\be
\frac{d\sigma_{\nu p}}{dx \ dy} = \frac{G_F^2 m_p E_\nu}{ \pi} \frac{m_{W,Z}^4}{(Q^2+m_{W,Z}^2)^2} \times  \big[ x f_{q}(x,Q^2) + x f_{\bar{q}}(x,Q^2)(1-y)^2 \big ] \ ,
\ee
where $x$ is the fraction of the proton's momentum carried by the quark in the initial state, $y$ is the fraction of the neutrino momentum transferred to the hadronic system, $Q^2= 2 E_\nu m_p x y$ is the transferred four-momentum, and $f_{q}(x,Q^2)$ are the proton parton distribution functions. For neutrino energies $E_\nu \sim \tev$, the cross section peaks at $x \sim 0.1$, $y\sim 0.5$, and $Q^2 \sim (10~\gev)^2$. In this regime, $Q^2 \ll m_{W,Z}^2$, the corresponding interaction cross section is roughly proportional to the neutrino energy $\sigma_{\nu N} \sim E_\nu$. This results in an energy spectrum of interacting neutrinos that peaks at higher energies. 

FASER$\nu$ will measure the interaction cross section of neutrinos with a tungsten nucleus, $\sigma_{\nu N}$. Ignoring nuclear effects in heavy nuclei, this cross section is given by $\sigma_{\nu N} = n_p \sigma_{\nu p} + n_n \sigma_{\nu n}$, where $n_p$ and $n_n$ are the number of protons and neutrons in the nucleus. However, for neutrino scattering on heavy nuclei, nuclear effects such as shadowing, anti-shadowing, and the EMC effect, become important. Although these effects can in principle be taken into account by using nuclear parton distribution functions~\cite{Eskola:2016oht, Kusina:2015vfa, AbdulKhalek:2019mzd}, we note that they only lead to small (few \%) changes in the cross section, and we therefore do not include then in this analysis~\cite{FASERvTP}.

In the left panel of \figref{interacting}, we show the DIS cross section, which we calculate using the \textsc{NNPDF3.1nnlo} parton distribution functions~\cite{Ball:2017nwa}. At lower energy, $E_\nu\lesssim 10~\gev$, the neutrino cross section will be dominated by additional quasi-elastic scattering and resonant production processes, while at higher energies, $E_\nu \gtrsim 10~\tev$, the momentum transfer will surpass the weak boson masses, $Q^2 > m_{W}^2$, and the cross section is expected to fall as $\sigma \sim Q^{-4} \sim E_\nu^{-2}$. Although the $\nu_e$ and $\nu_\mu$ cross sections are the same, the $\nu_\tau$ CC cross section experiences an additional kinematic suppression due to the $\tau$ lepton mass being close to its production threshold~\cite{Jeong:2010nt}; this suppression disappears by $E_\nu\sim 1~\tev$. The anti-neutrino interaction cross sections are additionally suppressed due to helicity by roughly a factor $2$.

\begin{figure}[tbp]
\centering
\includegraphics[width=0.323\textwidth]{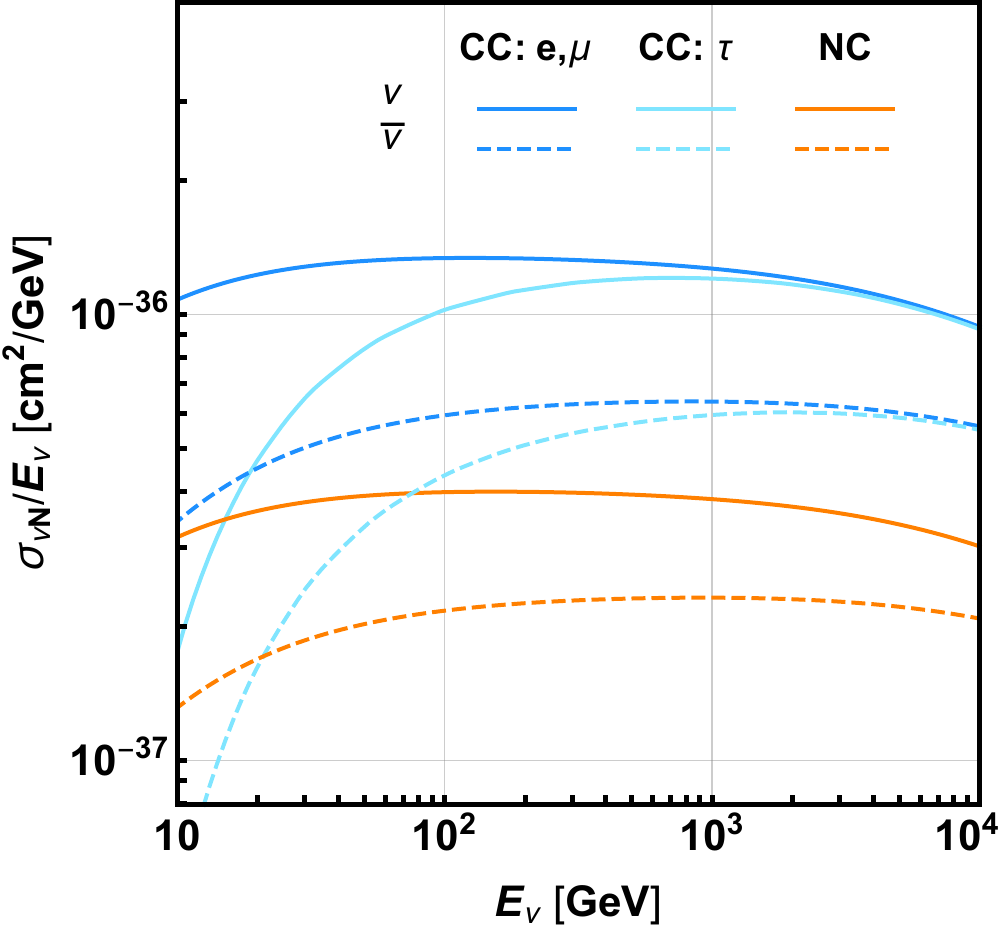}
\includegraphics[width=0.320\textwidth]{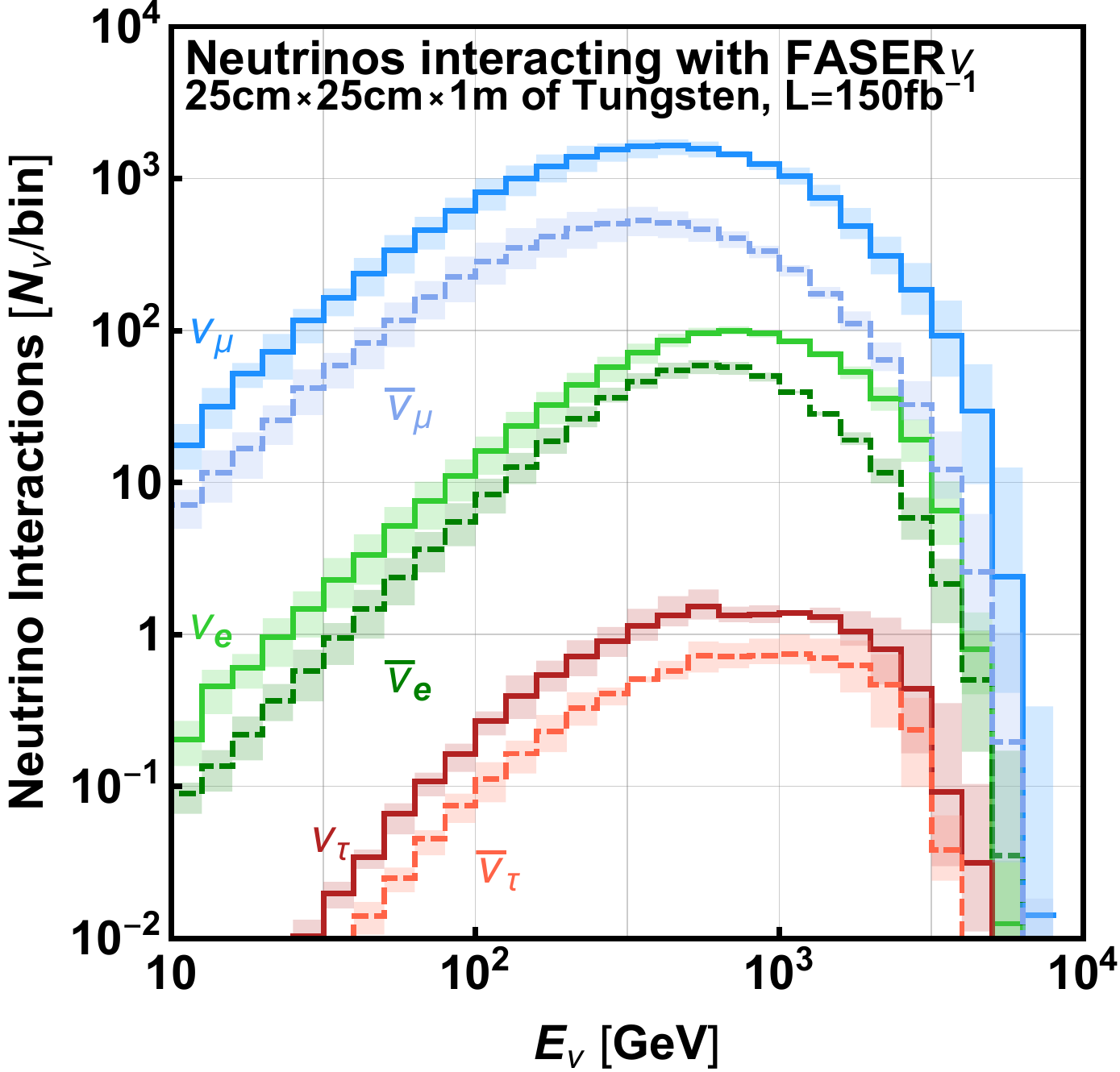}
\includegraphics[width=0.314\textwidth]{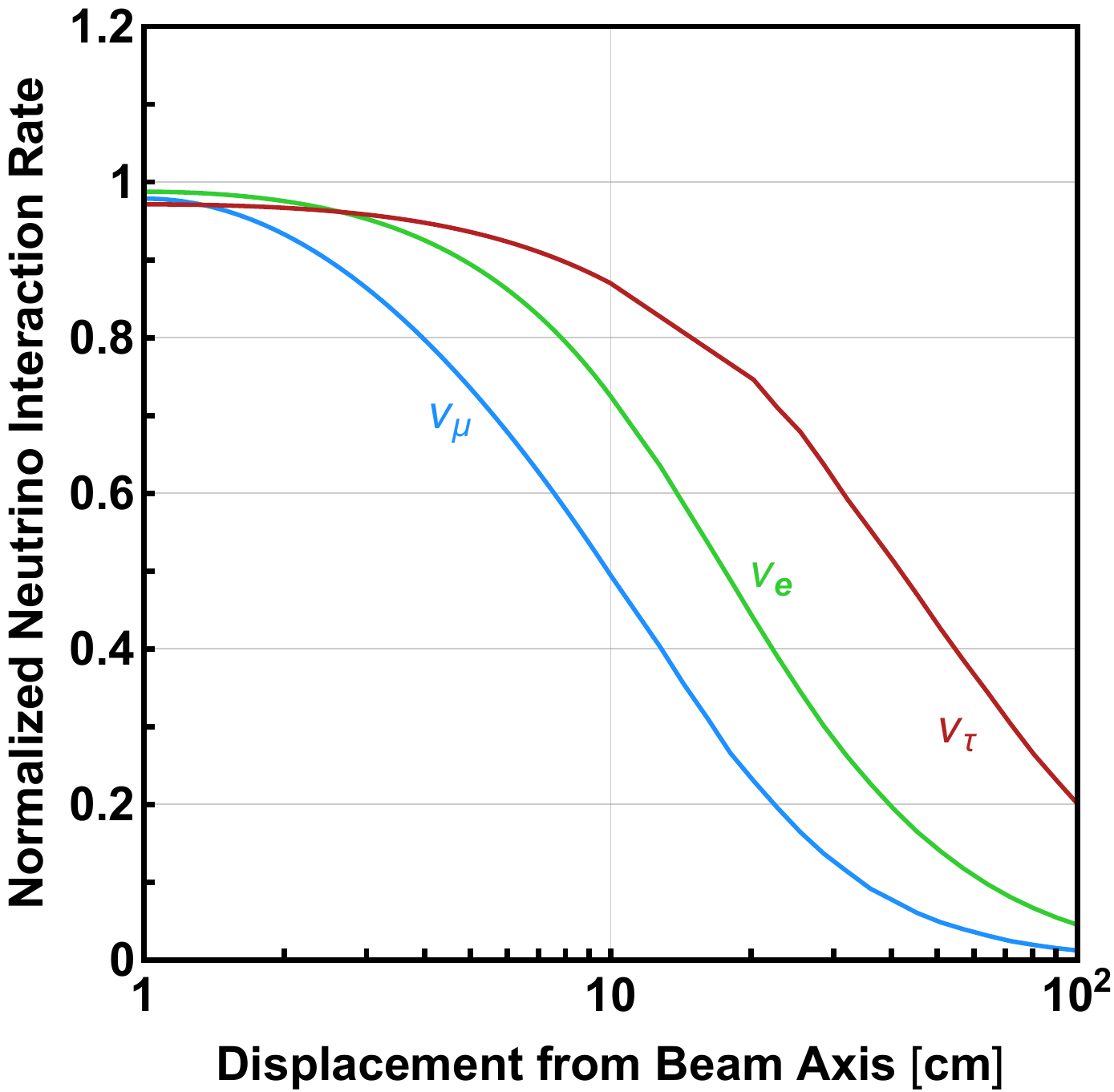}
\caption{\textbf{Left}: The $\nu N$ (solid) and $\bar{\nu} N$ (dashed) DIS cross sections, where $N$ is a tungsten nucleus, calculated with the NNPDF3.1 PDF~\cite{Ball:2017nwa}. \textbf{Center}: The energy spectrum of neutrinos with CC interactions in a 1-ton tungsten detector with dimensions $25\,\cm \times 25\,\cm \times 1\,\m$ centered on the beam collision axis at the FASER location at the 14 TeV LHC with $150~\ifb$. \textbf{Right}: The neutrino interaction rate per unit area normalized to the prediction at the beam collision axis for a detector with large radius.  
} 
\label{fig:interacting}
\end{figure}

In the following, we consider a benchmark detector of pure tungsten with dimensions $25\,\cm \times 25\,\cm \times 1\,\m$, which has a density $\rho=19.3~\g/\cm^3$, resulting in a total detector mass $m_\text{det}=1.2$ tons. (The actual FASER$\nu$ detector has the same total tungsten target mass, but is extended by 35 cm in length by the inclusion of emulsion films.) The probability of a neutrino to interact with the detector is then given by
\begin{equation}
P = \frac{\sigma_{\nu N} \times \text{Number of Nuclei}}{\text{Detector Area}}
= \frac{\sigma_{\nu N}}{A} \frac{m_\text{det}}{m_N} 
= \sigma_{\nu N} \ \frac{ \rho L }{ m_N} \ , 
\end{equation}
where $N$ is the target nucleus, $L$ is the detector length, and $m_N$ is the mass of the target nucleus. The number of neutrinos interacting with FASER$\nu$, which again is calculated as function of the neutrino energy and and angle, is given by the product of the number of neutrinos passing through FASER$\nu$ and the interaction probability.

In the central panel of \figref{interacting}, we show the energy spectrum of neutrinos interacting in the benchmark detector, assuming an integrated luminosity of $150~\ifb$ for Run 3 at the 14 TeV LHC. Here we have combined all the different neutrino production modes. In \tableref{intrate} we present the expected event rates at FASER$\nu$. In the second column we show the expected number of neutrinos with energy $E_\nu>100~\gev$ that interact with the detector. In the third column we additionally take into account the acceptance rate to reconstruct a vertex by requiring the interaction to have at least 5 charged tracks (see discussion in \secref{features}). The last column shows the mean energy of the neutrino that interact in FASER$\nu$.

\begin{table}[t]
\centering
\begin{tabular}{|c|c|c|c|}
\hline
& Number of CC Interactions & Number of Reconstructed Vertices & Mean Energy  
\\
\hline
$\nu_e$+$\bar{\nu}_e$ & $1296^{+77}_{-58}$ & $1037^{+52}_{-36}$ & 827~GeV \\ 
\hline
$\nu_\mu$+$\bar{\nu}_\mu$  &$20439^{+1545}_{-2314}$ & $15561^{+1103}_{-1514}$  & 631~GeV\\ 
\hline
$\nu_\tau$+$\bar{\nu}_\tau$ & $21^{+3.3}_{-2.9}$ & $ 17^{+2.6}_{-2.6}$ & 965~GeV  \\ 
\hline
\end{tabular}
\caption{The expected number of neutrinos with $E_{\nu} > 100~\gev$ interacting through CC processes in FASER$\nu$, the expected number of reconstructed vertices in FASER$\nu$ requiring $n_{\text{tr}} \ge 5$, and the mean energy of neutrinos that interact in FASER$\nu$.  Here we assume a benchmark detector made of tungsten with dimensions $25\,\cm \times 25\,\cm \times 1\,\m$ at the 14 TeV LHC with an integrated luminosity of $L=150~\ifb$. Reductions in the number of reconstructed vertices from the geometrical acceptance and lepton identification efficiency have not been included.  The uncertainties correspond to the range of predictions obtained from different MC generators. 
}
\label{table:intrate}
\end{table}

In the right panel of \figref{interacting} we consider a detector with very large radius and show the expected neutrino interaction rate per unit area in this detector as a function of the distance from the beam collision axis. The rates are normalized to the interaction rate per area at the beam collision axis. We can see that the $\nu_{\mu}$ beam is the most collimated with an event rate that falls off steeply for distances from the beam axis larger than about $10~\cm$. Electron neutrinos are mainly produced in kaon decay and inherit a slightly broader spectrum. Tau neutrinos are produced in the decay of heavy flavor hadrons, resulting in the least collimated beam.\footnote{The measurement of the radial distribution of neutrino events might provide additional physics insight, but it is also strongly affected by the forward LHC magnets. A dedicated study is needed to investigate if the observed radial distribution of neutrino events can be related to the $p_T$ distribution at the IP.} 

In contrast, the flux of high-energy muons increases away from the beam collision axis. These muons, coming from the direction of the IP, are deflected by the LHC magnets causing an effective ``shadow" along the beam collision axis and hence at the FASER location. As we will see below in \secref{bg-environment}, a FLUKA background study predicts an increase in the muon rate (and muon-associated backgrounds) of roughly a factor of $10-100$ between the beam collision axis and a location $\sim 1~\m$ displaced from it. We therefore see that a small detector, with a radial size of $\mathcal{O}(10~\cm)$ placed on the beam collision axis has many advantages for neutrino searches.

\section{The FASER\texorpdfstring{$\nu$}{nu} Detector}
\label{sec:detector}

\subsection{Detector Design}
\label{sec:design}

To study all three neutrino flavors, the detector should be able to identify the $e$, $\mu$, and $\tau$ leptons produced by CC neutrino interactions. For this, we choose an emulsion detector. Emulsion detectors~\cite{Nakamura:2006xs} have spatial resolutions down to 50 nm, better than all other particle detectors. A typical emulsion detector consists of silver bromide crystals with diameters of 200 nm dispersed in gelatin media. Each crystal works as an independent detection channel. The high density of detection channels (of the order of $10^{14}/\cm^3$) makes emulsion detectors unique for detecting short-lived particles. Emulsion detectors have been successfully employed by several neutrino experiments such as CHORUS~\cite{Eskut:2007rn}, DONuT~\cite{Kodama:2007aa}, and OPERA~\cite{Acquafredda:2009zz, Agafonova:2014ptn, Agafonova:2015jxn, Agafonova:2018auq}. The same technique has also been used in hadron experiments, e.g., WA75~\cite{Albanese:1985wk} and E653~\cite{Kodama:1993wg} for beauty particle studies, and recently DsTau~\cite{Aoki:2017spj, Aoki:2019jry} for charmed particle measurements.

The \FASERnu neutrino detector will be placed in front of the FASER main detector, as shown in \figref{detector_location}. It will be on the collision axis to maximize the number of neutrino interactions. \Figref{module_structure} shows a magnified view of the neutrino detector modules. The detector is made of a repeated structure of emulsion films interleaved with 1-mm-thick tungsten plates. The emulsion film is composed of two emulsion layers, each 70 $\mu$m thick, that are poured onto both sides of a 200-$\mu\text{m}$-thick plastic base; the film has an area of $25\,\cm \times 25\,\cm$. The tungsten target was chosen because of its high density and short radiation length, which helps keep the detector small and also localizes electromagnetic showers in a small volume. The whole detector has a total of 1000 emulsion films, with a total tungsten target mass of 1.2 tons, and its length corresponds to 285 radiation lengths $X_0$ and 10.1 hadronic interaction lengths $\lambda_\text{int}$.

\begin{figure}[tpb]
\centering
\includegraphics[width=0.9\textwidth]{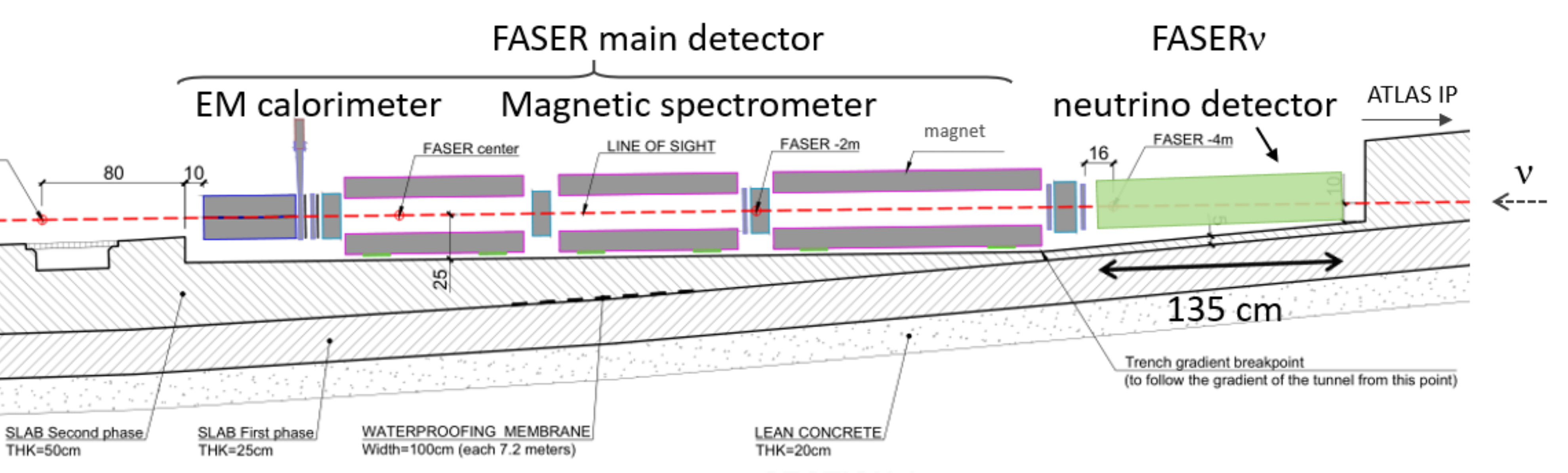}
\caption{Side view of the FASER main detector and FASER$\nu$ in side tunnel TI12.
}
\label{fig:detector_location}
\end{figure}

\begin{figure}[tpb]
\centering
\includegraphics[width=0.85\textwidth]{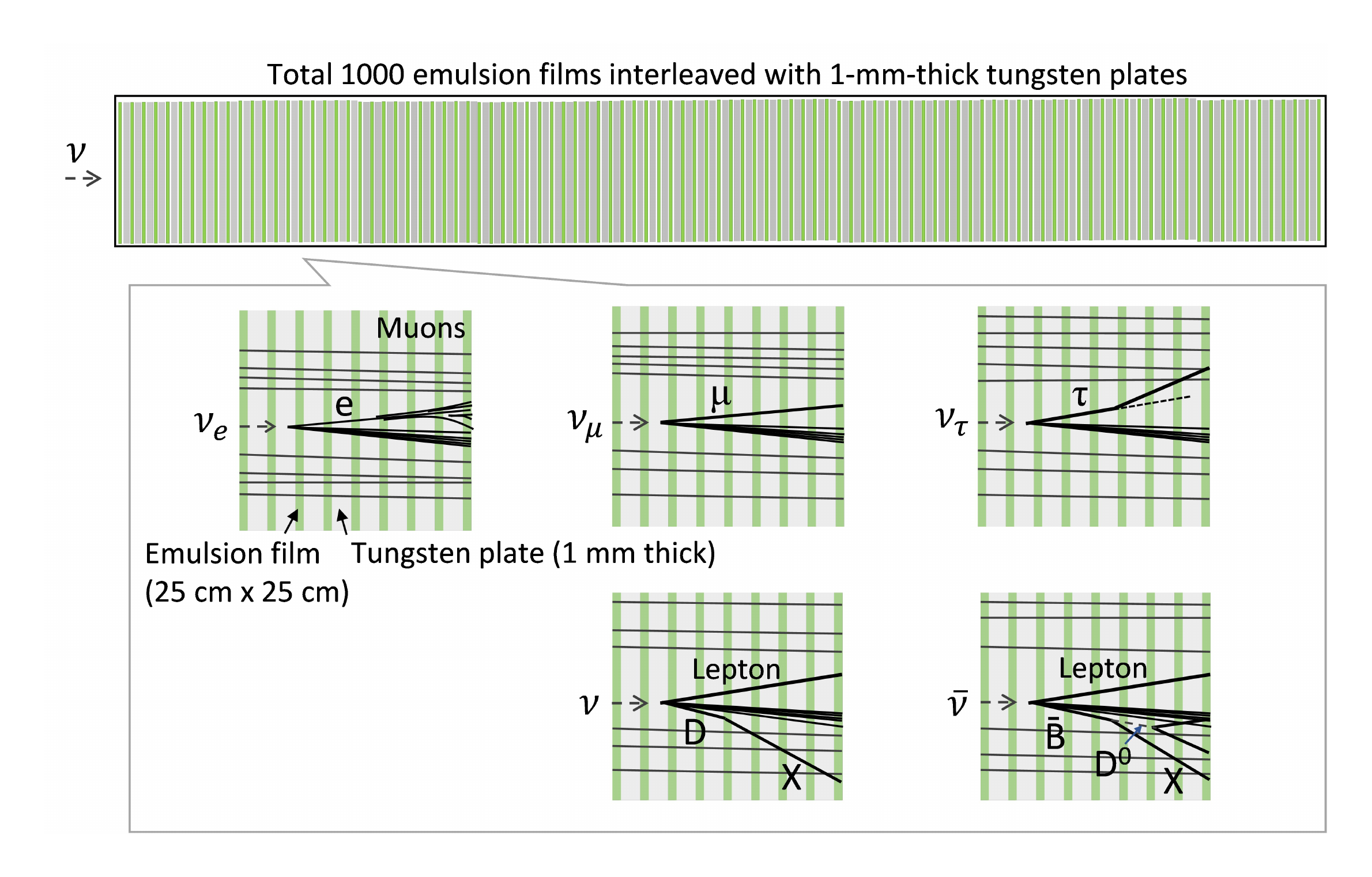}
\caption{Schematic of the detector structure and the topology of various neutrino signal events that can be seen in the detector.
}
\label{fig:module_structure}
\end{figure}

As the emulsion detector doesn't have time resolution and records all charged particle trajectories, the pile up of events is an issue. The emulsion detector readout and reconstruction works for track densities up to $\sim 10^6$ tracks/\si{cm^2}. To keep the detector occupancy low, the emulsion films will be replaced during every Technical Stop of the LHC, which will take place about every 3 months. This corresponds to  $10-50~\ifb$ of data in each data-taking period.  We performed \textit{in situ} measurements in 2018 (see \secsref{bg}{firstnu}), which measured a charged particle flux of $\Phi \approx 3 \times 10^4~\fb/\cm^2$ at the FASER location. When removed, the track density of the emulsion detectors will be roughly $0.3-1.5 \times 10^6~\text{tracks}/\cm^2$. Our experience with the \textit{in situ} measurements in 2018 further demonstrated that we can analyze the emulsion detector in this detector environment.  Assuming seven replacements during LHC Run 3 (one in 2021 and three replacements in each of 2022 and 2023), a total emulsion surface area of $440~\m^2$ will be used. The detector is being designed for easy transport across the LHC beamline and into and out of the TI12 trench, given that the emulsion detector will have to be replaced in four days or less.

On the other hand, thanks to the high density of TeV-energy muons, the emulsion films can be aligned precisely. Experience from the DsTau experiment, which has a similar track density of 400 GeV protons, shows that the position resolution of each hit in the emulsion detector is $\sigma_{\text{pos}}=0.4~\mu\text{m}$~\cite{Aoki:2019jry}. The angular resolution $\sigma_{\text{angle}}$ in the FASER$\nu$ detector depends on the track length $L_{\text{tr}}$, and it at least reaches
\be
\sigma_{\text{angle}} = \sqrt{2}\; \sigma_{\text{pos}} / L_{\text{tr}} \ .
\label{eq:angular_resolution}
\ee
For a particle with $L_{\text{tr}}=1~\cm$, the angular resolution would be better than $\sigma_{\text{angle}}=0.06~\mrad$.

The neutrino event analysis will be based on readout of the full emulsion detectors by the Hyper Track Selector (HTS) system~\cite{Yoshimoto:2017ufm}. HTS has a readout speed of  $0.5~\m^2$/hour/layer, which makes it possible to analyze $1000~\m^2$/year of double-sided emulsion films effectively. After reading out the full area of the emulsion films, a systematic analysis will be performed to locate neutrino interactions. 

As shown in \figref{module_structure}, the classification of $\nu_{e}$, $\nu_{\mu}$, and $\nu_{\tau}$ CC interactions is made possible by identifying the $e$, $\mu$, and $\tau$ leptons produced in the interactions. Electrons are identified by detecting electromagnetic showers along a track. If a shower is found, the first film with activity will be checked to see if there is a single particle (an electron) or an $e^+ e^-$ pair (from the conversion of a $\gamma$ from a $\pi^0$ decay). The separation of single particles from particle pairs will be performed based on the energy deposit measurements.\footnote{In addition to the track position and direction, emulsion detectors can measure the energy deposit $dE/dx$, which is nearly linearly proportional to the number of dots along the trajectory. Therefore a single minimum ionizing particle (MIP) can be separated from multiple MIPs.} Muons are identified by their track length in the detector.  Since the detector has a total nuclear interaction length of $10.1\lambda_{\text{int}}$, all the hadrons from the neutrino interactions will interact, except for hadrons created in the far downstream part of the detector. Tau leptons are identified by detecting their short-lived decays.  Charm and beauty particles, with $c\tau \sim 100-500~\mu\text{m}$, are also identified by their decay topology. 

Neutral current (NC) events are primarily defined as zero-lepton events. Those are partly contaminated by CC events with a misidentified lepton or by neutral hadron interactions. NC events require further study, and in the following discussion, we will mainly focus on CC events.

\subsection{Features of High-Energy Neutrino Interactions}
\label{sec:features}

As we have seen before, the typical energy of neutrinos interacting within FASER$\nu$ will be between a few $100~\gev$ and a few $\tev$. Let us therefore first look at some key features of high-energy neutrino interactions, which we can then use to identify a neutrino interaction and measure the neutrino energy.

For the studies presented in this work, the \textsc{Genie} Monte Carlo v2.12.10 simulation package~\cite{Andreopoulos:2009rq, Andreopoulos:2015wxa} was used to generate events with the required neutrino flux and flavor spectrum interacting with a tungsten target nucleus. The ``ValenciaQEBergerSehgalCOHRES'' model configuration was used, but we note that the differences between \textsc{Genie} model configurations are focused on low-energy transfer physics ($\nu \sim 1$ GeV), not on the high-energy transfers that make up the majority of interactions in \FASERnu. In \textsc{Genie}, the deep-inelastic scattering model, which dominates the cross section at the relevant neutrino energies, is based on Bodek-Yang~\cite{Bodek:2002ps}, which uses parton distribution functions tuned to data from $e$/$\mu$ scattering on hydrogen and deuterium targets. Nuclear corrections are included based on charged lepton and neutrino scattering data on iron targets~\cite{Bodek:2002vp}. Hadronization is governed by \textsc{Genie}'s ``AKGY'' model~\cite{Yang:2009zx}, which uses PYTHIA6~\cite{Sjostrand:2006za} for the invariant masses $W \geq 3$ GeV that are most relevant for \FASERnu. The high-$W$ model is tuned to BEBC data on measured neutrino--hydrogen particle multiplicities~\cite{Allen:1981vh}, although other authors have found better agreement with other datasets by retuning the model~\cite{Katori:2014fxa}. Final State Interactions, where particles produced at the interaction vertex can re-interact before leaving the nucleus, are modelled with \textsc{Genie}'s INTRANUKE ``hA'' intranuclear cascade model~\cite{intranuke}, which is tuned to $\pi$--iron and $\pi$--nucleon scattering data, and is then extrapolated to other targets with $A^{2/3}$ scaling, where $A$ is the atomic number. An important deficiency of \textsc{Genie}'s intranuclear cascade is that particles are transported through the nucleus independently, so cannot feel the potential from other interaction products. A comparison of the various available simulation packages and $\pi$--nucleus scattering data can be found in Ref.~\cite{PinzonGuerra:2018rju}. \textsc{Genie} provides default cross section splines that range up to 100 GeV. To generate neutrino interactions at higher energies, the cross section splines were pre-computed by using a dedicated tool implemented in \textsc{Genie} (gmkspl).

In \figref{genie_plots} we show distributions of topological and kinematic features of high-energy $\nu_{\mu}$ interactions simulated by \textsc{Genie}. The upper two panels of \figref{genie_plots} show the multiplicities of charged tracks $n_{\text{tr}}$ and gamma rays $n_\gamma$ with momentum $p_{\text{tr}}>0.3~\gev$ and angle $\theta<45^\circ$ (relative to the neutrino direction) associated with a neutrino interaction vertex. These quantities are known to be correlated with $W$ through the relation $n_{\text{tr}} \propto \text{const} + \log W^2$ (the so-called Feynman scaling~\cite{GrosseOetringhaus:2009kz}).  Above 100 GeV, the track multiplicity grows slowly with the neutrino energy. The middle left panel shows the inverse of the emission angle $\tan\theta$, also called the slope, of the lepton with respect to the incoming neutrino direction. The average transverse momentum of the lepton in neutrino interactions is expected to be $\langle P_\perp \rangle \simeq 30$ GeV at $E_\nu=1$ TeV. The lepton emission angle follows an inverse relation $\tan\theta\sim \left< P_\perp \right> /E_\nu$. The middle right panel shows the inverse of the median of slopes of all charged tracks, which has a behavior similar to the lepton slope.  The muon and sum of hadron momenta, shown in the lower panels of \figref{genie_plots}, are typically proportional to $E_\nu$. 

\begin{figure}[htpb]
\centering
\includegraphics[height=5.cm]{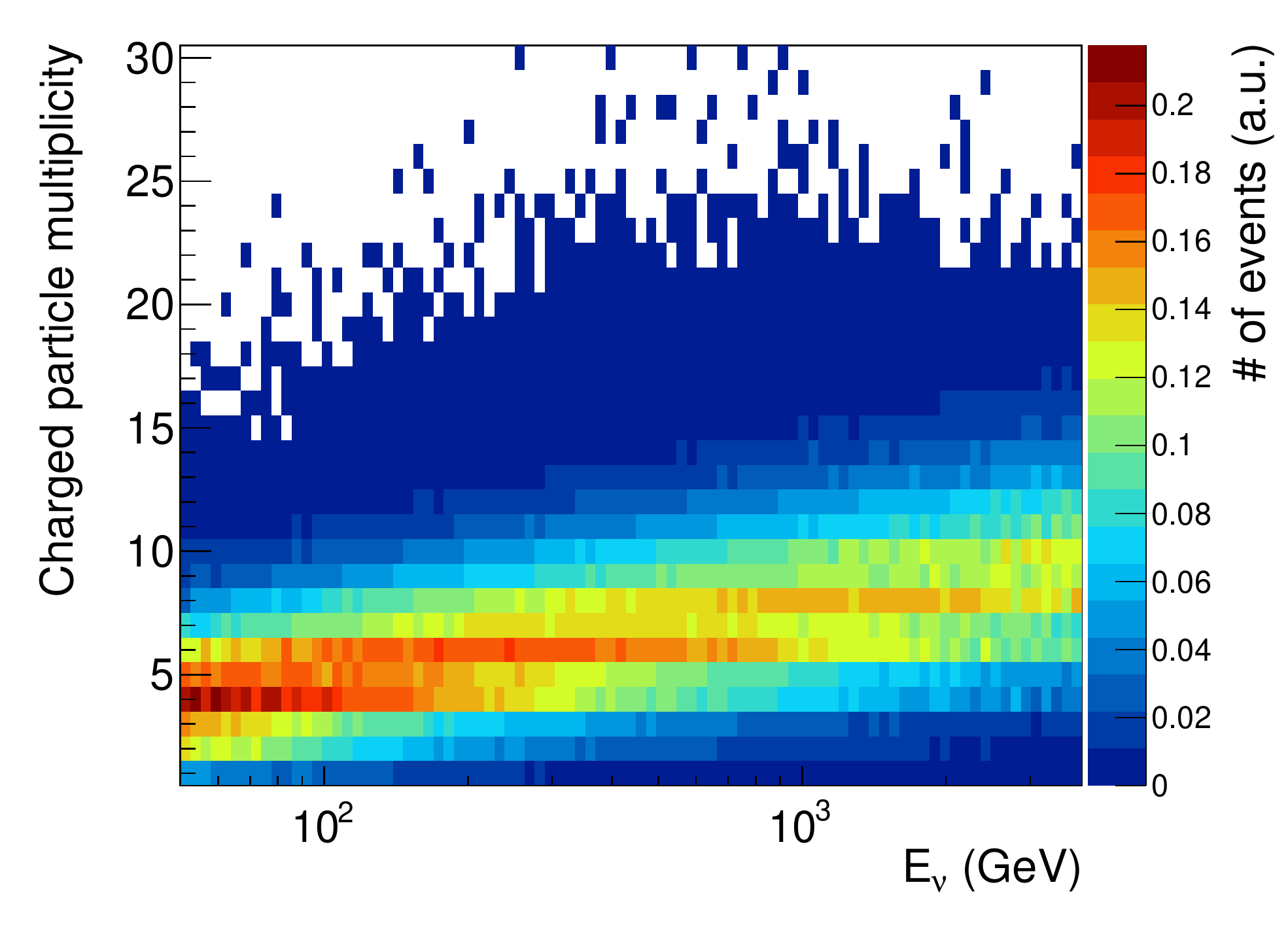}
\includegraphics[height=5.cm]{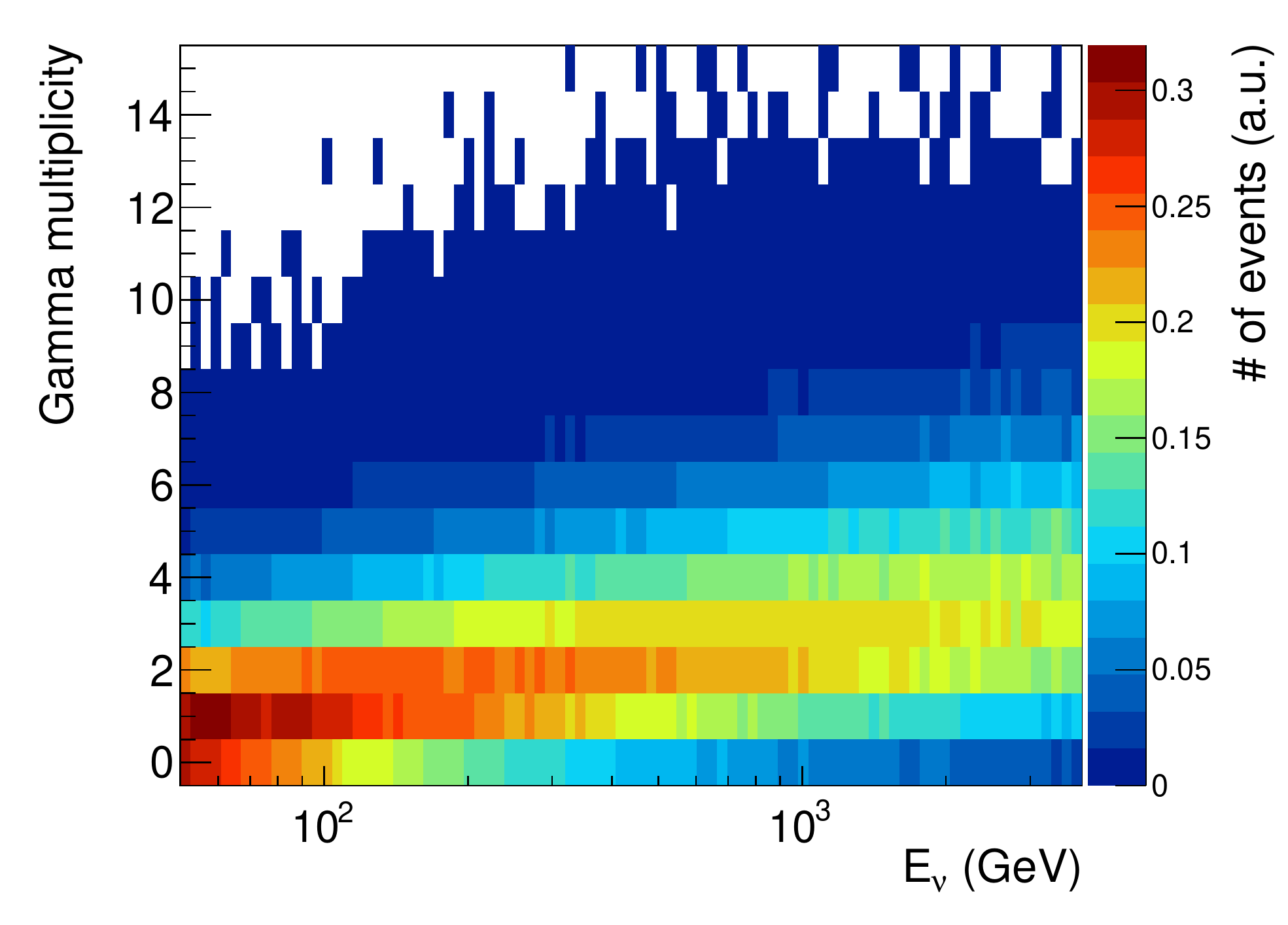}
\includegraphics[height=5.cm]{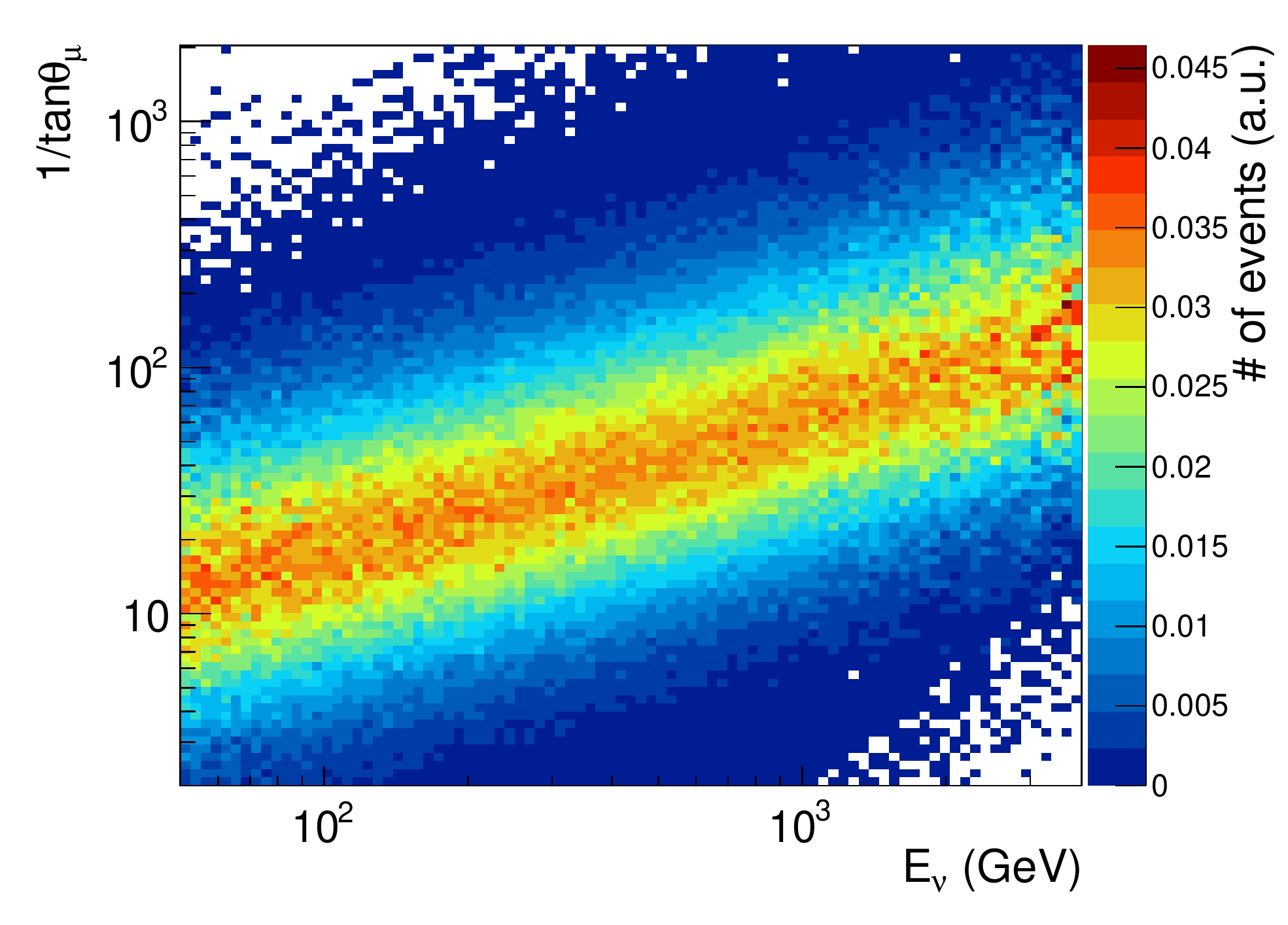}
\includegraphics[height=5.cm]{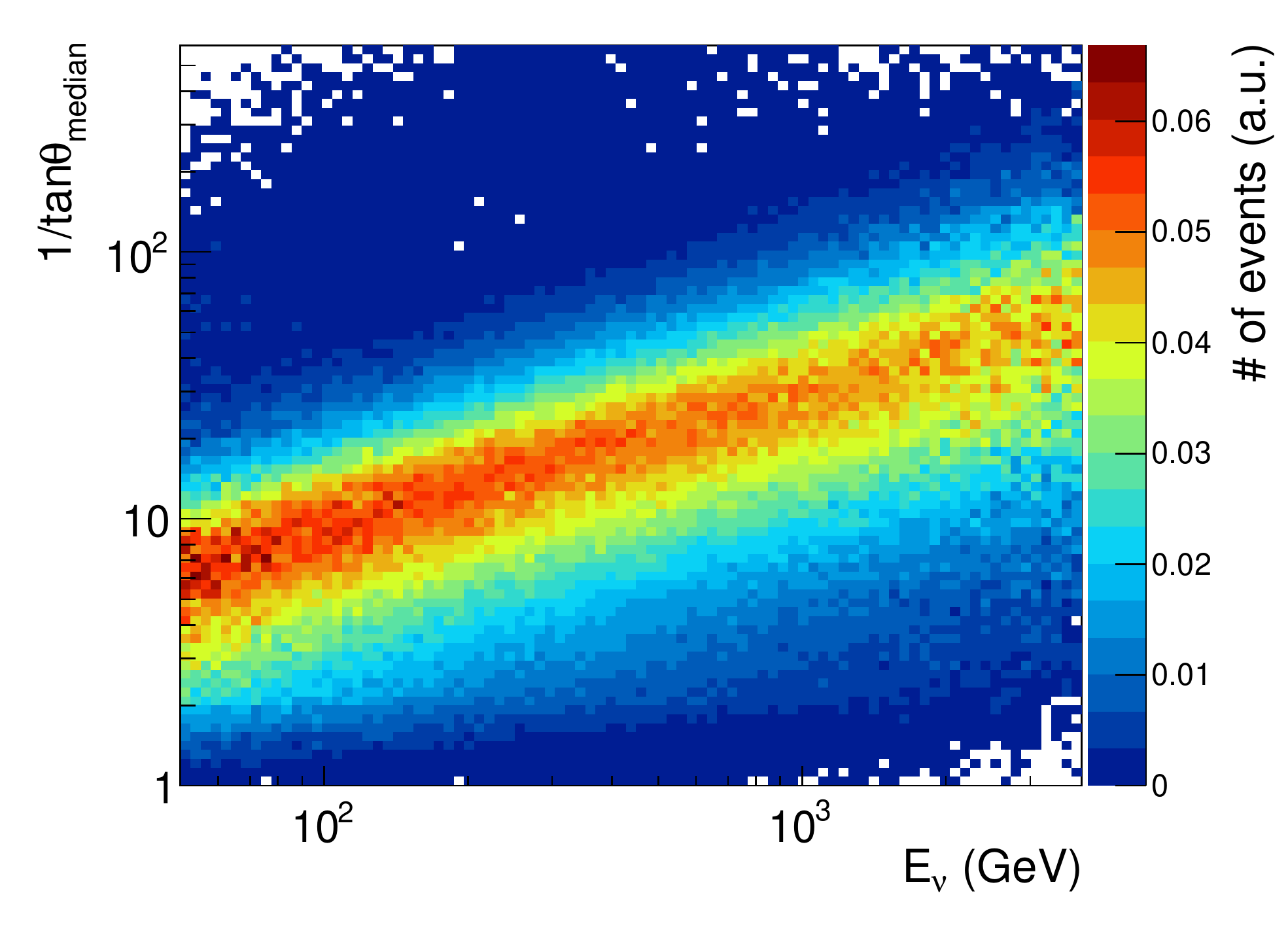}
\includegraphics[height=5.cm]{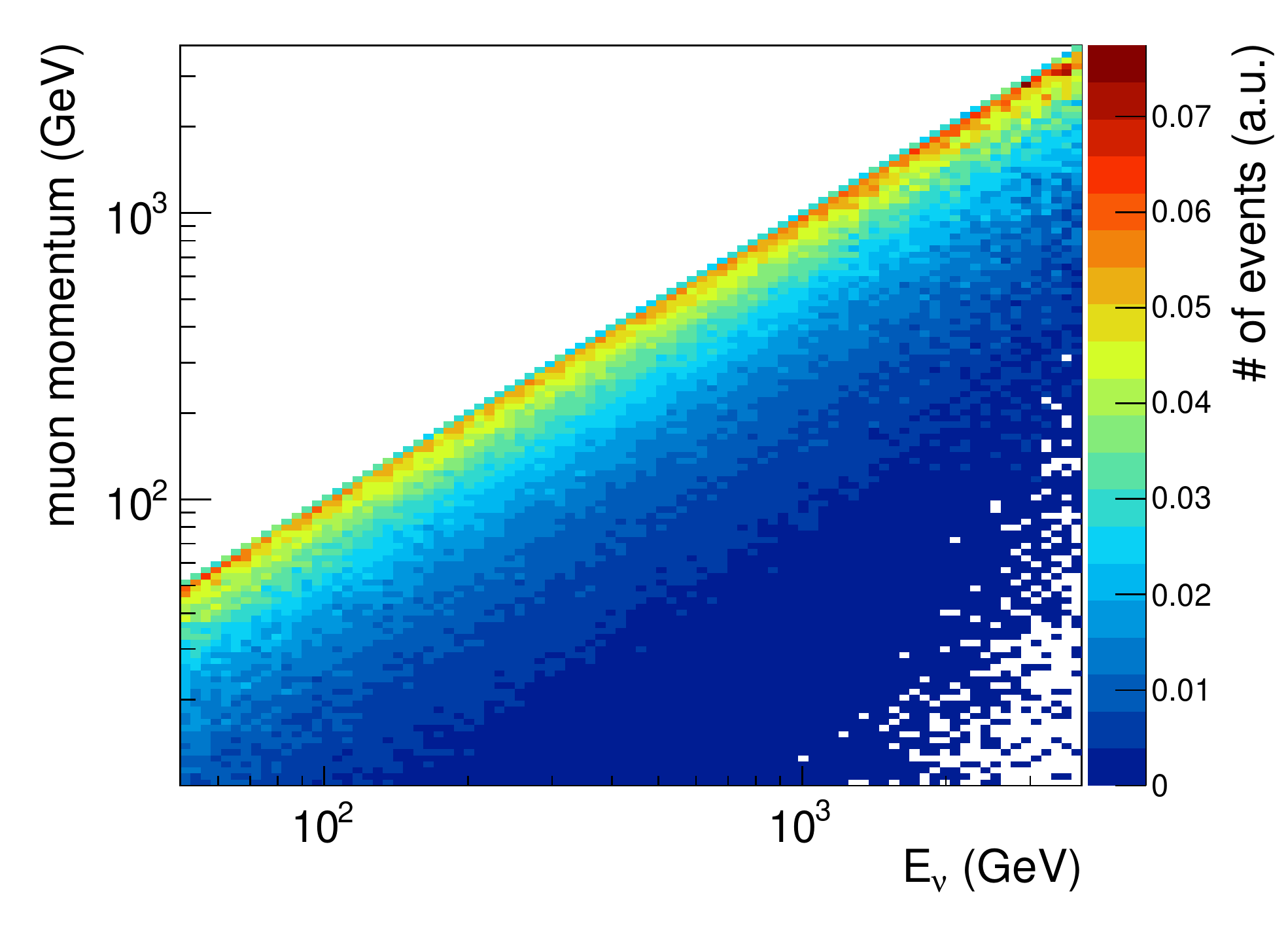}
\includegraphics[height=5.cm]{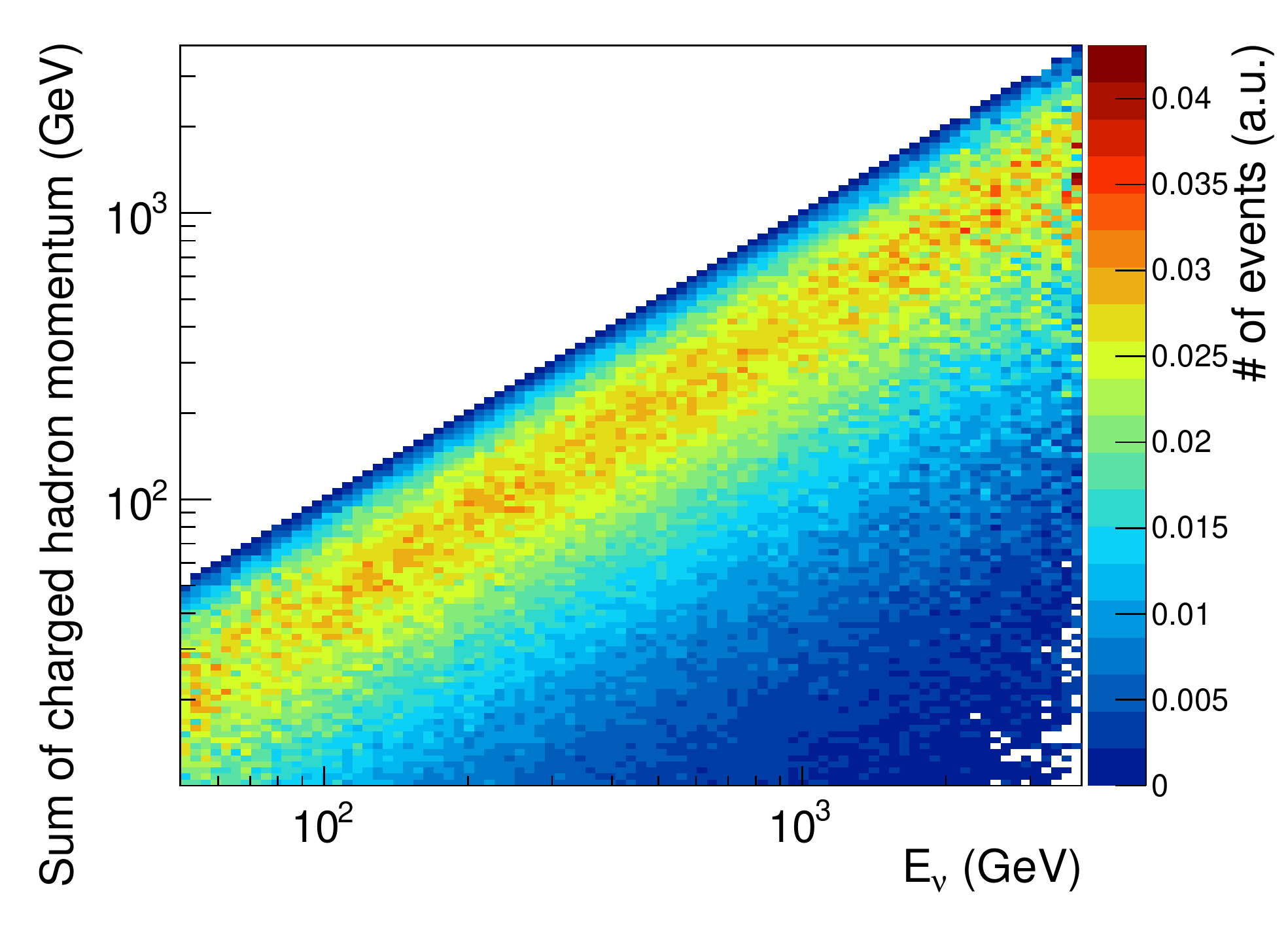}
\caption{Topological and kinematic features of neutrino interactions as a function of the simulated neutrino energy ("MC truth"). As an example, the distributions of $\nu_\mu$ CC interactions are shown. \textbf{Upper panels}: The multiplicity of charged particle tracks and $\gamma$ rays with momentum $p_{\text{tr}} > 0.3~\gev$ and angle $\theta<45^\circ$ relative to the neutrino direction. \textbf{Middle left}: Inverse of lepton slope.  \textbf{Middle right}: Inverse of median of charged particle slopes. \textbf{Lower left}: Lepton momentum.  \textbf{Lower right}: Sum of charged hadron momenta.
}
\label{fig:genie_plots}
\end{figure}

As we have seen, high-energy neutrino interactions typically lead to a sizable number of charged tracks that emerge from the interaction vertex. Taking the expected high track density into account, when searching for neutrino interactions we require a vertex to have at least five charged tracks $n_{\text{tr}} \ge 5$.  The resulting vertex detection efficiency as a function of the neutrino energy is shown in \figref{detection_eff} on the left. We see that for high-energy neutrino interactions with energies above $500~\gev$, the detection efficiency is above 80\%. 

\begin{figure}[tpb]
\centering
\includegraphics[height=5.5cm]{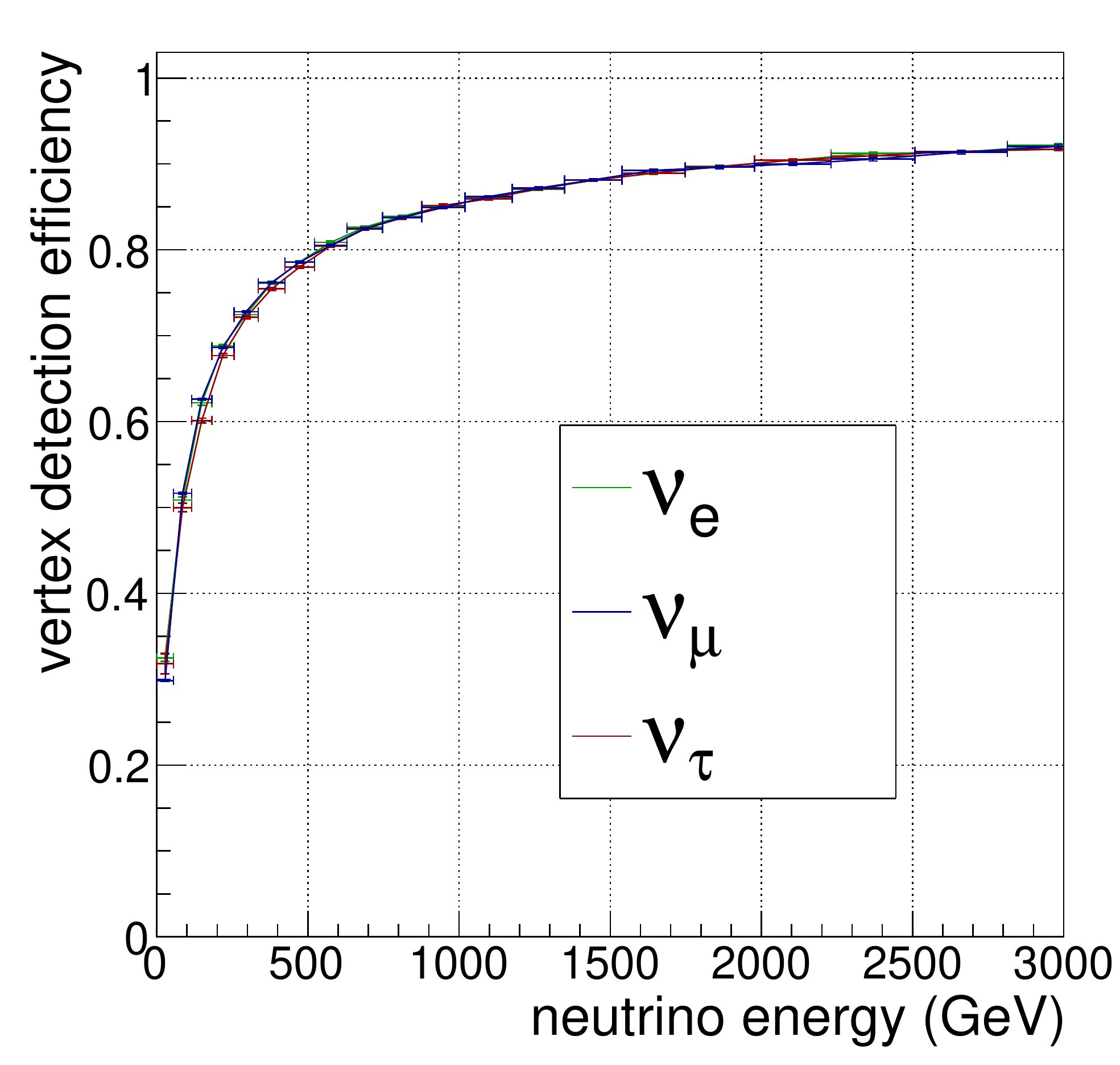}
\includegraphics[height=5.5cm]{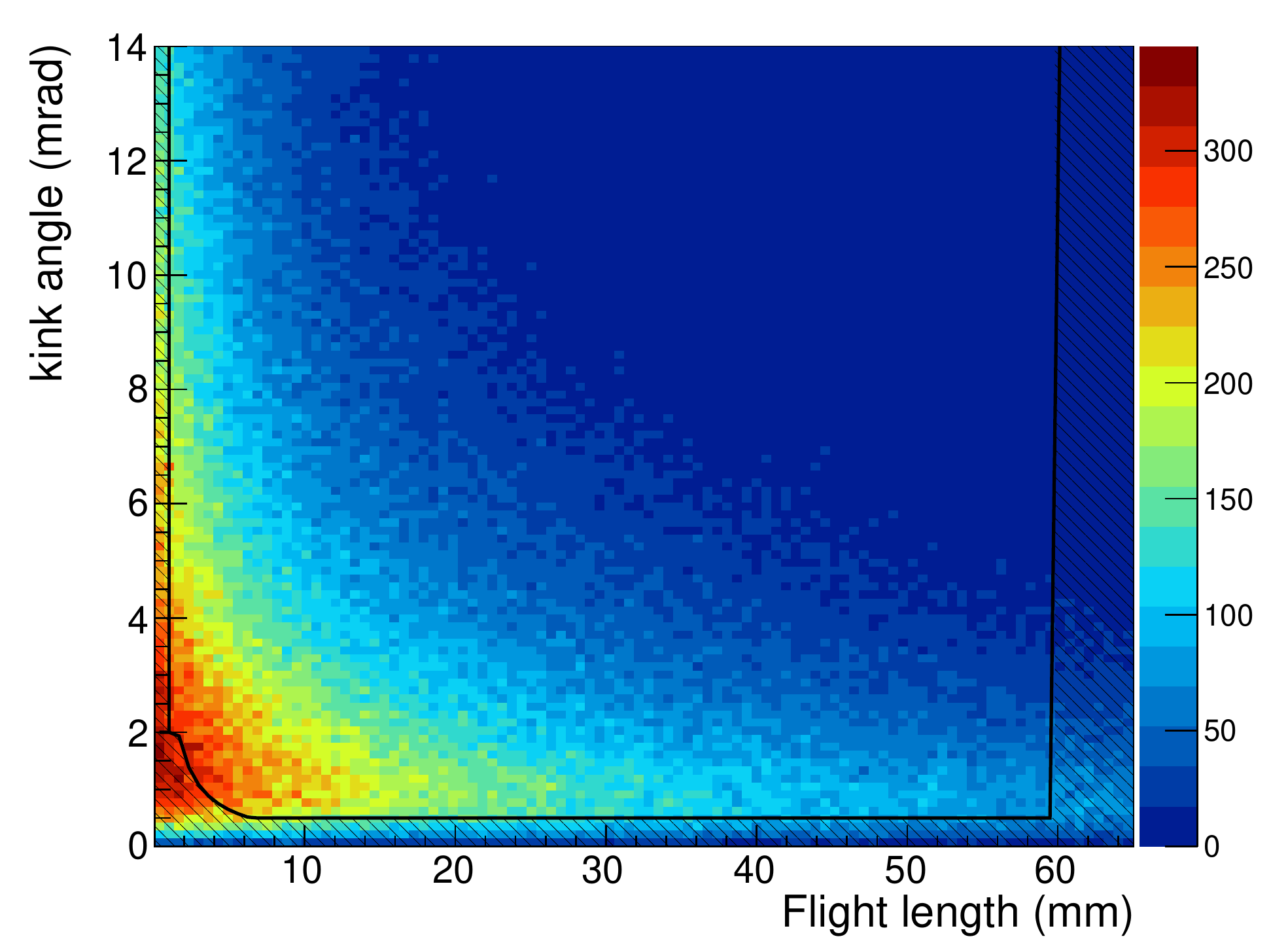}
\caption{\textbf{Left}: Vertex detection efficiency after requiring at least 5 charged particles at a neutrino interaction vertex (CC and NC inclusive). Only the statistical uncertainties of the generated \textsc{Genie} samples are shown. \textbf{Right}: The distribution of events in the $(\tau~\text{flight length}, \text{kink angle})$-plane, where the events are $\nu_{\tau}$ CC interactions in FASER$\nu$ that produce tau leptons that decay through 1-prong decays. The black hatched region is excluded by the cuts described in the text; requiring the event to be in the unhatched region leads to a 75\% $\tau$ detection efficiency. 
}
\label{fig:detection_eff}
\end{figure}

We now consider $\nu_{\tau}$ CC interactions in FASER$\nu$. These produce $\tau$ leptons, which have 1-prong decays 85\% of the time.  In the right panel of \figref{detection_eff} we plot the distribution of these events in the $(\tau~\text{flight length}, \text{kink angle})$-plane.  The mean $\tau$ flight length is $3~\cm$. To detect a kink, we require that the $\tau$ crosses at least one emulsion film, the kink angle is bigger than 4 times the angular resolution given in \eqref{angular_resolution} and more than $0.5~\mrad$, and the flight length is less than 6 cm, where the last requirement is implemented to reduce hadronic backgrounds.  With these requirements, the detection efficiency for 1-prong $\tau$ events is estimated to be 75\%. The detection efficiency for the 15\% of $\tau$ leptons that decay to 3-prong decays is expected to be similar to or higher than the detection efficiency in the 1-prong channel.

\subsection{Neutrino Energy Reconstruction}
\label{sec:energyresolution}

High-energy neutrino interactions via DIS have a large energy transfer to the nucleus, resulting in significant hadronic energy. To estimate the neutrino energy, we therefore need to know both the leptonic and hadronic energies in the event, since $E_\nu = E_\ell + E_{\text{had}}$. In the following discussion, we illustrate how the neutrino energy can be reconstructed and obtain a first estimate for the energy resolution that could be achieved at FASER$\nu$.

Due to its high spatial resolution, the FASER$\nu$ detector will precisely measure topological variables, such as track multiplicity and the slopes of tracks. As we will discuss below, the combination of these topological measurements allows one to estimate kinematic quantities, e.g., the charged particle momentum by multiple Coulomb scattering (MCS) and the electromagnetic shower energy. Those topological and kinematic variables can then be assembled with a multivariate method to estimate the neutrino energy.

\begin{description}

\item [Topological Variables] At the neutrino interaction vertex, the charged particle multiplicity ($n_{\text{tr}}$) can be measured. Given tungsten's short radiation length $X_0=3.5~\mm$, most of photons start developing electromagnetic showers within $1~\cm$ and can be associated with the vertex ($n_\gamma$). In contrast, given tungsten's much larger hadronic interaction length $\lambda_{\text{int}}=10~\cm$, neutral hadrons are much more difficult to associate with the neutrino vertex.

The particle emission angle with respect to the neutrino direction is $\theta\approx P_\perp/P$. Assuming that the muon transverse momentum is distributed around typical values $\langle P_\perp \rangle$, we obtain an estimator for the muon momentum $P_\mu \propto 1/\theta_\mu$. Similarly, the
sum of inverses of the hadron track angle, $\sum \left|1/\theta_{\text{had}}\right|$, is typically roughly proportional to the total hadronic energy. Additionally, the median of the track angles of all charged particles is inversely-related to neutrino energy. 

\item [Momentum Measurement using MCS]  The momenta of charged particles can be measured by multiple Coulomb scattering (MCS) in the detector~\cite{Kodama:2002dk}.  The emulsion detector's excellent position resolution will allow one to measure the MCS of TeV-energy particles. The fringe fields of the FASER magnets at the end of \FASERnu ($B\sim 10$ mT) give a negligible effect for the MCS measurement.

The difference of a particle's hit position and what is expected of a straight track from two previous hits is $s = y_2 - 2y_1+y_0$. (See \figref{mcs}.) For a given track and a fixed value of $x$, the half length of the trajectory under evaluation, the RMS value of $s$ is calculated using every possible initial hit location.  This is done for every possible value of $x$.  The resulting distribution of points in the $(x, s^\text{RMS}(x))$ plane are then fit with the functional form 
\begin{equation}
s^\text{RMS}\left(x\right)^2=\left( \sqrt{\frac{2}{3}}\frac{13.6~\mev}{\beta P} x \sqrt{\frac{x}{X_0}} \right)^2 + \left(\sqrt{6}\ \sigma_\text{pos}\right)^2 \ ,
\label{eq:mcs}
\end{equation}
where $X_0$ is the material's radiation length, and $\sigma_\text{pos}$ is the position resolution of each hit.\footnote{As a simple example, for a particle with a momentum of 1 TeV crossing 10 cm of tungsten, the expected deviation from a straight line is $s^\text{RMS}=2~\mu\text{m}$.} The value of $P$ that gives the best fit is the reconstructed track momentum~\cite{Kodama:2002dk,OPERA:2011aa}.   

\begin{figure}[tbp]
\centering
\includegraphics[height=4.2cm]{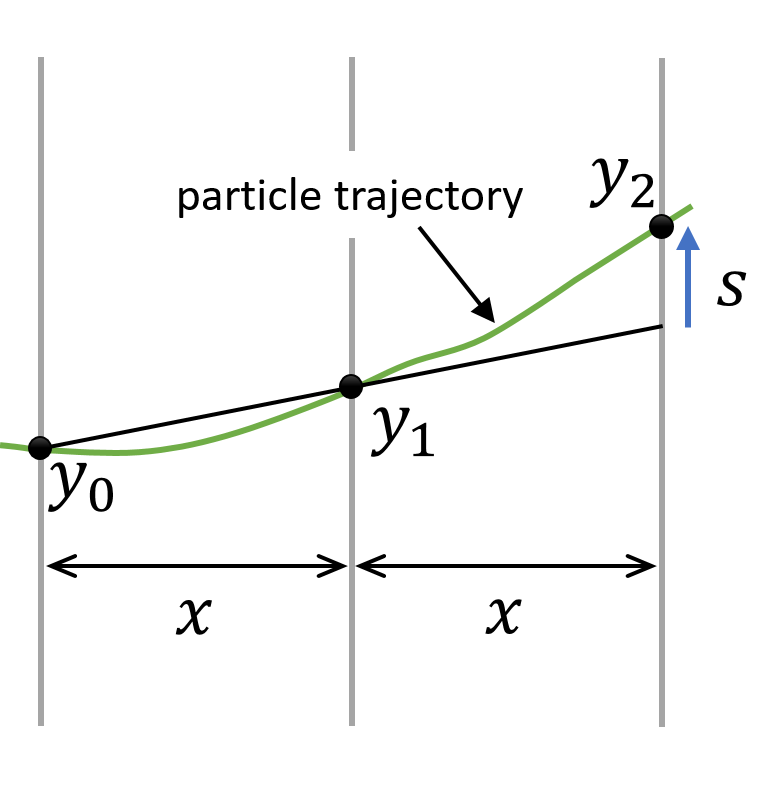}
\includegraphics[height=5.1cm]{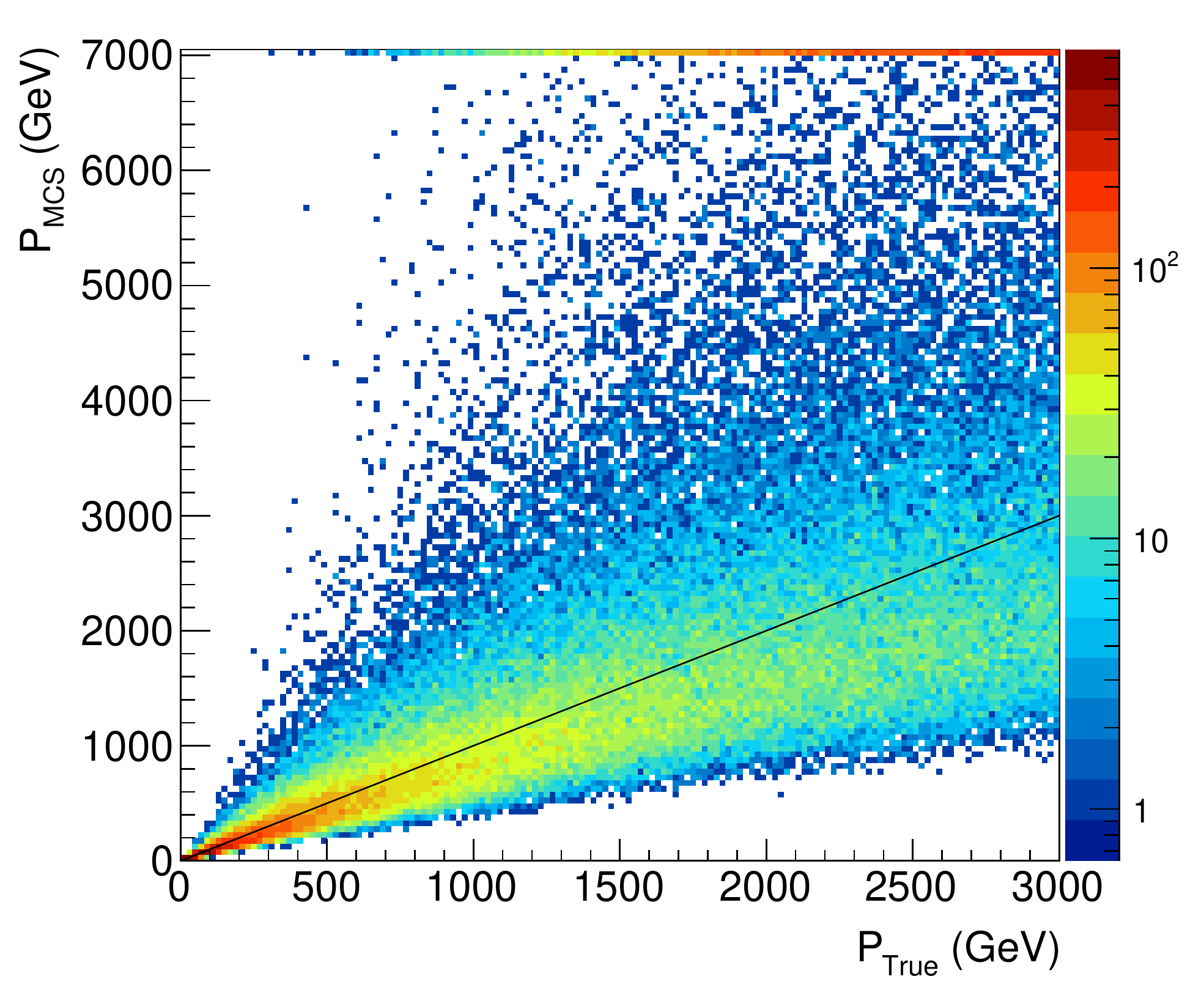}
\includegraphics[height=5.1cm]{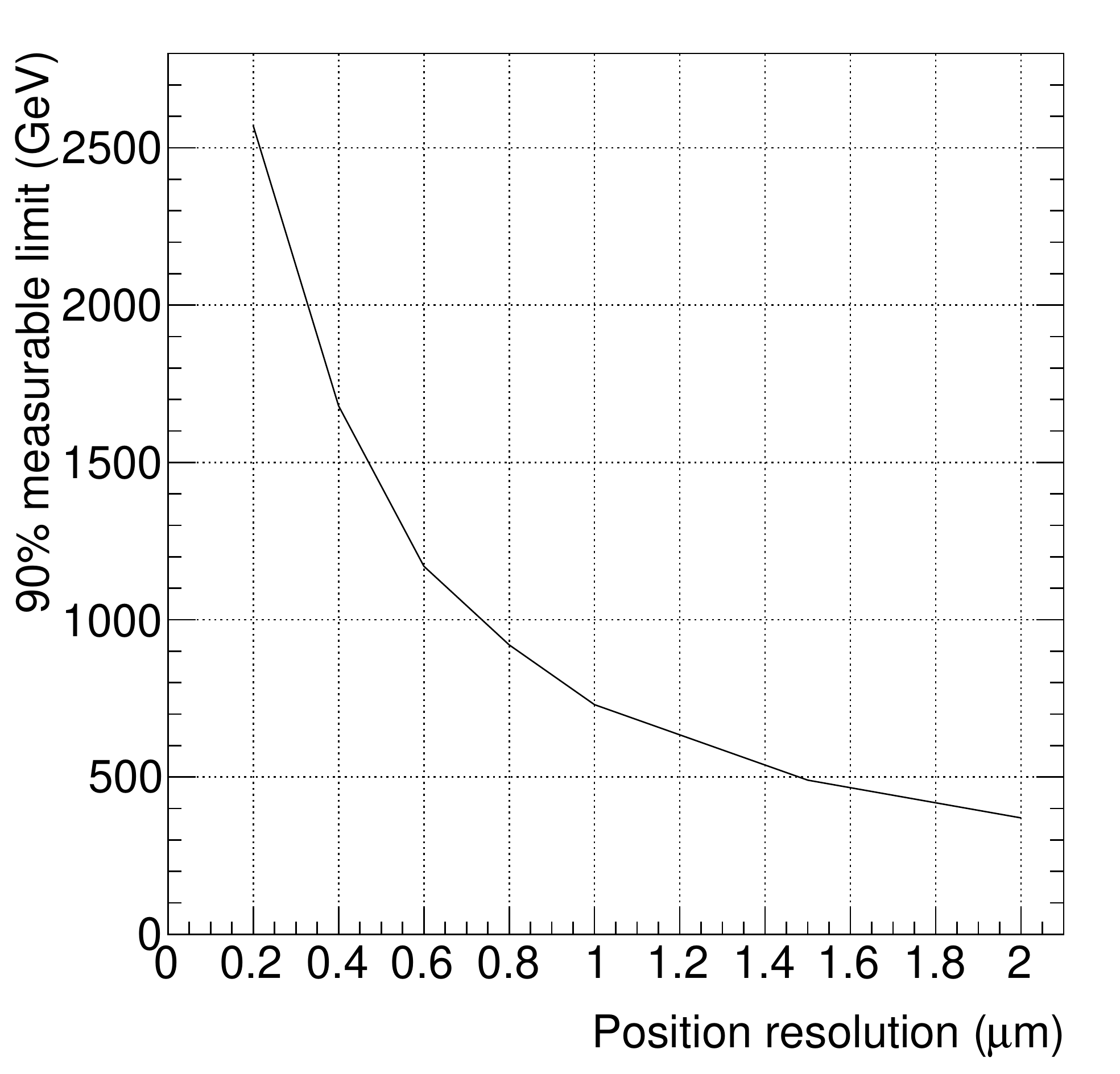}
\caption{\textbf{Left}: Schematic of the MCS measurement. \textbf{Center}: MC study of momentum reconstruction with the MCS method, assuming a position resolution of $\sigma_{\text{pos}} = 0.4~\mu\text{m}$ and a track length of 100 1-mm-thick tungsten plates. Events with $P_{\text{MCS}}>7~\tev$ are shown at $P_{\text{MCS}} = 7~\tev$ so that they can be seen in the figure. On the black line, $P_{\text{true}} = P_{\text{MCS}}$. \textbf{Right}: The largest momentum for which the MCS method is applicable as a function of position resolution. The largest momentum is defined to be the momentum for which 90\% of the particles' momenta are reconstructed to have $P_{\text{MCS}} < 7~\tev$. 
}
\label{fig:mcs}
\end{figure}

The central panel of \figref{mcs} shows a MC study of momentum reconstruction with the MCS method, assuming a position resolution of $0.4~\mu\m$ and that the charged particles cross 100 layers of 1-mm-thick tungsten plates (28.5 $X_0$ and 1.0 $\lambda_{\text{int}}$).  A linear relationship between the reconstructed and true momentum holds for momenta up to and beyond 1 TeV. The RMS momentum resolution obtained is 46\% at 200 GeV and 57\% at 1 TeV. Defining the largest momentum for which the MCS method can be used to be the momentum at which 90\% of the particles' momenta can be measured to have $P_{\text{MCS}}<7~\tev$, the largest momenta can be calculated as a function of position resolution, as shown in \figref{mcs} on the right. For the expected $\sigma_{\text{pos}} = 0.4~\mu\text{m}$, the largest momentum is around 1.5 TeV.

\item [Calorimetric Electromagnetic Energy Measurement]
Electromagnetic showers will be reconstructed in the emulsion detector. The shower energy can be measured by counting the number of track segments in the reconstructed shower and/or with a multivariate analysis combining additional variables such the shower's longitudinal and lateral profile. The energy resolution of such an algorithm is estimated to be $\sigma_E=50\%/\sqrt{E}$ for $E$ in GeV~\cite{Juget:2009zz}, implying an energy resolution of 25\% for $E = 4~\gev$, and much better for higher energies. Another study with a similar detector technology reported energy resolutions below 15\% for 50 and 200 GeV electron data~\cite{Kobayashi:2012jb}.  However, in the case of FASER, a large number of low-energy background electron tracks produced by high-energy muons would limit the energy resolution. Therefore, we conservatively assume a constant $\gamma$ energy resolution of 50\% for further calculations.

\end{description}

As a primary approach, a simple sum of the visible energy, which has the least dependence on the interaction generator, already gives a good estimate of neutrino energy, as shown in the left panel of \figref{ann}. The resolution in the right panel deteriorates and becomes unstable at high energy ($E_\nu>1$ TeV) because of the difficulty of providing an upper bound on energy for the largest momenta in the MCS method.  

To improve the neutrino energy reconstruction resolution, an artificial neural network (ANN) algorithm was built with inputs from the topological and kinematic observable shown in \tableref{ann_inputs}, using the MLP package in the CERN ROOT framework~\cite{Antcheva:2009zz}. For this study, $\nu_\mu$ CC events were simulated by \textsc{Genie}, smeared by the MCS and shower reconstruction resolutions,\footnote{Charged particle momenta are smeared with a look-up table of the MCS measurement. Shower energies are smeared with a Gaussian resolution.} and used to train and evaluate the ANN algorithm. The reconstructed neutrino energy is shown in \figref{ann}.  An energy resolution of about 30\% was obtained for the entire energy range relevant for \FASERnu. Although this result is promising, more careful study of the interaction generator is needed to understand and quantify the uncertainties of the energy estimation, for example due to the modeling of the nuclear corrections, hadronization, and final state interactions. The currently used modeling in
\textsc{Genie} has not been validated for neutrino interactions with tungsten at the energies relevant for \FASERnu. Such a validation would include a cross-check between the visible energy method and ANN method at lower energy $E<1~\tev$. Additionally, so far the energy measurement with the ANN algorithm has been tested only for $\nu_\mu$ CC events. We will extend this analysis to $e$ and $\tau$ neutrinos and to NC interactions in the future. 

\begin{table}[t]
\centering
\begin{tabular}{|c|l|c|}
\hline
\hline
\multicolumn{2}{|l|}{\textbf{Topological Variables}} & related to \\
\hline
$n_{\text{tr}}$ & Multiplicity of charged tracks at the neutrino interaction vertex  & $E_{\text{had}}$\\
& with momentum $p_{\text{tr}}>0.3~\gev$ and angle $\tan\theta_{\text{tr}}>0.3$ & \\
\hline
$n_{\gamma}$ & Photon multiplicity &$E_{\text{had}}$\\
\hline
$\left|1/\theta_{\ell}\right|$ & Inverse of lepton angle with respect to neutrino direction &$E_{\ell}$\\
\hline
$\sum \left|1/\theta_{\text{had}}\right|$ & Sum of inverse of hadron track angles&$E_{\text{had}}$\\
\hline
$1/\theta_{\text{median}}$ & Inverse of the median of the track angles of all charged particles&$E_{\text{had}}, E_{\ell}$\\
\hline
\multicolumn{2}{|l|}{\textbf{Track Momentum via MCS}} & \\
\hline
$p^\text{MCS}_{\ell}$ & Estimated lepton momentum from MCS &$E_{\ell}$\\
\hline
$\sum p^\text{MCS}_{\text{had}}$ & Sum of estimated charged hadron momenta from MCS &$E_{\text{had}}$\\
\hline
\multicolumn{2}{|l|}{\textbf{Energy in Showers}} & \\
\hline
$\sum E_{\gamma}$ & Sum of energy in photon showers  &$E_{\text{had}}$\\
\hline
\hline
\end{tabular}
\caption{Inputs for the ANN algorithm.  For the momentum estimates using the MCS method, we assume a position resolution of $0.4~\mu\text{m}$. For the energy measurement in photon showers we assume an energy resolution of 50\%.  
}
\label{table:ann_inputs}
\end{table}

\begin{figure}[t]
\centering
\includegraphics[width=\textwidth]{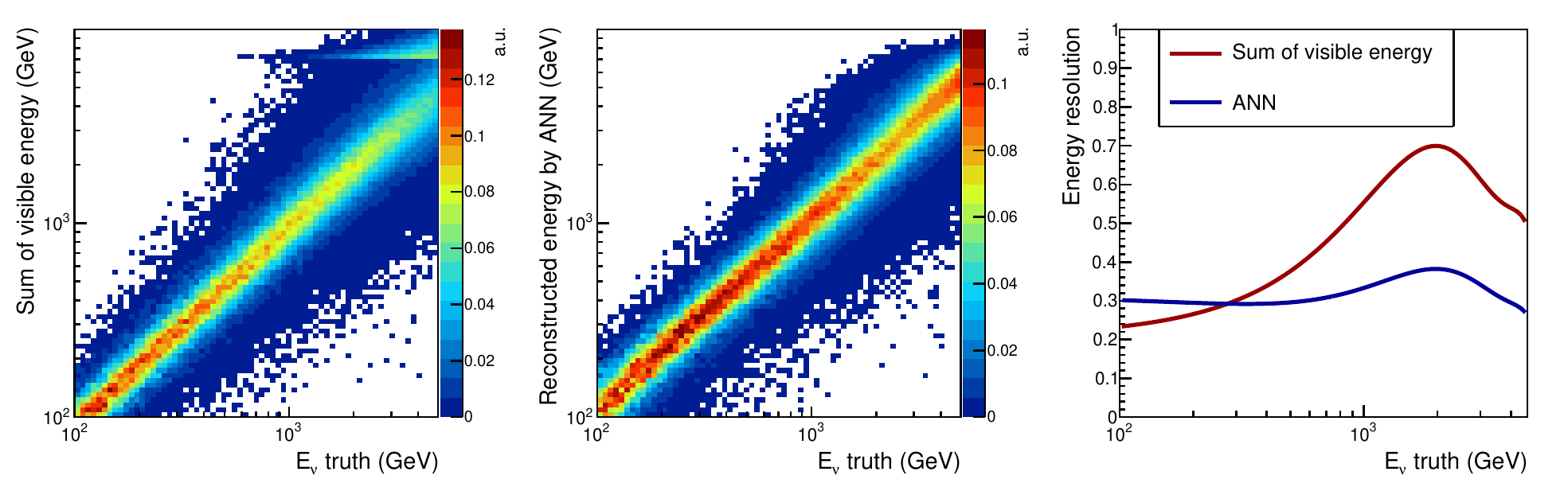}
\caption{\textbf{Left}: Neutrino energy and sum of visible energy (momentum of charged particles and energy of electromagnetic showers) for $\nu_\mu$ CC samples with at least five charged tracks $n_{\text{tr}} \geq 5$, with smearing (MC). The effect of the 7 TeV upper limit used in the MCS method is visible near the top of the figure.
\textbf{Center}: Neutrino energy reconstruction based on the ANN. The observables listed in \tableref{ann_inputs} are used as the inputs for the ANN algorithm. \textbf{Right}: $\Delta E_{\nu}^{\text{ANN}} / E_{\nu}^{\text{true}}$ for the same sample. An energy resolution of 30\% (RMS) was obtained for the energies of interest.
}
\label{fig:ann}
\end{figure}

\subsection{Global Reconstruction with the FASER Detector}

So far, we have considered the capabilities of FASER$\nu$ alone for the discovery and study of neutrinos.  However, there may be significant advantages to coupling \FASERnu to the FASER main detector (spectrometer) by using additional silicon tracker layers at the interface of FASER$\nu$ with the main detector, as shown in \figref{detector_upgrade}. Such a hybrid emulsion/counter detector was successfully used in detecting neutrino interactions in a series of experiments~\cite{Eskut:2007rn, Kodama:2007aa, Acquafredda:2009zz, Kodama:1993wg, Albanese:1985wk}. The FASER spectrometer can provide charge information and is also useful to improve the energy resolution.  Moreover, a scintillator that is upstream of the emulsion detector provides an additional opportunity to identify an incoming muon, and so aids in differentiating muon-induced backgrounds from neutrino interaction events.  

\begin{figure}[tpb]
\centering
\includegraphics[width=0.8\textwidth]{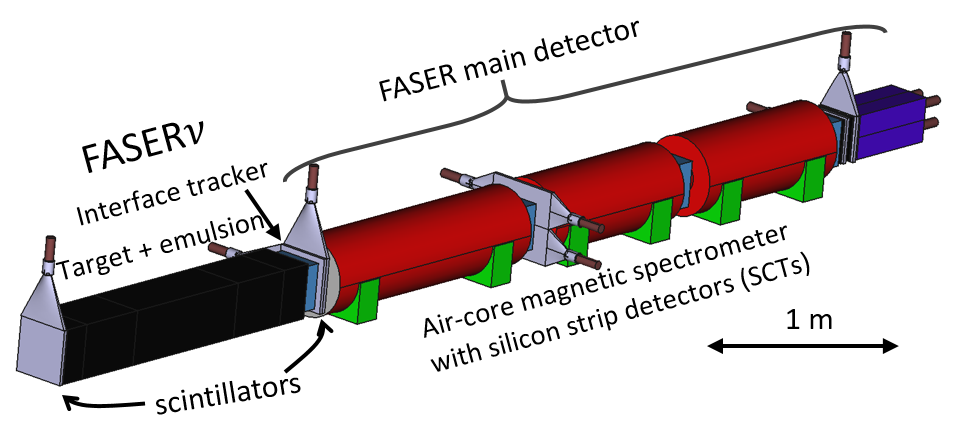}
\caption{Plan of the detector upgrade to couple FASER$\nu$ to the FASER main detector. The components include the FASER$\nu$ emulsion detector (black), scintillators (gray), tracking layers (blue), magnets (red), and electromagnetic calorimeter (purple). 
}
\label{fig:detector_upgrade}
\end{figure}

To find the correspondence between tracks in emulsion detectors and those in the electronic detectors, an interface tracker with high spatial resolution is required. The tracks emerging from neutrino interaction vertices have an angular spread of $\simeq75~\mrad$ (rms, $p_{\text{tr}}>1~\gev$), corresponding to a position spread of $7.5~\mm$ at $10~\cm$ downstream of the vertex. The interface tracker should have a spatial resolution of the order of 0.1 mm to reconstruct several tracks in this scale. The ATLAS silicon strip detector (SCT~\cite{ATLAS:1997ag, ATLAS:1997af}, $80~\mu\text{m}$ pitch),  which is being used in the FASER spectrometer, and monolithic silicon pixel sensors~\cite{Paolozzi:2018flw} are promising candidates. Alternative detector technologies and structures, including interleaving the emulsion detector with some layers of electronic detectors, are currently being investigated.

In a combined FASER/FASER$\nu$ detector, events can be reconstructed by matching multiple tracks between the emulsion and interface detector in position and angle.  The trigger rate in FASER due to background muons is $\sim 100$ Hz, and so event pile-up in a single event timing is rare: for the anticipated LHC bunch crossing frequency of 40 MHz, we expect an $\mathcal{O}(10^{-6})$ probability for a muon to coincide with an event in \FASERnu. In each chunk of \FASERnu data ($\sim 30~\ifb$), about $4 \times 10^8$ muons are expected (see \secref{bg}). Most of these are single-track events. On the other hand, we expect $\mathcal{O}(10^4)$ neutrino and hadron events that are mostly multi-track events. The fake matching between a track in \FASERnu and the tracks from another events in the interface detector can be estimated. Assuming we require positional matching to 1 mm in both dimensions of the $25\,\cm \times 25\,\cm$ detector and angular matching of 10 mrad in the 75-mrad angular spread of background neutrino-related particles, the fake matching rate is roughly $(1/250)^2 \times (10/75)^2 = 3\times 10^{-7}$ for each track. By requiring that two or more tracks match, the correspondence between events in FASER$\nu$ and the FASER main detector will be uniquely identified.

\section{Signal and Background Characteristics}
\label{sec:bg}

The signal of collider neutrinos scattering through CC interactions in FASER$\nu$ is defined by the following characteristics:
\begin{enumerate}
\item A vertex is identified in the detector with at least 5 charged tracks associated with it.
\item The vertex is a neutral vertex, with none of the charged tracks connected to it entering the detector from the direction of the ATLAS IP. 
\item The event points back to the ATLAS IP within the angular resolution of the detector, which, as discussed below, is of the order of $3-5~\mrad$ for $\tev$ neutral hadrons and neutrinos undergoing CC interactions in FASER$\nu$.
\item The total energy in the event is very high, with mean energies of 600 GeV to 1 TeV, depending on the neutrino flavor.  As discussed in \secref{energyresolution}, the resolution for reconstructing the neutrino energy is approximately 30\%, and so events with energies in the 10 to 100 GeV range can be effectively distinguished.
\item Among the charged tracks leaving the vertex is a high-energy lepton, identifying it as a CC neutrino interaction.
\end{enumerate}
These characteristics make the neutrino signal a spectacular one that cannot be easily mimicked by other SM particles at the FASER location. 

In this section, before discussing background processes in detail and ways to differentiate them from the signal, we start in \secref{bg-environment} with an extended discussion of the detector environment.  We discuss the particles present in TI12, their sources, and their fluxes, and describe the results of FLUKA and GEANT simulations and the validation of these results by the {\em in situ} measurements conducted in 2018.  Given this context, we then turn in \secref{bg-vertex} to a discussion of processes that can appear as a neutral vertex in FASER$\nu$ and ways these can be differentiated from the neutrino signal.  In \secref{bg-neutrino}, we then discuss the final requirement of identifying hard leptons leaving from the vertex.  We show that the backgrounds may be reduced to very low levels, and we discuss the prospects for differentiating the high-energy leptons to disentangle the $\nu_e$, $\nu_{\mu}$, and $\nu_{\tau}$ signals.  

\subsection{Detector Environment}
\label{sec:bg-environment}

As noted above, FASER$\nu$ will be placed in side tunnel TI12 about $480~\m$ away from the ATLAS IP. As shown in \figref{infrastructure}, between the IP and FASER$\nu$ there are many elements of the LHC infrastructure, including the TAN $\sim 140~\m$ downstream from the IP, the dispersion suppressor region, and finally 10 m of concrete and 90 m of rock, which stops all hadronic particles that originate from the IP. In addition to this shielding, magnets deflect away the majority of charged particles produced in the forward direction. 

The LHC infrastructure therefore naturally shields the FASER location from most particles produced at the IP. The remaining particles at FASER$\nu$ can be grouped into four main categories, depending on their origin:

\begin{enumerate}

\item Muons produced either at the ATLAS IP or further downstream. Aside from neutrinos, muons are the only SM particles that can efficiently transport energy through 100 m of concrete and rock. 

\item Secondary particles produced by muons in the rock in front of the detector or within the detector.  Such secondary particles may create both electromagnetic and hadronic showers that mimic the signal.

\item Particles may be produced in beam-gas collisions, as well as proton-loss-induced showers in the dispersion suppressors at the transition between the LHC insertion and arc regions.  Particles from beam-related processes require accurate modeling of the LHC infrastructure and beam optics. This has been done for the FASER Collaboration by the CERN Sources, Targets, and Interactions (STI) group~\cite{FLUKAstudy}, which performed for this purpose state-of-the-art simulations based on the FLUKA code~\cite{Ferrari:2005zk,Battistoni:2015epi}.  The conclusion was that beam-related backgrounds are also negligible: the excellent vacuum in the LHC means beam-gas interaction rates are extremely small, and the dispersion function close to the FASER location minimizes proton losses in this region. These backgrounds do not contribute to the high-energy particle flux of most interest here, and the resulting particles also have the wrong directionality.

\item Cosmic rays may pass through FASER$\nu$.  However, this potential background can be suppressed to negligible levels by requiring that they have high energy and point back to the IP within the angular resolution of the detector.

\end{enumerate} 

The particles of most interest are, then, muons and secondary particles.  These have been investigated by the FLUKA simulation mentioned above, and the FLUKA results have been validated by {\em in situ} measurements with pilot emulsion detector data from 2018.  We now describe the results of these simulations and the experimental data.

\paragraph{Results from pilot emulsion detectors.} The pilot emulsion detectors were placed in TI12 and TI18 in Technical Stops in 2018. Two small detectors to check the detector environment and a pilot neutrino detector (see \secref{firstnu}) were used. One of the small detectors had a heterogeneous structure with successive layers consisting of film/film/film/ tungsten/film/ tungsten/film, which is different from the detector structure for Run 3.  The first three films were used to reconstruct tracks with a low energy threshold of $E \agt 50~\mev$. In the last three films, low energy tracks were suppressed due to scattering in the 0.5-mm-thick tungsten plates, and these films were therefore used to reconstruct tracks with a higher energy threshold of $E \agt 1~\gev$.

The first results from this emulsion detector are shown in the left panel of \figref{projection_ti12BG}, while a more detailed discussion of the results of pilot emulsion detectors is postponed to a future study~\cite{FASERmuondata}.  The angular distributions of charged particles entering the detector are shown for the two energy thresholds of $E \agt 50~\mev$ and $E \agt 1~\gev$.  A clear peak can be observed from the direction compatible with the ATLAS IP ($\theta_x = 0$).  The tracks at larger angles are mostly composed of low-energy particles below $1~\gev$, and they, therefore, will not contribute to background for the high-energy signal of our interest.

\paragraph{Comparison with the simulations.} According to the simulations, more than 99.999\% of particles with $E>100~\gev$ are expected to be muons and related electromagnetic components. The muon flux predicted by the FLUKA simulations is $\Phi \simeq 2\times 10^4$ $\fb/\cm^{2}$ for energies $E_\mu> 10~\gev$, while the measured total charged particle flux was $(3.0\pm 0.3)\times 10^4~\fb/\cm^{2}$, which also includes softer particles with energies above $50~\mev$. The measured flux within the main peak (within $10~\mrad$, corresponding to the high-energy component) was $(1.9\pm 0.2)\times 10^4~\fb/\cm^{2}$. The results of the simulations and measurements agree remarkably well, given the empirical and theoretical uncertainties that, in the latter case, can be as large as a factor of a few. In particular, the subdominant peak for $\tan\theta_x\gtrsim 0.5$ corresponds to particles entering the detector from the other side with respect to the IP1, as discussed in Ref.~\cite{Ariga:2018pin}.

\begin{figure}[tpb]
\centering
\includegraphics[width=0.4\textwidth]{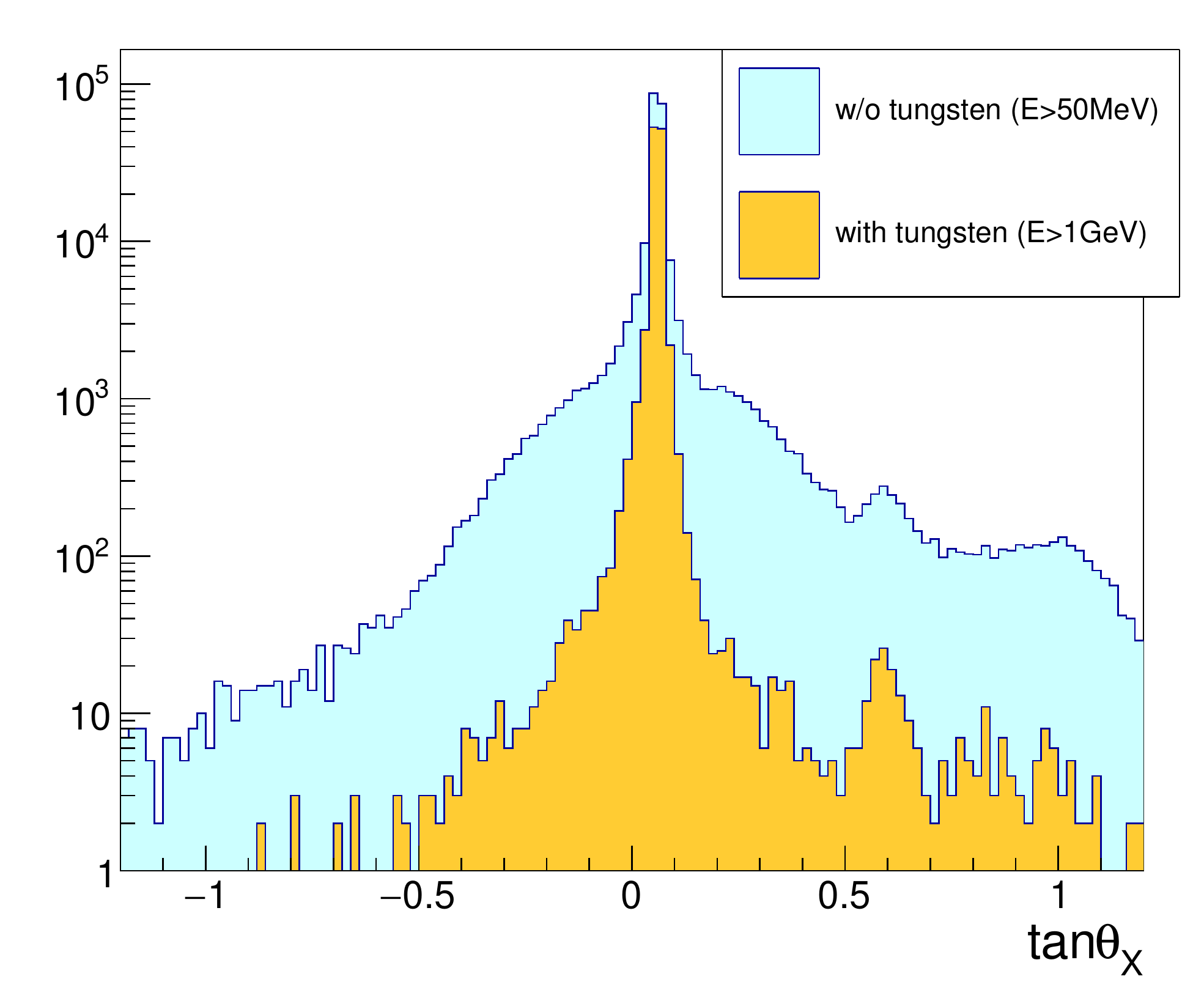}
\includegraphics[width=0.59\textwidth]{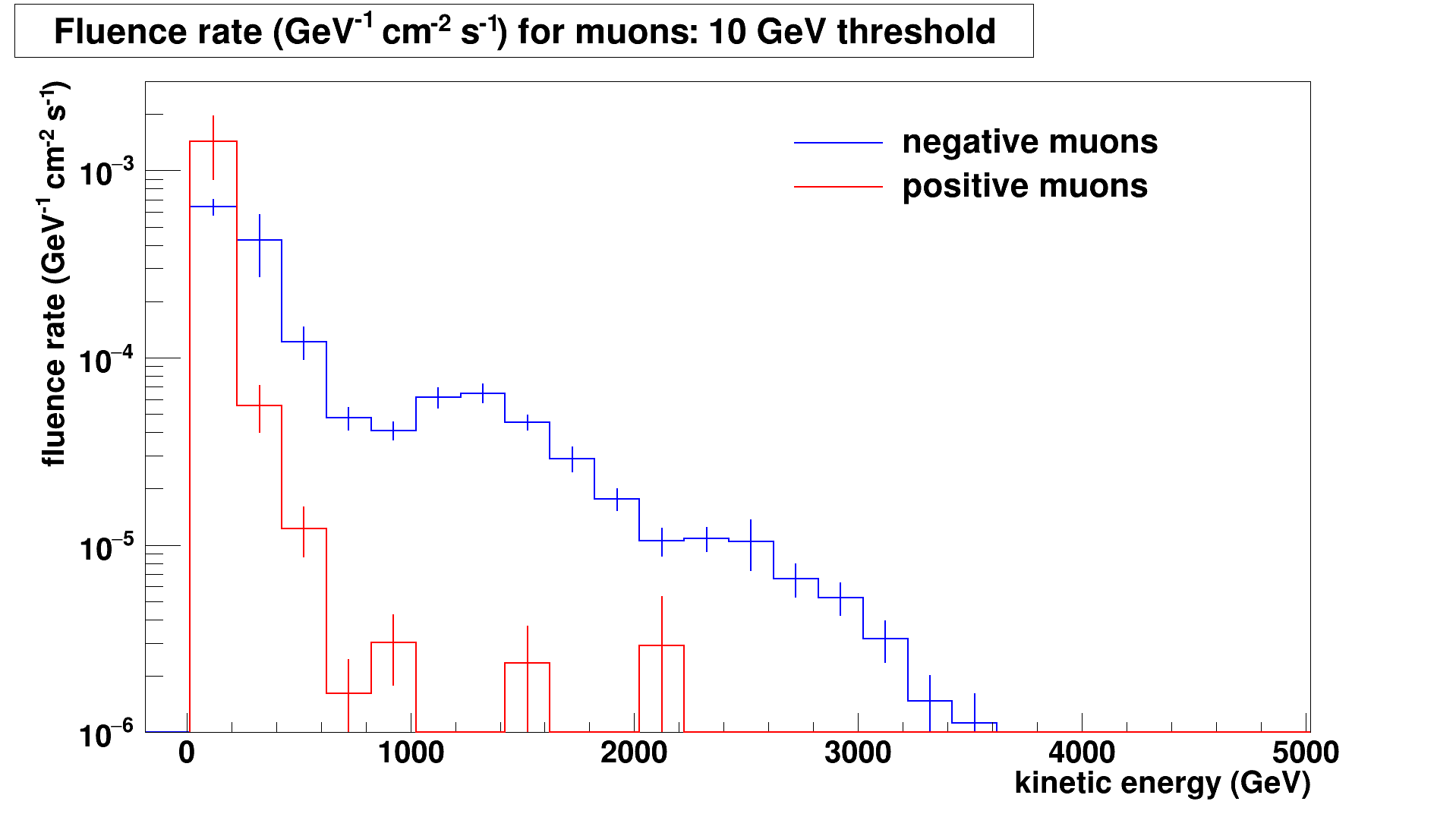}
\caption{\textbf{Left}: Angular distributions of charged particles measured by the emulsion films with and without tungsten plates, corresponding to energy cutoffs of about 1 GeV and 50 MeV, due to multiple Coulomb scattering, respectively. \textbf{Right}: Fluxes of positive and negative muons at the FASER$\nu$ location predicted by the FLUKA simulations and normalized to an instantaneous luminosity of $2\times 10^{34}~\cm^{-2}~\s^{-1}$. From Ref.~\cite{Ariga:2018pin}.}
\label{fig:projection_ti12BG}
\end{figure}

Given the aforementioned muon flux of $\Phi\simeq 2\times 10^4~\fb/\cm^2$ going through FASER$\nu$, an LHC Run 3 integrated luminosity of 
$L=150~\ifb$, and a FASER$\nu$ area of $\mathcal{A}=25\,\cm\times 25\,\cm = 625\,\cm^2$, we expect a total number of $N_\mu = \Phi L \mathcal{A} \simeq 2 \times 10^9$ muon tracks with $E_\mu>10~\gev$.  This is a large flux, but, of course, single muons do not mimic neutrinos scattering in the detector.

\paragraph{Muon-induced background.} However, when muons undergo photo-nuclear interactions in the rock in front of FASER$\nu$, as well as within the detector volume, they can produce secondary neutrons and other neutral hadrons that can mimic neutrino signatures.  To estimate these particle fluences, we have performed additional extensive FLUKA studies, using the spectra of muons going through the detector obtained by the CERN STI group as shown in the right panel of \figref{projection_ti12BG}. The results obtained for neutrons emerging from rocks have been additionally compared with simulations employing GEANT4~\cite{Agostinelli:2002hh,Allison:2006ve,Allison:2016lfl}, with general agreement between the two. The total muon flux for $E_\mu>10~\gev$ can be divided into roughly equal contributions from positive and negative muons, while the high-energy part of the spectrum is dominated by $\mu^-$. The difference between the positive and negative muon fluxes is due to the complicated impact of the LHC optics on muon trajectories on their way to FASER$\nu$. As a result, when high-energy neutrinos are studied, background induced by positive muons plays a subdominant role and we subsequently focus on negative muons. Notably, the FLUKA interaction and transport code relevant for the production of muon-induced neutrons in rocks has been validated in the past against the experimental data collected for cosmogenic muons~\cite{Empl:2014zea}. A similar study with GEANT4 can be found, e.g., in Ref.~\cite{Lindote:2008nq}.

It is important to note that the aforementioned impact of the LHC optics would lead to a significantly larger muon flux if FASER$\nu$ were displaced from the beam collision axis. We illustrate this in \figref{muonheatmap}, where we present the distributions of both negative and positive muons crossing the tunnel TI18, which is in a symmetric position but on the opposite side of the ATLAS IP with respect to the tunnel TI12. The results are based on the FLUKA simulations performed by the CERN STI group~\cite{FLUKAstudy}. As can be seen, the on-axis position of FASER$\nu$ is actually close to a local minimum of the flux for both types of muons, which makes it a particularly promising place to study the interactions of high-energy neutrinos.

\begin{figure}[tpb]
\centering
\includegraphics[clip,trim = 1cm 0 3cm 0 , width=0.49\textwidth]{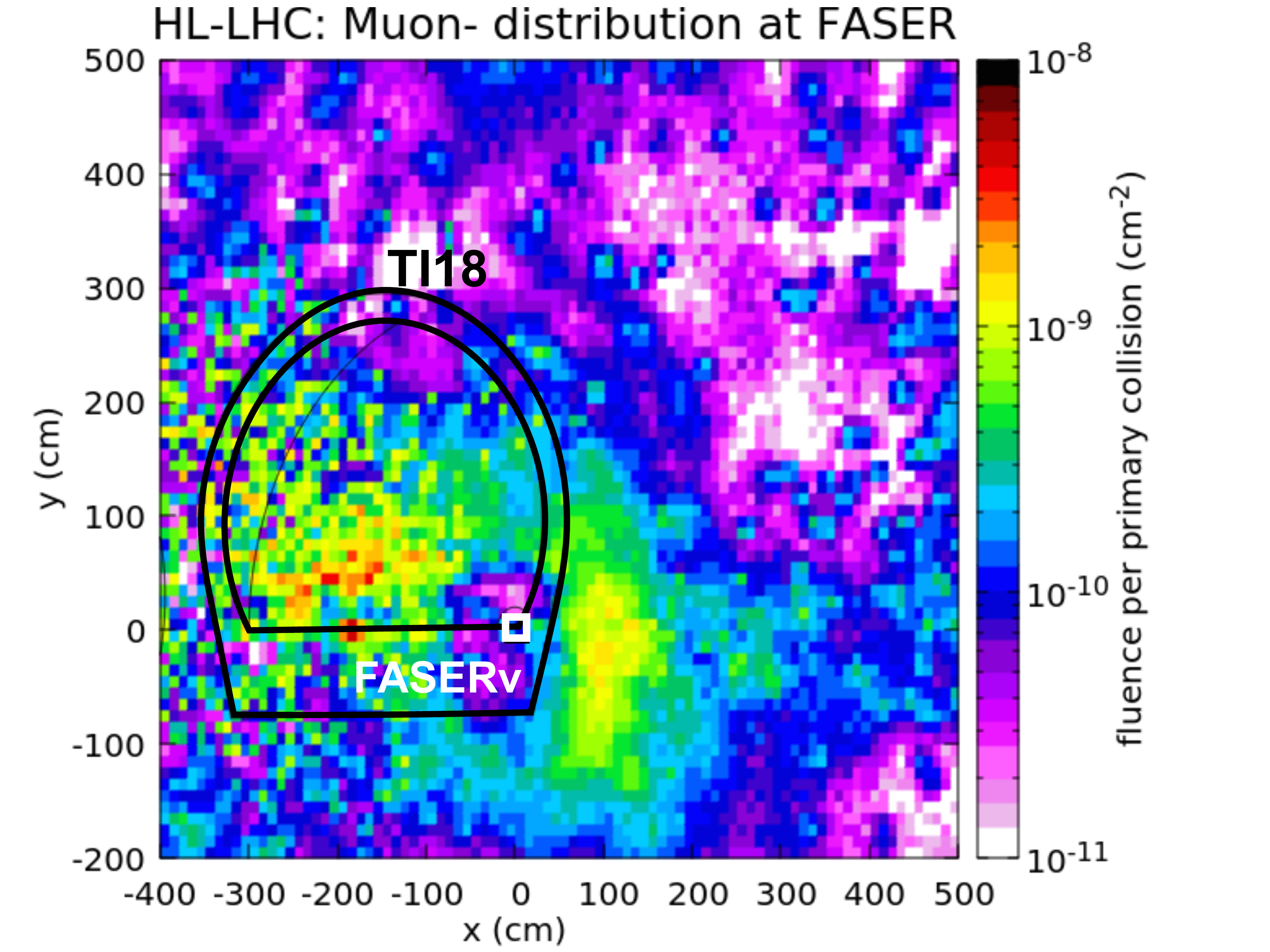}
\includegraphics[clip,trim = 1cm 0 3cm 0 , width=0.49\textwidth]{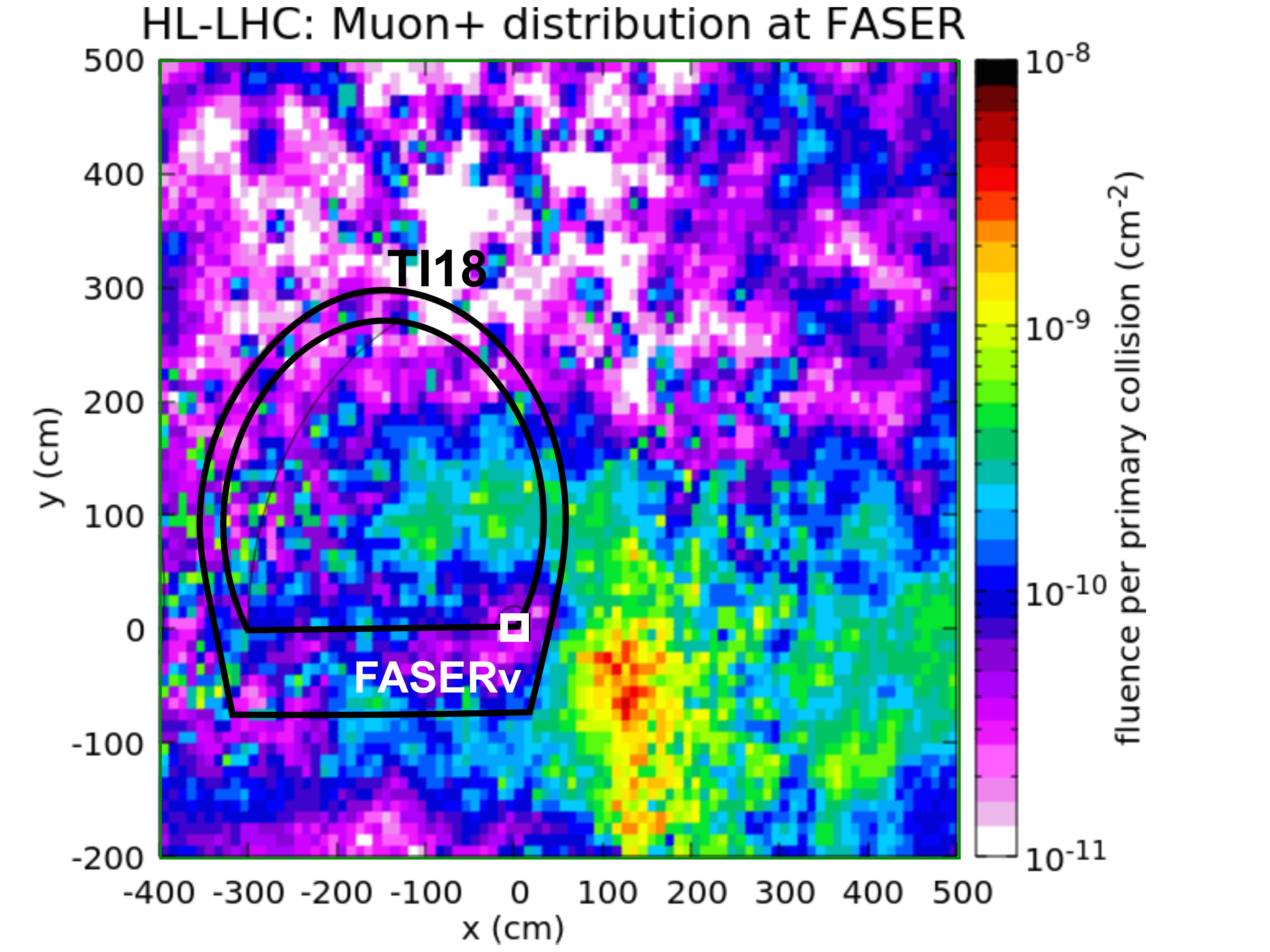}
\caption{Distributions of negative (left) and positive (right) muons crossing the tunnel TI18, which is in a symmetric position on the opposite side of the ATLAS IP with respect to FASER$\nu$. The position of FASER$\nu$ in the center of the coordinate system is indicated by a white square. From Ref.~\cite{Ariga:2018pin}.}
\label{fig:muonheatmap}
\end{figure}

In \tableref{BGtable}, we show the number of negative-muon-induced neutral particles that emerge from the rock in front of FASER (the first number in each entry) and pass through FASER$\nu$ in Run 3 for several energy ranges: $E>10,\ 100,\ 300$ and $1000$~GeV. As can be seen, among neutral hadrons, the most numerous are muon-induced neutrons and anti-neutrons, followed by $\Lambda$ baryons and a smaller number of kaons and other mesons. Most of these neutral hadrons are expected to interact in the first part of FASER$\nu$, since the hadronic interaction length in the tungsten is only $\sim 10~\cm$. These simulations show that roughly 3600, 440, and 10 neutral hadrons are expected with energies above 100, 300, and 1000 GeV, respectively, that will typically interact in the upstream part of the detector.  For all of these energy ranges, the number of neutral hadrons is far below the number of muon neutrino interactions expected.

\begin{table}[t]
\centering
\begin{tabular}{|c|c|c|c|c|c|}
\hline
\hline
\multirow{2}{*}{Particle} & \multicolumn{4}{c|}{Expected number of particles passing through FASER$\nu$} \\
\cline{2-5}
 & $E>10$~GeV &  $E>100$~GeV &  $E>300$~GeV & $E>1$~TeV \\ 
\hline
Neutrons $n$ & $27.8$k / $138$k & $1.5$k / $11.5$k & $150$ / $1.1$k & $2.2$ / $42$ \\
\hline
Anti-neutrons $\bar{n}$ & $15.5$k / $98$k & $900$ / $9$k & $110$ / $1.5$k & $2.8$ / $46$ \\
\hline
$\Lambda$ & $5.3$k / $36$k & $390$ / $4.1$k & $39$ /  $800$ & $0.9$ / $58$\\
\hline
Anti-$\Lambda$ & $3.4$k / $31$k & $290$ / $3.5$k & $31$ / $200$ & $0.6$ / $14$ \\
\hline
$K^0_S$ & $1.3$k / $30$k & $240$ / $6.8$k & $52$ / $390$ & $1.8$ / $6.2$ \\
\hline
$K^0_L$ & $1.6$k / $31$k & $270$ / $5.7$k & $55$ / $500$ & $1.2$ / $18$ \\
\hline
$\Xi^0$ & $240$ / $1.3$k & $13$ / $190$ & $2.3$ / $12$ & $0.1$ / $-$ \\
\hline
Anti-$\Xi^0$ & $150$ / $1$k & $10$ / $200$ & $1.4$ / $19$ & $-$\\
\hline
Photons $\gamma$ & $2.2$M / $62$M & $160$k / $16.3$M & $38.2$k / $6.3$M & $5.9$k / $1.1$M \\
\hline
\hline
$\nu_\mu+\bar\nu_\mu$ (signal int.) & 23.1k& 20.4k & 13.3k & 3.4k \\
\hline
\hline
\end{tabular}
\caption{The expected number of negative-muon-induced particles passing through FASER$\nu$ in LHC Run 3, as estimated by a dedicated FLUKA study.  In each entry, the first number is the number of particles emerging from the rock in front of FASER$\nu$, and the second is the number of particles produced in muon interactions in the tungsten plates in FASER$\nu$.  $2\times 10^9$ muons are expected to pass through FASER$\nu$ in Run 3. We note that the statistical uncertainties of the numbers presented in this table can reach even factors of a few, especially for the less abundant neutral hadrons.
\label{table:BGtable}}
\end{table}

On the other hand, the dominant source of secondary particles passing through FASER$\nu$ is muon interactions within the detector volume.  The rates are shown as the second number in each entry in \tableref{BGtable}.  This is especially relevant, if muon photo-nuclear interactions in some of the tungsten plates produce neutral hadrons that travel a typical distance of the hadronic interaction length in tungsten before interacting. Given a possible spatial separation between the parent muon track and secondary hadron interaction vertex, as well as the large total track density in the emulsion films, it is not possible to properly associate neutral hadrons with the parent muons going through the detector in most cases. We estimate that the relevant total contribution will be about an order of magnitude larger than the one coming from secondaries produced in rock. Importantly, however, secondary interactions of neutral hadrons produced in muon scatterings in the tungsten plates will be more uniformly distributed along the entire detector length. Similarly to the other sample, however, for all energy ranges above 300 GeV, the number of neutral hadrons is below the number of muon neutrino interactions expected.

In the left panel of Fig.~\ref{fig:BGspectrum}, we show the fitted energy spectrum of all neutral hadrons emerging from rock or produced within the detector volume. For comparison, we also show the spectrum of neutrinos interacting in FASER$\nu$. As can be seen, the number of signal neutrino interactions drops slowly with increasing energy, while the number of neutral hadrons drops exponentially to rates below the signal rate. 

\begin{figure}[tpb]
\centering
\includegraphics[width=0.48\textwidth]{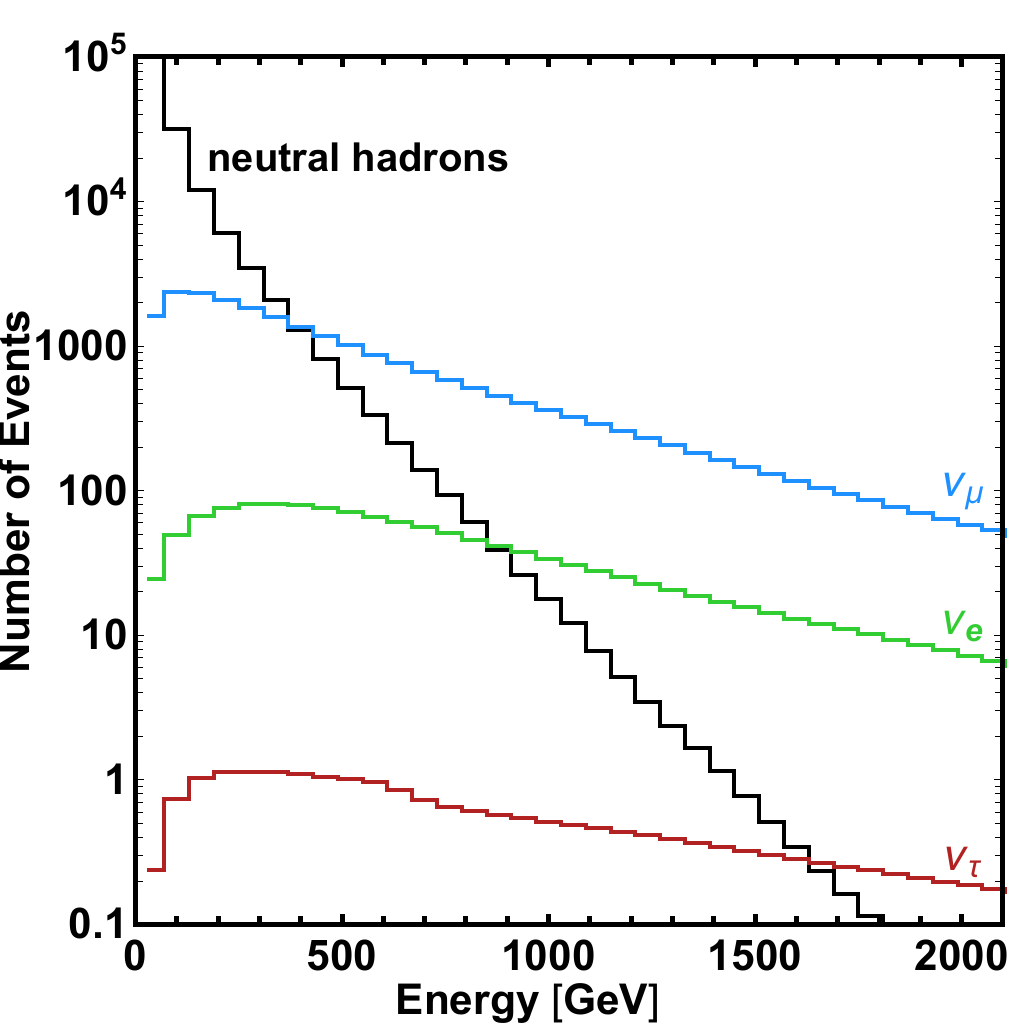}
\includegraphics[width=0.48\textwidth]{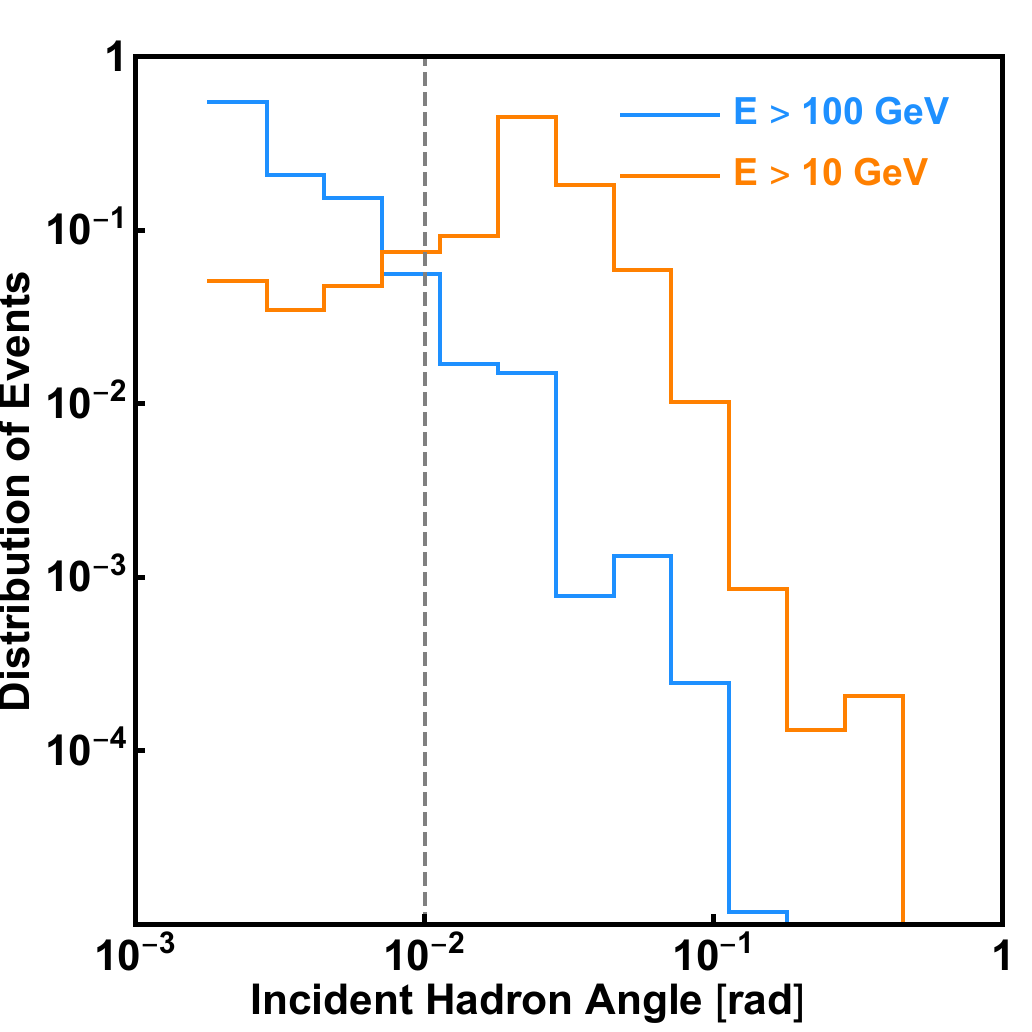}
\caption{The energy and angular distributions of neutral hadrons that are produced by negative muon interactions and pass through FASER$\nu$ in Run 3.  \textbf{Left}: The energy spectrum of neutral hadrons produced in the rock in front of FASER$\nu$ and produced within the detector (black) as well as neutrinos interacting with the detector for an integrated luminosity of $150~\ifb$: $\nu_e$ (red), $\nu_\mu$ (blue) and $\nu_\tau$ (green). \textbf{Right}: The angular distribution of neutral hadrons produced in the rock in front of FASER$\nu$ and passing into the detector for energies $E>10~\gev$ (orange) and $E>100~\gev$ (blue). The angle is given with respect to the beam collision axis.  The estimated angular resolution of a $\sim 100~\gev$ hadron is about $10~\mrad$, as indicated by the vertical dashed line. }
\label{fig:BGspectrum}
\end{figure}

The angular distribution of neutral hadrons emerging from the rock in front of FASER$\nu$ and entering FASER$\nu$ is shown in the right panel of Fig.~\ref{fig:BGspectrum}. The angular distribution of neutral hadrons produced within the detector volume is not included here, but is expected to be similar. We find that most low-energy neutral hadrons will be produced with angles larger than the directional resolution of the detector, while high-energy ones are more collimated along the beam collision axis. The expected resolution for direction reconstruction of $\tev$ neutral hadrons and neutrinos undergoing CC interactions in FASER$\nu$ is approximately $3-5~\mrad$, while for $\sim 100~\gev$ neutral hadrons, it is closer to $10~\mrad$, as indicated by the vertical dashed line. As a result, we estimate that more than 90\% of high-energy ($>100~\gev$) neutral hadrons will have a reconstructed direction consistent with coming from the ATLAS IP. On the other hand, only about $20\%$ of neutral hadrons with lower energy, $E\sim 10~\gev$, will be produced with $\theta\alt 10~\mrad$.

\subsection{Background for Neutral Vertex Search}
\label{sec:bg-vertex}

In the above discussion, we have estimated the flux of high-energy particles going through FASER$\nu$ that could interact within the detector volume. We will now discuss in more details the corresponding expected interaction rates in the detector, as well as ways to discriminate between any possible background and the neutrino-induced signal events characterized by at least $5$ charged tracks emerging from a single vertex.

On top of these discrimination methods, it is also important to mention that any muon-induced background could be significantly reduced when combined with electronic detectors, which can actively tag the parent muon and provide additional timing information about the entire event if other tracks are also detected. In the discussion below, we focus on differentiating signal from background with FASER$\nu$ alone, but an interface with the FASER spectrometer could further improve this discrimination in the future.

\begin{description}

\item[Muons] By far the largest non-neutrino flux of high-energy particles going through FASER$\nu$ is muons from the direction of the ATLAS IP. As discussed above, muons can source other background particles interacting within the detector volume. However, the parent muon in such events typically accompanies the secondary particles and passes through FASER$\nu$.  The efficiency for detecting muon tracks exceeds $90\%$ per segment of an emulsion detector. As a result, the number of events in which a muon passes the first $n$ segments undetected is less than $N_\mu \times 0.1^n$. We can see that after $n=10$ layers (out of the total 1000), corresponding to the first centimeter of the detector, none of the $N_\mu \sim 2\times 10^9$ muons passing through FASER$\nu$ in Run 3 is expected to remain undetected.
    
Even when the first few layers of the detector are considered, to be misreconstructed as neutrino interactions, a muon interaction would have to produce at least 5 charged tracks in the detector that satisfy respective energy and angular cuts. The probability of this to happen is highly suppressed. We have actually not seen any such event in the FLUKA simulation with $10^9$ muons impinging on a 1-mm-thick tungsten layer.

On the other hand, if these muons are attached to a neutrino interaction vertex, the event would look like a charged particle interaction. This will introduce an inefficiency in the neutral vertex search. The expected number of background muons per neutrino interaction vertex is $N_\mu^\text{exp}=\pi r^2 \rho$, where $r$ is the radius of the parent charged particle search, and $\rho$ is the muon track density (per area).
For $r=2~\mu\text{m}$ and $\rho=6\times 10^5~\cm^{-2}$ (corresponding to 30/fb of data), $N_\mu^\text{exp}=0.075$. 7.5\% of events could be discarded from the neutrino vertex candidates. The inefficiency could be reduced by requiring a significant scattering angle of parent track at the vertex. If the parent candidate is just a penetrating muon, it will not bend a single mrad at the vertex. 

\item[Photons] The next most abundant high-energy particles that could interact in FASER$\nu$ are muon-induced photons. Photons emerging from the rock in front of FASER$\nu$ will typically interact in the front layers of the detector or its edges, while the ones produced within the detector volume will often shower in the vicinity of the parent muon track. Importantly, EM showers are easily distinguishable from neutrino vertices as they typically correspond to only two charged tracks in the initial vertex from $e^+e^-$ pair-production and more rarely to three charged tracks from photon scattering in the electron field. We have additionally studied photo-nuclear processes in the tungsten layers using FLUKA, generating two samples, each corresponding to $10^7$ incident photons with energies $E_\gamma=100~\gev$ and $1~\tev$. We have found no events with 5 or more tracks that had enough energy and an angular distribution to mimic the neutrino signal.

\item[Neutral Hadrons] A potentially important source of background is associated with a subdominant flux of muon-induced neutral hadrons that could mimic neutrino signatures in the detector when scattering in the tungsten layers. Although a detailed analysis of how to discriminate between this background and neutrino signatures is left for future study and dedicated multivariate analyses, it is useful to briefly summarize several features of the signal events to be employed for this purpose. 

The ratio of the total number of high-energy neutrino signal events to the number of neutral hadron background events is expected to be $S/B\sim 3$ by requiring a reconstructed energy $E>300~\gev$, as can be deduced from \tableref{BGtable}. A further substantial improvement in the signal identification, as well as determination of the neutrino flavor for CC events, will be achieved by tagging an outgoing charged lepton in the interaction vertex as discussed below.

It is also possible for neutral hadrons crossing FASER$\nu$ to leave a decay signature. However, for energetic neutral hadrons, the decay-in-volume probability is typically suppressed with respect to the probability of scattering in the tungsten layers. In addition, such decays are typically two-body decays that will not constitute background for neutrino searches, since they will not be identified as vertices with $n_{\text{tr}} \ge 5$ charged tracks.\footnote{Note that such decays happening in the FASER main detector will also not contribute to background for searches for light long-lived particles, since they will be rejected based on the detection of an accompanying parent muon in the front veto layers.}

\end{description}

\subsection{Background for Flavor-Specific Neutrino Interactions}
\label{sec:bg-neutrino}

A further improvement in background rejection will be achieved by identifying charged leptons at the interaction vertex. This will reject most of the neutral hadron events. We discuss this briefly below for each individual neutrino flavor, beginning with the first target of FASER$\nu$, muon neutrino CC events.

\begin{description}

\item[$\nu_\mu$ CC events]  In the case of a CC interaction of a muon neutrino, an outgoing muon will leave an extended track in the emulsion detector that can be easily discriminated from background. The muon track will be identified by determining its length in the detector and by analyzing the momentum distribution of the tracks coming from the vertex. According to our \textsc{Genie} simulations, the muon is the highest momentum particle (HMP) in 72\% of $\nu_\mu$ CC interactions with at least five charged tracks. This number increases to 93\% after a distance of two hadronic interaction lengths, $\lambda_{\text{int}}$, from the vertex, as hadrons are rejected based on their interactions in the tungsten layers. The HMP without any hadronic interaction is selected as a muon candidate. 

The background to $\nu_\mu$ CC events are $\nu$ NC or neutral hadron events with a high-energy hadron mis-identified as a muon. To discriminate between these two, we consider two simple test variables. One is the angle $\phi$ between the HMP and the vectorial sum of momenta of all the other tracks, measured in the two-dimensional transverse plane. As the lepton and hadron systems have back-to-back kinematics in this plane, $\phi$ is expected to peak at 180$^\circ$. The second variable we can consider is the momentum fraction of the HMP with respect to the sum of all visible tracks, $P_{\text{HMP}}/\Sigma_i^n P_i$. The HMP is selected among the charged particles that pass at least 2$\lambda_{\text{int}}$, which eliminates 86\% of hadrons. As a result, for NC and neutral hadron interactions, the ratio $P_{\text{HMP}}/\Sigma_i^n P_i$ peaks at small values. The scatter plot of these variables (including smearing discussed in \secref{energyresolution}) is shown in \figref{ccnc} for muon neutrino interactions. The same plot for neutron interactions ($E_n>100$ GeV) is also shown on the right.

By applying the simple cuts $\phi>\frac{\pi}{2}$, $P_{\text{HMP}}>20~\gev$, and $P_{\text{HMP}}/\Sigma P>0.15$, the efficiency for identifying $\nu_\mu$ CC events with energies $E_\nu>100~\gev$ within the fiducial volume is $\epsilon_{\mu,\text{CC}}=86\%$. Here the fiducial volume is defined to be the front 80\% of the FASER$\nu$ detector, which leaves at least two interaction lengths after the vertex for muon identification.  In addition, a high event sample purity, weighted by the expected event rates, can be achieved, with $N_{\text{CC}} / (N_{\text{CC}} + N_{\text{NC}} + N_{\text{neutral hadron}})=86\%$. The purity increases for the case where the muon traverses more than $2\, \lambda_{\text{int}}$. Also, the higher the energy, the lower the contamination from neutron background.

\begin{figure}[t]
\centering
\includegraphics[width=0.49\textwidth]{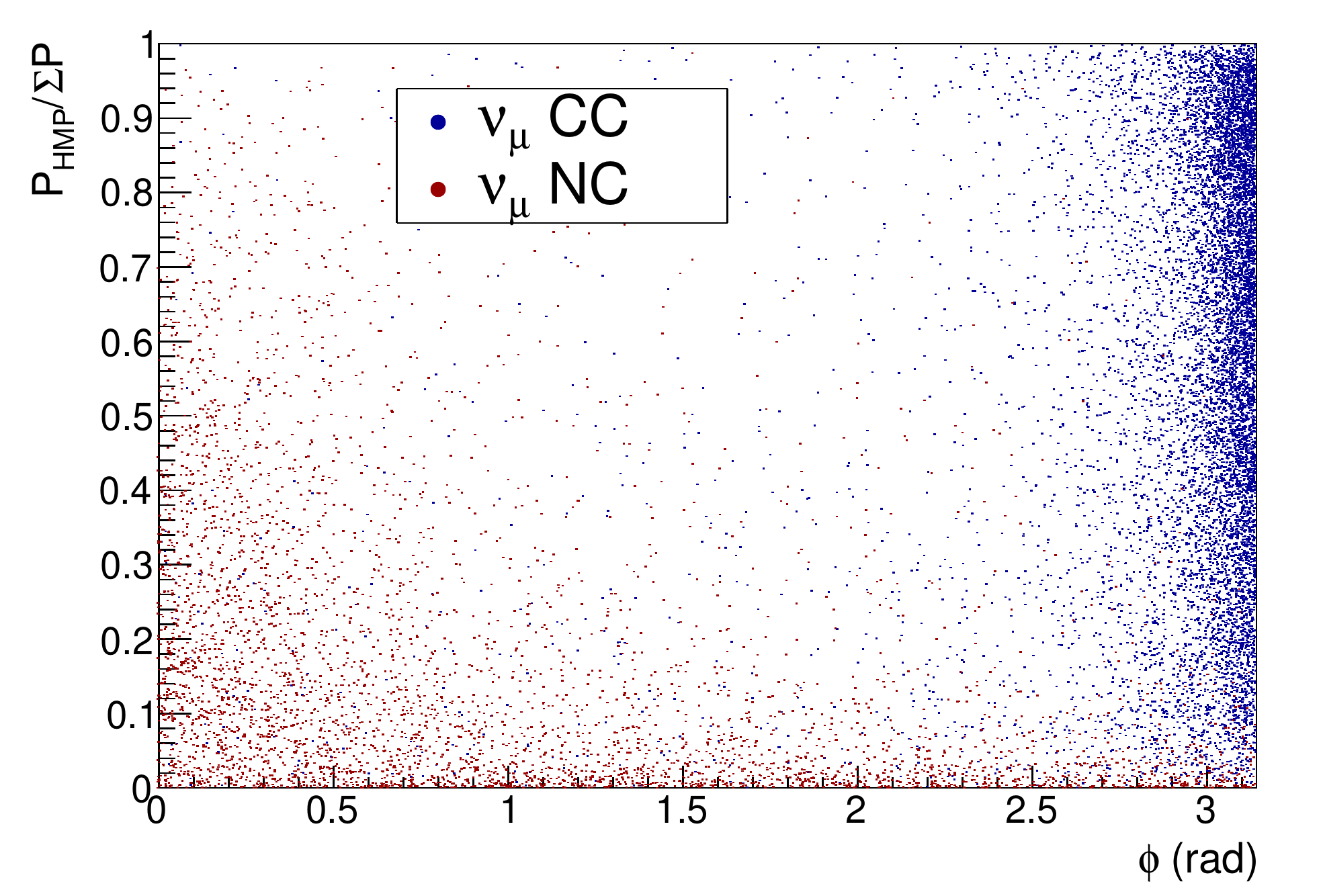}
\includegraphics[width=0.49\textwidth]{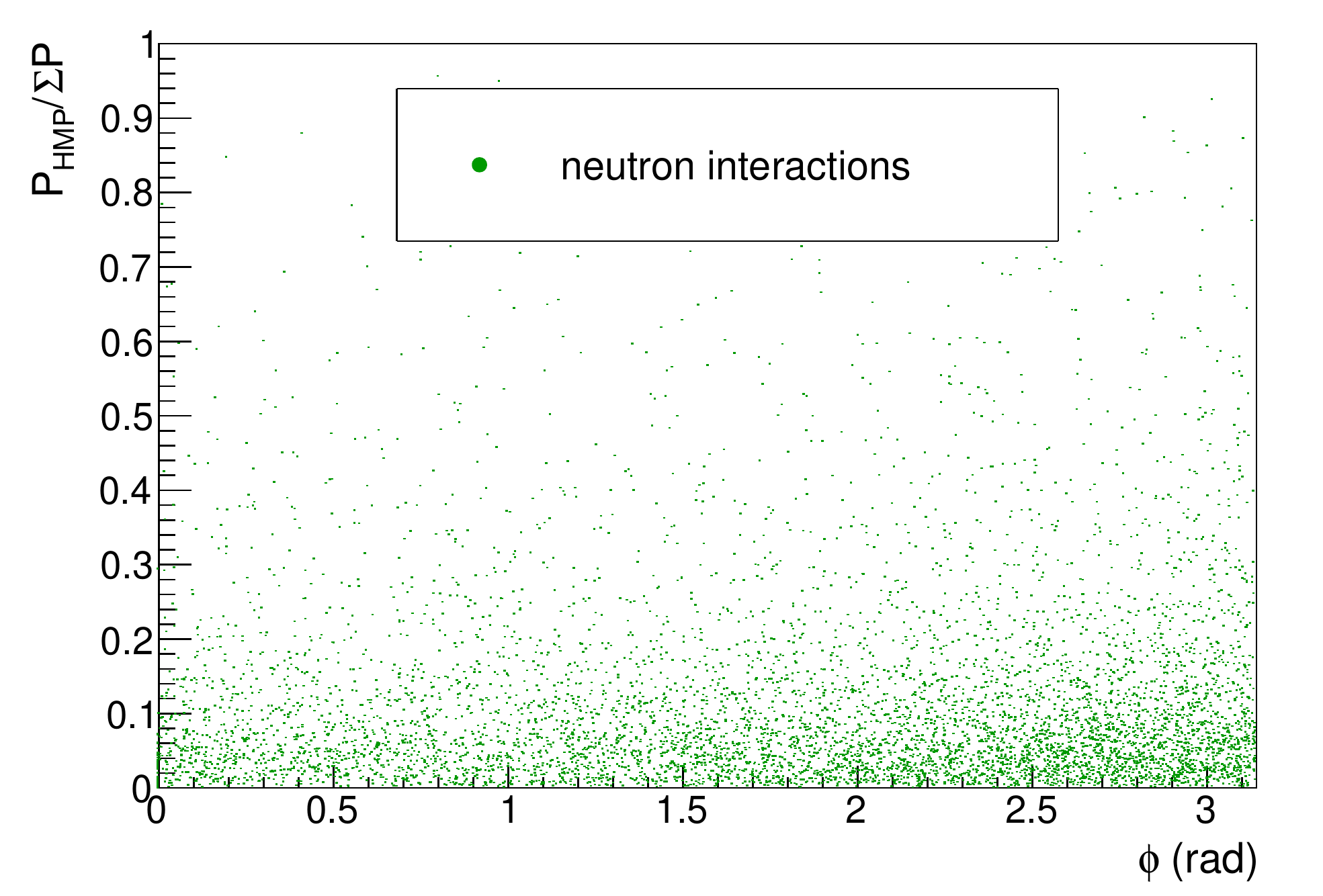}
\caption{\textbf{Left}: Distribution of the angle $\phi$ between the HMP and the vectorial sum of momenta of all tracks, in the two-dimensional transverse plane, and the momentum fraction of the HMP, for $\nu_\mu$ CC (blue) and NC (red) interactions. The MC sample is generated by \textsc{Genie} with the FASER $\nu_\mu$ spectrum, and the same smearing as in \secref{energyresolution} is applied.
\textbf{Right}: Same plot for neutral hadron backgrounds.
} 
\label{fig:ccnc}
\end{figure}

\item[\boldmath{$\nu_e$} CC events]  An outgoing electron produced in the CC interaction of a high-energy electron neutrino will promptly initiate an EM shower, which can be easily distinguished from other charged tracks coming from the vertex. This greatly reduces the impact of any potential neutral-hadron-induced background for $\nu_e$ CC interactions. In particular, high-energy EM showers could be initiated by decays of neutral pions, $\pi^0 \to \gamma \gamma$, when pions are produced in scatterings of neutral hadrons. However, these can be efficiently discriminated from single-electron-initiated showers based on the corresponding shower development and energy deposition in consecutive emulsion films. At the same time, we have verified, using simulations with the \textsc{Qgsjet-ii-04} MC generator, that backgrounds from neutron interactions to $\nu_e$, in which a high-energy electron with $E_e>100~\gev$ emerges from a neutron interaction vertex, are negligible, of order one per mille for 1 TeV neutron interaction. In addition, most of such background electrons come from 3-body decays of final-state pions, $\pi \to \gamma e e$, produced in neutron scatterings. Therefore, they will once again initiate EM showers distinct from the $\nu_e$-induced signal. 

An additional potential background is $\nu_\mu$ CC interactions, where an outgoing muon track overlaps with an EM shower from, e.g., a bremsstrahglung photon. However, if such a $\nu_\mu$ CC interaction takes place within the fiducial volume of the detector, the outgoing muon will be efficiently identified based on its extended track. An example of such an identification strategy based on HMP found among particles surviving $2\,\lambda_{\textrm{int}}$ was mentioned above. Note that, for the high energies of our interest, the probability of a muon to lose much of its energy within the detector before it travels far enough to be identified, due to, e.g., catastrophic energy loss, is below $10^{-4}$. Muon-neutrino CC interactions are therefore expected to be a negligible background to $\nu_e$ CC events.

\item[\boldmath{$\nu_\tau$} CC events]  In case of CC interactions of $\tau$ neutrinos, it is essential to exploit an additional signal feature associated with the displaced secondary vertex coming from decays of $\tau$ lepton. Possible sources of background for such a signature include NC interactions of other (dominant) neutrino flavors or neutral hadrons producing hadrons that could lead to a secondary displaced interaction vertex in the detector. However, these secondary vertices are typically characterized by a large track multiplicity, making it less probable for them to mimic, e.g., a 1-prong (high-energy) $\tau$ vertex. Another source of background is CC muon/electron neutrino interactions with failed muon/electron identification, especially when there is associated charm production in the neutrino interaction vertex. Notably, this was an important source of background for the $\nu_\tau$ searches performed by the OPERA experiment, because of the inefficiency of muon identification for low-energy muons. On the other hand, muon identification is much improved thanks to long high-energy $\mu$ tracks in FASER$\nu$, which allows one to significantly suppress this background.  In addition, the aforementioned $\phi$ angle is again a powerful discriminator of $\nu_\tau$ from the NC, charm, and neutral-hadron-induced backgrounds, as it is typically large for outgoing $\tau$ leptons from the neutrino scattering events in contrast to the HMPs produced in background events.

\end{description}

\section{Collider Neutrino Detection and Cross Section Measurements}
\label{sec:measurements}

Given the large number of high-energy neutrinos produced at the LHC, \FASERnu will be able to detect collider neutrinos for the first time and measure their interaction cross section at very high energies.  In this section we describe these physics goals for FASER$\nu$ in LHC Run 3 from 2021-23.

\subsection{First Detection of a Neutrino at the LHC}
\label{sec:firstnu}

So far, no experiment has detected neutrinos from colliders. Although the other experiments at the LHC are equipped with massive detector components, the high background fluence of hadronic particles hasn't allowed them to identify neutrinos. In contrast, thanks to the hadron shielding of 100 m, the fluence of hadronic particles at the FASER location is extremely low. At the same time, FASER$\nu$ is in a position where the high-energy neutrino flux is high, and \FASERnu will be the first experiment at the LHC with sensitivity to neutrinos. 

In 2018 we installed a pilot neutrino detector, which consisted of two 15 kg modules. One module had 100 layers of 1-mm-thick lead plates and emulsion films, and the other module had 120 layers of 0.5-mm-thick tungsten plates and emulsion films. The transverse dimensions were $12.5\,\cm \times 10\,\cm$ for each module. The modules were installed in plastic boxes and placed side by side on the beam collision axis, as shown in the left panel of \figref{pilotrun}. The pilot neutrino detector collected $12.5~\ifb$ of data from  September to October 2018. During this time, about 40 $\nu_\mu$ CC interactions were expected to occur in the detector. The central panel of \figref{pilotrun} shows a part of the reconstructed data. A track density of $\simeq 3\times 10^5$ tracks/$\cm^2$ was observed.  To find vertices in this data, we required multiple charged particles emerging from one point, and several vertex candidates were selected in a sub-sample analysis. The right panel of \figref{pilotrun} shows a vertex candidate that was found in a sub-sample of analyzed data, which has no charged incoming particle. Such events demonstrate the capability of FASER$\nu$ to observe LHC neutrinos. Further details will be reported in future publications~\cite{FASERmuondata}.

\begin{figure}[t]
\centering
\includegraphics[height=4cm]{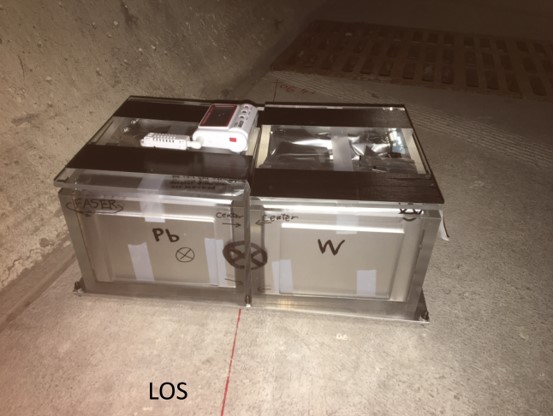}
\includegraphics[height=4cm]{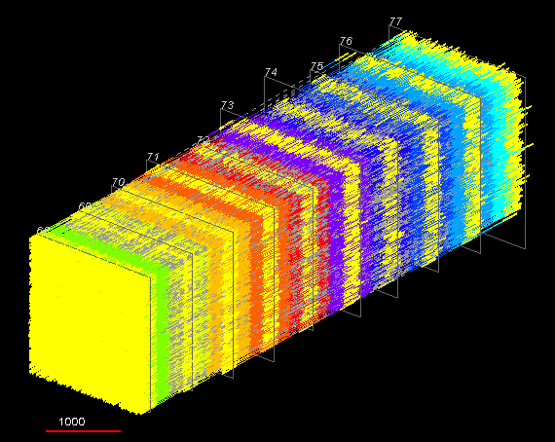}
\includegraphics[height=4cm]{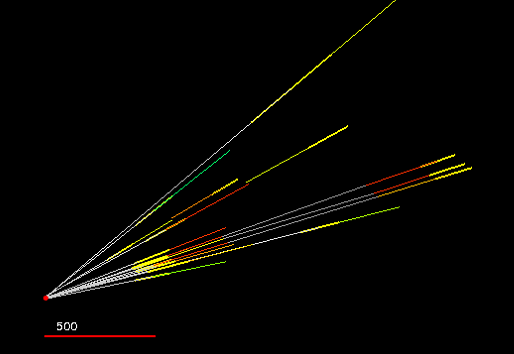}
\caption{\textbf{Left}: The 30 kg pilot neutrino detector that was installed in the TI18 tunnel in 2018. It collected 12.5 fb$^{-1}$ of data. \textbf{Center}: Reconstructed tracks in 2 mm $\times$ 2 mm $\times$ 10 emulsion films. About 13,000 tracks were observed, corresponding to about $3\times 10^5$ tracks/\si{cm^2}. \textbf{Right}: A vertex found in the detector with no incoming charged track. The vertex axis is compatible with the beam direction. The red scale bars in the center and right figures are $1000~\mu\text{m}$ and $500~\mu\text{m}$ long, respectively.}
    \label{fig:pilotrun}
\end{figure}

\subsection{Tau Neutrino Detection}

Of the seventeen particles in the standard model of particle physics, the tau neutrino is the poorest measured. Directly detecting a $\nu_\tau$ requires that the neutrino beam has enough energy to produce a $\tau$ lepton ($E_\nu > 3.5$ GeV), which must then be identified. As $\tau$ leptons are short-lived ($c\tau = 87~\mu\text{m}$) and their decays always involve tau neutrinos, which escape measurement, the identification of $\tau$ leptons is extremely difficult. DONuT and OPERA have observed $\sim 10$ $\nu_\tau$ events each~\cite{Kodama:2007aa, Agafonova:2018auq}, and these datasets provide the primary information about tau neutrinos at present. Although SuperKamiokande and IceCube have recently reported higher statistics $\nu_\tau$ appearance in atmospheric oscillations~\cite{Li:2017dbe,Aartsen:2019tjl}, care must be taken when interpreting the physics in these cases as identification is only made through statistical means. In addition, those measurements rely on knowing the flux of atmospheric neutrinos, which has large uncertainties.  As a result, despite the larger number of events, the resulting $\nu_\tau$ cross section constraints from SuperKamiokande and IceCube are at the $\sim 30\%$ uncertainty level and comparable to OPERA. One also has to take into account that the measurements from oscillated $\nu_\tau$ neutrinos are at comparatively low energies ($E_\nu<70~\gev$), while DONuT observed neutrino interactions in the fully DIS regime. In the future, the proposed SHiP beam dump experiment at CERN~\cite{Anelli:2015pba} may provide high statistics $\nu_\tau$ measurements up to $E_\nu \simeq 150~\gev$.

During LHC Run 3 with an integrated luminosity of $150~\ifb$, FASER$\nu$ will accumulate $\sim 21~\nu_\tau$ CC interactions peaked at energies $\sim 1~\tev$, as shown in \tableref{intrate}, and about 11 $\nu_\tau$ events will be identified after taking into account geometrical acceptance, vertex detection and tau identification efficiencies.  This will significantly increase the world's supply of reconstructed $\tau$ neutrinos and will allow them to be studied at much higher energies.

\subsection{Charged Current Cross Section Measurements}

\FASERnu will be able to both identify neutrino events and estimate the corresponding neutrino energy. Assuming no new physics contribution to neutrino production, the observed neutrino spectrum at \FASERnu can be used to measure the neutrino interaction cross section.  Additionally, the measured neutrino events and kinematics could provide valuable input for the tuning of MC tools used to simulate high-energy neutrino events.  

\begin{figure}[t]
\centering
\includegraphics[width=0.32\textwidth]{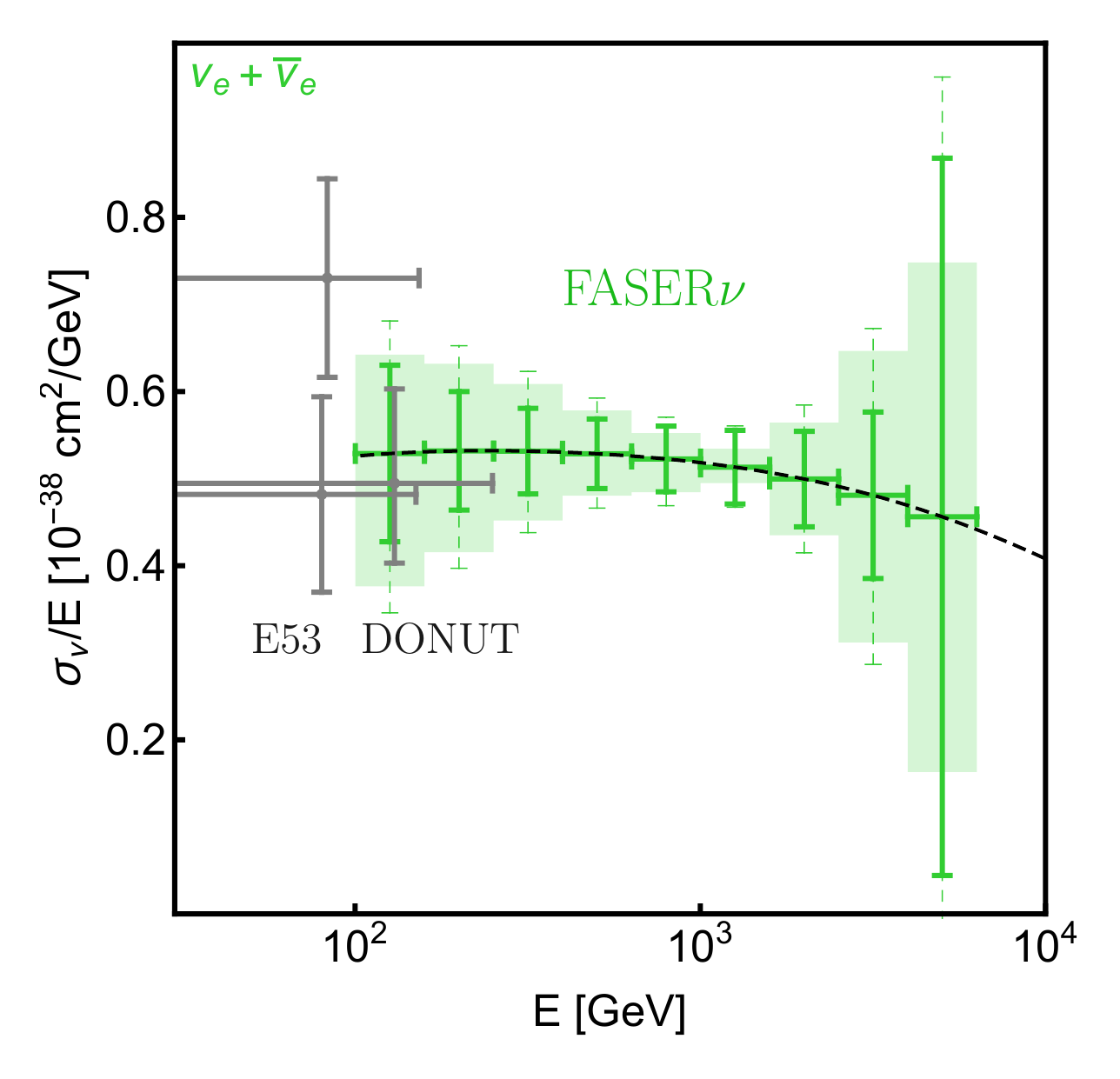}
\includegraphics[width=0.32\textwidth]{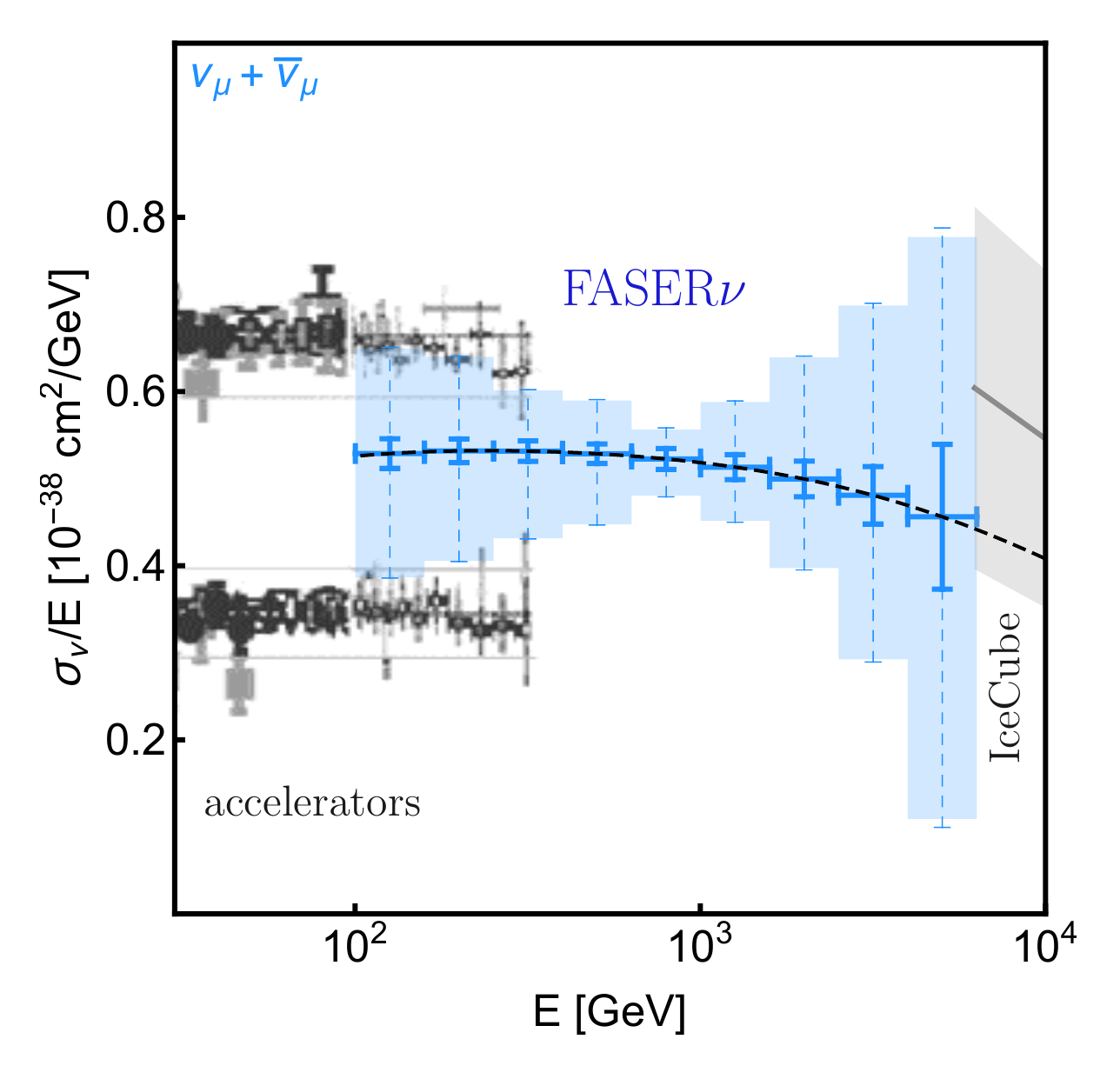}
\includegraphics[width=0.32\textwidth]{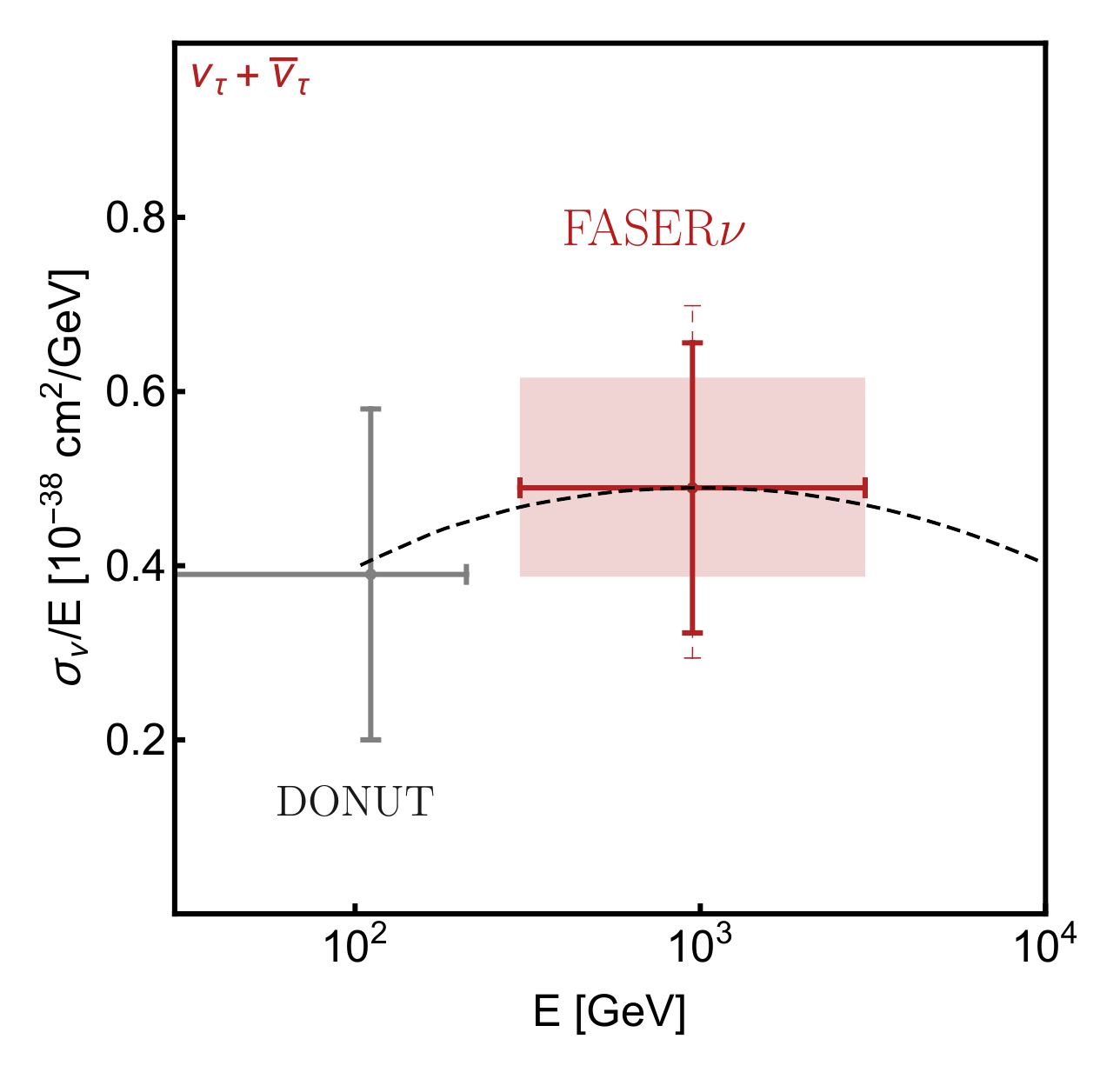}
\caption{FASER$\nu$'s estimated $\nu$-nucleon CC cross section sensitivity for $\nu_e$ (left), $\nu_{\mu}$ (center), and $\nu_\tau$ (right) at Run 3 of the 14 TeV LHC with an integrated luminosity of $150~\ifb$ collected from 2021-23. Existing constraints~\cite{Tanabashi:2018oca} are shown in gray for $\sigma_\nu$ and $\sigma_{\bar\nu}$ at accelerator experiments and for their weighted average at IceCube. The black dashed curve is the theoretical prediction for the average DIS cross section per tungsten-weighted nucleon, as introduced in \eqref{average_xs}. The solid error bars correspond to statistical uncertainties, the shaded regions show uncertainties from neutrino production rate corresponding to the range of predictions obtained from different MC generators, and the dashed error bars show their combination.  Here we include the geometrical acceptance, vertex detection efficiency and lepton identification efficiency as discussed in the text.
}
\label{fig:xs_measurement}
\end{figure}

Without an interface to the FASER spectrometer, \FASERnu will not be able to distinguish between the lepton charges. \FASERnu will therefore constrain the average of the neutrino and anti-neutrino cross sections 
\be
\langle\sigma\rangle= \frac{\phi_\nu \sigma_{\nu }+\phi_{\bar\nu}\sigma_{\bar\nu }}{\phi_\nu +\phi_{\bar\nu}} \approx \frac{\sigma_{\nu }+\sigma_{\bar\nu }}{2} \ ,
\label{eq:average_xs}
\ee
where $\sigma_{\nu}$ and $\sigma_{\bar\nu}$ are the neutrino and anti-neutrino cross sections, and $\phi_\nu$ and $\phi_{\bar\nu}$ are the neutrino and anti-neutrino fluxes, respectively. In the last step we have taken the fluxes to be roughly equal $\phi_\nu \approx \phi_{\bar\nu}$, consistent with the results in \secref{rates}. The expected SM cross section is shown as a dashed black line in \figref{xs_measurement}. Existing measurements are shown in gray. Note that \FASERnu is sensitive to the tungsten-weighted cross section. Other experiments report the isoscalar-weighted cross section, but since tungsten has a very high neutron fraction ($Y_n=1.48$), a direct comparison of \FASERnu's measurements with those of other experiments must account for this difference.

In \figref{xs_measurement} we show FASER$\nu$'s expected sensitivity to constraining neutrino CC cross sections. The solid error bars show the sensitivity considering only statistical uncertainties, while the shaded region shows the systematic uncertainties from the range of neutrino production rates predicted by different MC generators. The combination of statistical and production rate uncertainties, added in quadrature, is shown as the dashed error bars.  When estimating the sensitivity, we (i) require the neutrino to interact $2\,\lambda_{\text{int}}$ before the end of the detector to permit muon identification and the development of electromagnetic showers needed for energy reconstruction, resulting in a geometrical acceptance of $80\%$; (ii) apply the energy-dependent vertex detection efficiency obtained in \secref{features}; (iii) apply a 86\% muon identification efficiency (see \secref{bg-neutrino}) and a 75\% tau identification efficiency (see \secref{features}); and (iv) assume the measurement is background free. Here we consider 5 energy bins per decade for $\nu_e$ and $\nu_\mu$, corresponding to an effective energy resolution of $\Delta E_{\text{bin}}/E_{\text{bin}}=45\%$ for a bin. For $\nu_\tau$ we only consider one energy bin, given the low number of expected $\nu_\tau$ events. As discussed in \secref{energyresolution}, our simulations suggest that an energy resolution for \FASERnu of about 30\% is achievable. 

We can see that \FASERnu significantly extends the neutrino cross section measurements to higher energies for both electron and tau neutrinos. For muon neutrinos, \FASERnu will close the gap between existing measurements using accelerator experiments and IceCube. Note that an additional interface between \FASERnu and the FASER spectrometer will allow one to distinguish $\nu_{\mu}$ and $\bar{\nu}_{\mu}$. The muon-neutrino measurement is limited by the production uncertainty, but the tau-neutrino measurement is statistically limited. In the electron-neutrino case, both contributions are roughly equally important. This stresses the importance of a precise and accurate modeling of forward neutrino production for \FASERnu measurements. 

Although \FASERnu can measure the interaction cross section, we note that it is difficult to identify viable models of BSM physics that significantly modify the total cross section at these energies. Uncertainties in the parton distribution functions used to calculate neutrino interaction cross sections only become relevant for $E_\nu \agt 10^7-10^8~\gev$~\cite{CooperSarkar:2011pa}. New physics models such as large extra dimensions could significantly increase the cross section above the scale of the new physics~\cite{AlvarezMuniz:2001mk}, but the relevant parameter space has constraints from the LHC~\cite{Sirunyan:2018ipj,Aaboud:2018bun}. Other new physics models that induce neutrino non-standard interactions could affect the overall rate.

\section{Additional Physics Studies}
\label{sec:studies}

In addition to detecting collider neutrinos of all three flavors and measuring their cross sections at higher energies than observed from any previous man-made source, FASER$\nu$ can explore the physics of neutrino production, propagation, and interaction at the energy frontier. This section sketches some of the diverse studies that will be undertaken. Thanks to copious production of neutrinos at the LHC, these studies may not be primarily limited by statistical precision, and therefore robust estimates of their physics reach must await a careful accounting of the detector performance and all relevant theoretical and experimental uncertainties. For the present survey, among the systematic effects neglected in whole or in part are: cross section uncertainties, including nuclear shadowing and anti-shadowing effects; final state interaction in the tungsten target nuclei; and biases, non-uniformity or other uncertainties in energy response and reconstruction, signal efficiency, and backgrounds. In what follows, only statistical errors and a rough estimate of neutrino production uncertainties (as characterized by the spread between several models) are included, unless otherwise stated.

\subsection{Heavy-Flavor-Associated Channels}

In addition to the inclusive CC cross section, we can also study specific neutrino interaction processes. One example is charm-associated neutrino interactions $\nu N \to \ell X_c + X$. This process has been studied indirectly at a variety of neutrino experiments~\cite{Rabinowitz:1993xx, Baroncelli:1981jya, Abramowicz:1982zr, Vilain:1998uw, Goncharov:2001qe, KayisTopaksu:2008aa}, which search for two oppositely-charged muons originating from the charged current interaction and the charm decay, respectively. In contrast, the CHORUS and OPERA experiments have demonstrated that an emulsion detector can be used to identify the secondary charm and tau decay vertices~\cite{KayisTopaksu:2011mx, Agafonova:2014khd}. This approach permits better background rejection and also makes it possible to distinguish neutral and charged charm hadrons. Similarly, as discussed in \secref{detector}, the \FASERnu detector will be able identify both charm and beauty hadrons based on their decay topology. 

The charm-associated neutrino interaction processes are best parameterized as the relative charm hadron production rate
\be
\frac{\sigma(\nu_\ell N \to \ell \, X_c + X)}{\sigma(\nu_\ell N \to \ell + X)} \ .
\ee
The systematic uncertainty from the neutrino flux normalization cancels in this ratio. In the left panel of \figref{charm}, we show the prediction for the weighted average of the fraction of charm associated (anti-)neutrino interactions, obtained using \textsc{Pythia~8} with the \textsc{NNPDF3.1nnlo} parton distribution function, for all (black), neutral (orange), and charged (blue) charm hadrons. We see that, depending on the energy, 10\% to 20\% of neutrino interactions at \FASERnu produce a charm hadron, leading to a large event sample for these processes. Previous measurements obtained in the CHORUS and E531 experiments, probing neutrino energies $E_\nu < 200~\gev$ below \FASERnu's sensitivity, are shown as gray error bars~\cite{KayisTopaksu:2011mx}. Note that since these measurements use a different target material and a beam with different neutrino/anti-neutrino ratio, and therefore cannot be directly compared with the \FASERnu predictions.

\begin{figure}[tbp]
\centering
\includegraphics[width=0.475\textwidth]{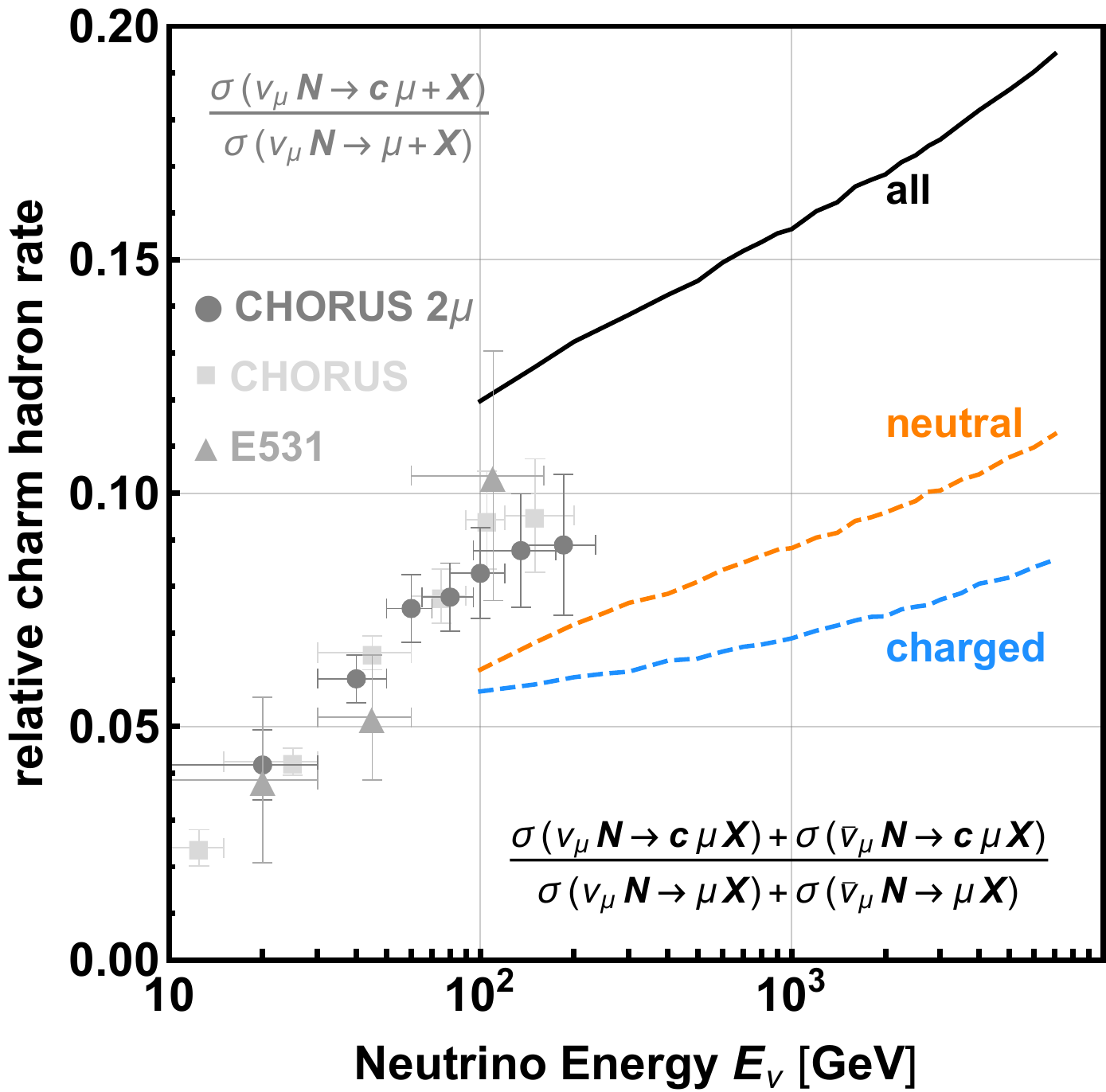}
\includegraphics[width=0.49\textwidth]{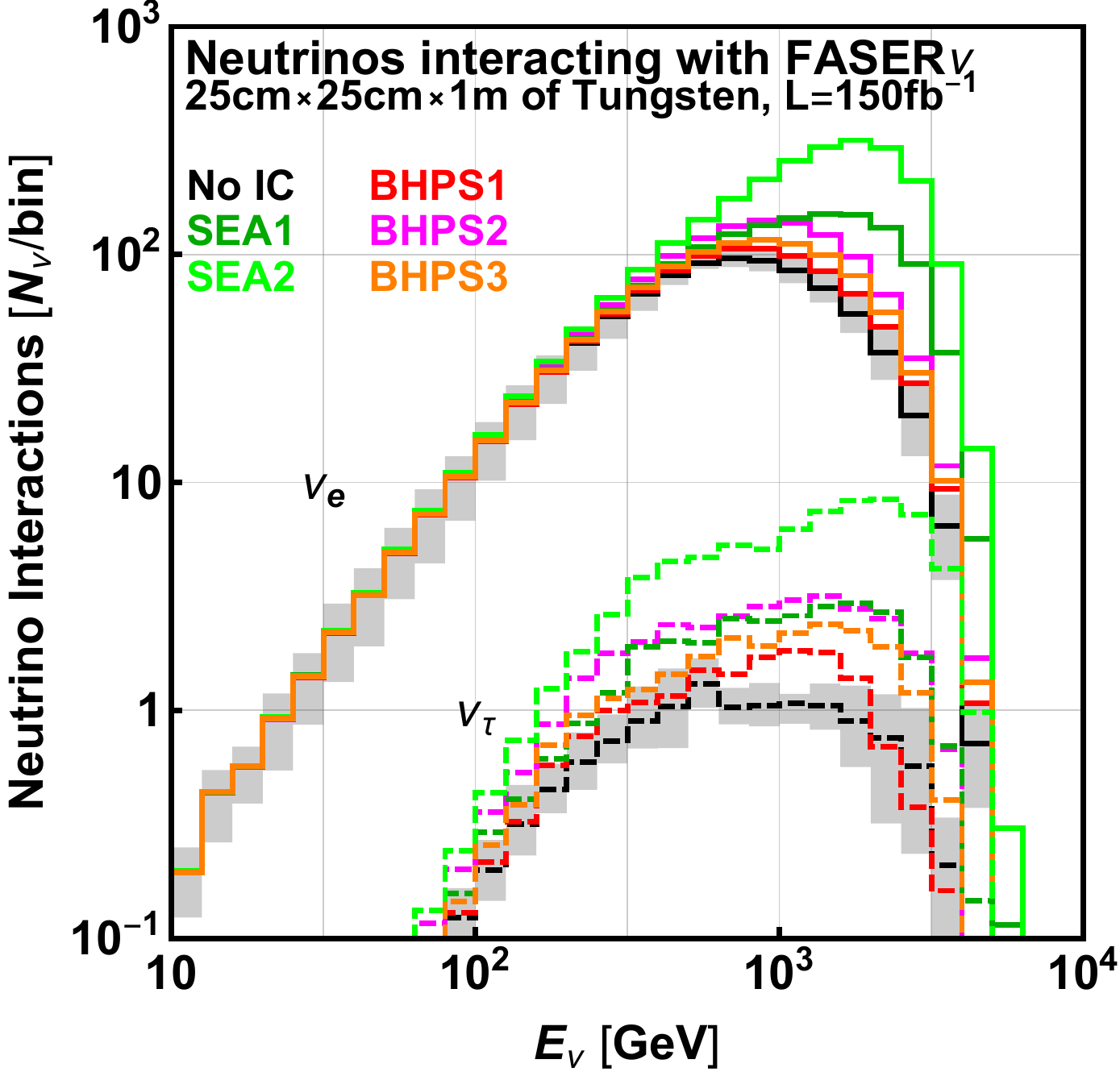}
\caption{\textbf{Left}: Relative charm hadron production rate in neutrino interactions. The solid black curve shows the expected fraction of charm-associated neutrino interactions in the \FASERnu detector as a function of neutrino energy, obtained with \textsc{Pythia~8}. The dashed blue and orange curve further separate the fraction into charged and neutral charm hadrons, respectively. The gray error bars show the results of previous measurements obtained in the CHORUS and E531 experiments~\cite{KayisTopaksu:2011mx}. Note that these results cannot be directly compared due to the different neutrino vs. anti-neutrino beam compositions and the different neutron fractions of the target material for CHORUS and \FASERnu. \textbf{Right}: Electron (solid) and tau (dashed) neutrino interactions in FASER$\nu$ using different intrinsic charm models in \textsc{CT14nnlo-IC}.}
\label{fig:charm}
\end{figure}

A similar search can be performed for beauty-hadron-associated neutrino interactions. In the SM, these processes are suppressed by $\mathcal{O} (V_{ub}^2) \simeq 10^{-5}$, and the expected number of beauty-associated neutrino interactions is $\mathcal{O}(0.1)$ in Run 3. A larger scale experiment would be needed to observe a sizable number of beauty events. 

It is important to note, however, that these heavy flavor event rates can be modified by BSM physics, such as $W'$ bosons, charged Higgs boson, and leptoquarks at the TeV scale. As mentioned in \secref{existing}, this possibility has been motivated by recent results from collider experiments, which provide tentative evidence of lepton universality violation in the decays $B \to D^{*} \ell \nu$~\cite{Lees:2013uzd,Hirose:2016wfn,Aaij:2017uff}, $B \to K^{*} \ell\ell$~\cite{Aaij:2017vbb}, and $B^+ \to K^+ \ell\ell$~\cite{Aaij:2019wad}. Four-point interactions from BSM physics that may be contributing to lepton universality violation are shown in the left panel of \figref{Bdiagrams}.  These may correspondingly impact neutrino interactions through the diagrams shown in the center panel of \figref{Bdiagrams}, enhancing otherwise exotic beauty-associated neutrino events, such as those shown in the right panel of \figref{Bdiagrams}.

\begin{figure}[t]
\centering
\includegraphics[width=0.8\textwidth]{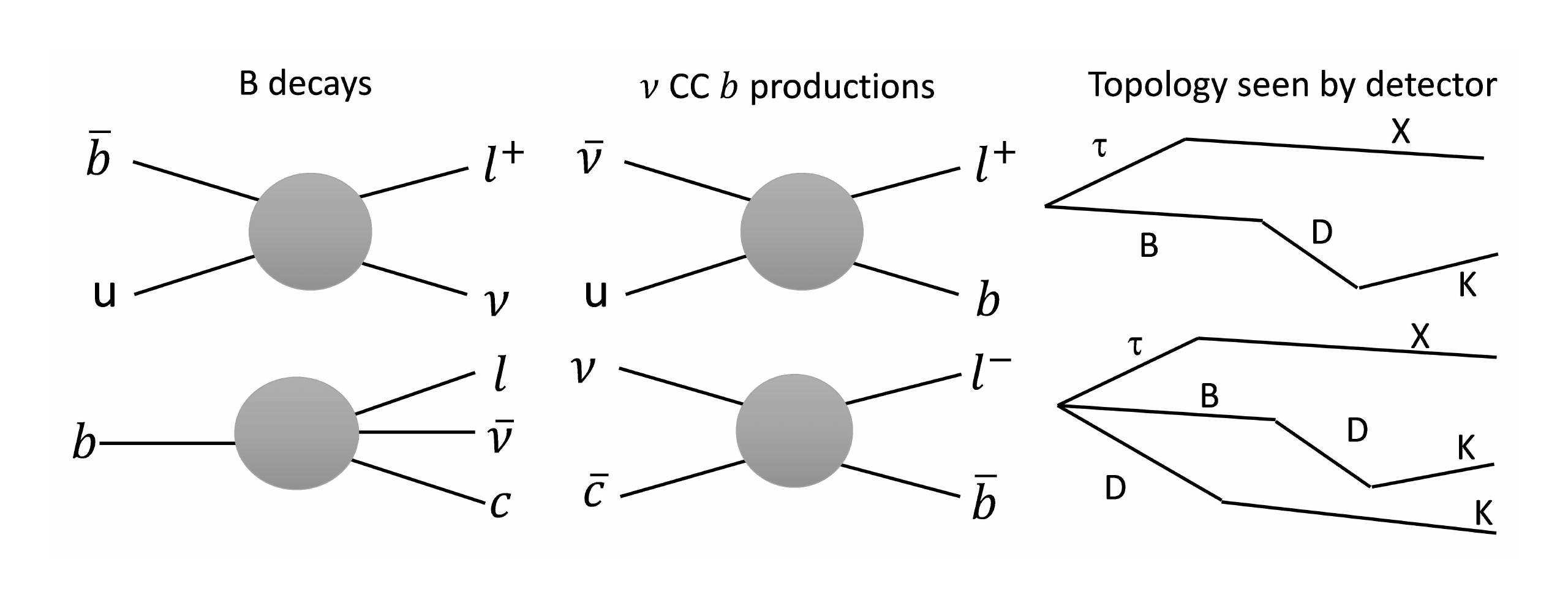}
\caption{\textbf{Left}: Diagrams of purely leptonic (upper) and semi-leptonic (lower) beauty decays. \textbf{Center}: The corresponding neutrino CC interactions with $b$ quark production. In the bottom diagram, perturbative charm quarks contribute to the neutrino interaction. \textbf{Right}: The event topology of $\nu_\tau$ CC interactions that produce beauty hadrons. For simplicity, the conjugate diagrams are omitted.}
    \label{fig:Bdiagrams}
\end{figure} 

Of course, the LHC and other colliders not only motivate beauty-associated BSM physics, but they also provide strong constraints~\cite{Aad:2012ej, Chatrchyan:2012gqa, Sirunyan:2019hkq, Takahashi:2019zsl}.  We leave a careful study of the complementarity of \FASERnu and LHC searches in this area to future work. 

\subsection{Neutrino Production and Intrinsic Charm}

Assuming that the neutrino interaction cross sections are SM-like, we can use neutrino measurements at FASER$\nu$ to analyze neutrino production in the very forward direction at the LHC.  As we have seen earlier, the different hadronic interaction models differ at the 10 - 50\% level, depending on the flavor and energy, and therefore measurements of the neutrino spectrum at FASER$\nu$ could help validate and improve these models. 

This is particularly interesting for neutrino production via heavy meson decays, which is sensitive to the poorly constrained charm parton distribution function. Besides the perturbative component described by DGLAP evolution, a sizable non-perturbative contribution to the charm parton distribution function is possible. We consider five parameterizations of this intrinsic charm component, as implemented in the \textsc{CT14nnlo-IC} parton distribution function~\cite{Hou:2017khm}. The \textsc{BHPS} model predicts a ``valence-like'' intrinsic charm component, peaking around momentum fraction $x\sim 0.3$, which will lead to an enhancement of forward high-energy charm-hadron production. The \textsc{SEA} model is parameterized by a ``sea-like'' non-perturbative function that is proportional to the light quark distributions.

In the right panel of \figref{charm} we show the $\nu_e$ (solid) and $\nu_\tau$ (dashed) interaction rates in FASER$\nu$. The black line corresponds to the expected rate using the perturbative charm and non-charm contributions, and the shaded region shows the corresponding range of predictions obtained by using different generators. The colored lines correspond to the five intrinsic charm parameterizations in \textsc{CT14nnlo-IC}~\cite{Hou:2017khm}. We see that an intrinsic charm component can significantly enhance the neutrino production rate in the regions where neutrino production in charmed hadron decay dominate: the high-energy regime of electron neutrino production, as well as tau neutrino production.  The intrinsic charm parameterization shown in the right panel of \figref{charm} predicts an enhancement of up to a factor of 10 for the tau neutrino rate.  In contrast, the flux of high-energy muon neutrinos is less sensitive to modifications of the charm production rate, which could be used to constrain the flux of neutrinos coming from kaon decays and therefore help to associate any possible excess of electron neutrinos with the charm component.

The measurement of forward neutrino production will be a key input for high-energy neutrino measurements by large-scale Cherenkov observatories, such as IceCube~\cite{Aartsen:2016nxy}, ANTARES~\cite{Collaboration:2011nsa}, Baikal-GVD~\cite{Avrorin:2013uyc}, and KM3NeT/ARCA~\cite{Adrian-Martinez:2016fdl}.  IceCube has measured an as-of-yet unidentified high-energy component of the neutrino flux~\cite{Aartsen:2013bka}. The dominant background to this new flux at lower energies is the conventional (from pion and kaon decays) atmospheric flux. There is also a prompt component of the atmospheric flux from the decay of heavy mesons that is expected to become dominant at higher energies, although no evidence of it has been found in the data so far as IceCube's constraints push down on theoretical models that have been very difficult to calculate~\cite{Enberg:2008te,Bhattacharya:2016jce,Aartsen:2016xlq}.  A direct measurement of the prompt flux by FASER would provide important data not only for IceCube, but also for all current and future high-energy neutrino telescopes.

\subsection{Sterile Neutrino Oscillations}

For neutrinos with energy $E \sim 800~\gev$ propagating a distance $L=480~\m$, no oscillations are expected within the SM to a very good approximation. Thus any oscillation signal, either appearance of extra neutrinos above the expected rate or disappearance below the expected rate, is evidence of a new neutrino mass difference $\Delta m^2$ that is inconsistent with the two mass differences that are already established.

By measuring a deficit or excess of neutrinos, FASER can identify sterile neutrino oscillations fairly independent of flux uncertainties by leveraging the distinct shape of the oscillation signal and the fairly broadband beam of neutrinos produced by the LHC.
Given typical energies of $E \sim 800~\gev$ and a baseline of $L=480~\m$, the sterile neutrino mass difference FASER will be sensitive to satisfies
\begin{equation}
\frac{\Delta m^2_{41}L}{4E}=\frac{\pi}{2} \ ,
\end{equation}
which implies $\Delta m^2_{41} \sim 2,000$ eV$^2$. Larger $\Delta m^2_{41}$'s can also be constrained by observing an overall deficit in the normalization, up to normalization uncertainties.

Neutrino oscillations in a $3+1$ framework are governed by the Hamiltonian
\begin{equation}
H=\frac1{2E}UM^2U^\dagger \ ,
\end{equation}
where $M^2\equiv\diag(0,\Delta m^2_{21},\Delta m^2_{31},\Delta m^2_{41})$, the mixing matrix is $U=U_{34}U_{24}U_{14}U_{\rm PMNS}$, and the PMNS matrix~\cite{Pontecorvo:1967fh,Maki:1962mu} is the usual mixing matrix, $U_{\rm PMNS}=U_{23}U_{13}U_{12}$.  We ignore the standard $CP$ phase, any new $CP$ phases, and the matter potential, as they do not affect oscillations in this case.  We parameterize the size of the mixing in each of the three sectors as $|U_{\alpha 4}|^2$ for disappearance measurements or $\sin^22\theta_{\alpha\beta}\equiv4|U_{\alpha4}|^2|U_{\beta4}|^2$ for appearance channels, where $\alpha,\beta\in\{e,\mu,\tau\}$.

In the two-flavor approximation, the disappearance and appearance oscillation probabilities are
\begin{align}
P(\nu_\alpha\to\nu_\alpha)&=1-4|U_{\alpha4}|^2(1-|U_{\alpha4}|^2)\sin^2\frac{\Delta m^2_{41}L}{4E} \ ,\\
P(\nu_\alpha\to\nu_\beta)&=\sin^22\theta_{\alpha\beta}\sin^2\frac{\Delta m^2_{41}L}{4E} \ .
\end{align}
The oscillation probability and expected flux are shown in \figref{sterile_prob} for the disappearance channel and representative sterile parameters. This figure shows that the strength of the constraint comes from comparing the high-energy bins, where there is no oscillation, to the bins at the oscillation minimum (600 GeV in this example), where there is a deficit. For the appearance channels, the effect is most significant in the cases where the larger fluxes oscillate to the smaller fluxes, so $\nu_\mu\to\nu_\tau$, $\nu_\mu\to\nu_e$, and $\nu_e\to\nu_\tau$.

\begin{figure}[t]
\centering
\includegraphics[width=0.6\textwidth]{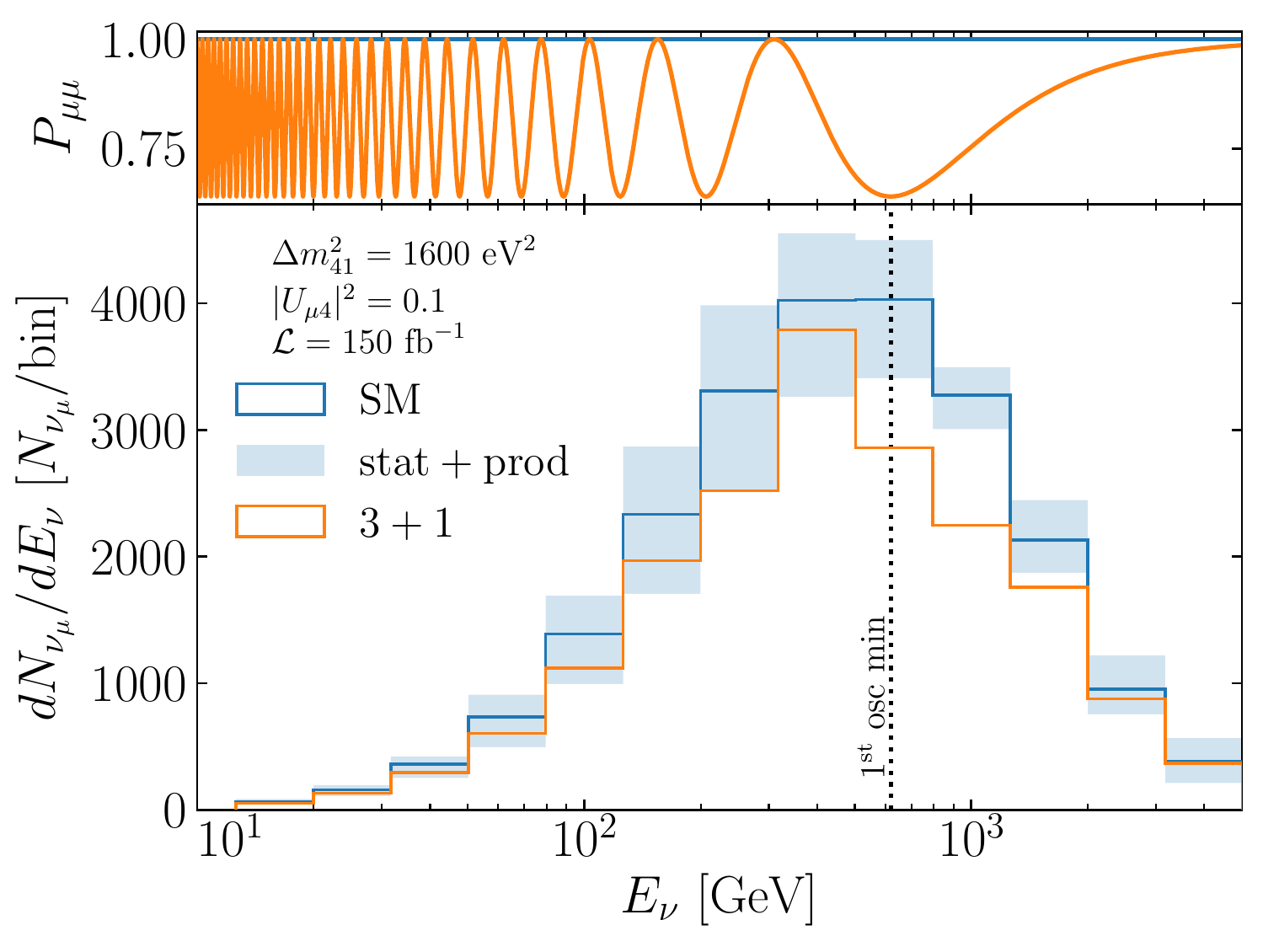}
\caption{\textbf{Upper panel}: The $\nu_\mu$ disappearance oscillation probability in the SM (blue) and in the presence of a sterile neutrino with $\Delta m^2_{41}=1,600$ eV$^2$ and $|U_{\mu4}|^2=0.1$ (orange). \textbf{Lower panel}: The expected $\nu_\mu$ event rate at FASER in the SM (blue) and in the presence of a sterile neutrino with the same parameters. The statistical and production uncertainties (added in quadrature) are shown in the blue shaded region. The location of the first oscillation minimum is marked with a dotted line.}
\label{fig:sterile_prob}
\end{figure}

Next, we calculate the sensitivity to sterile neutrino parameters. We use 5 bins per decade of energy, corresponding to an energy resolution of 45\%.  This is conservative compared the expected neutrino energy resolution of FASER$\nu$, which is approximately 30\% (see~\figref{ann}).  In addition, although the LHC is not a point neutrino source, the fact that neutrinos are created at different locations has an effect on $L/E$ that is smaller than the effect of the neutrino energy resolution and so is safely neglected.  We calculate the sensitivity to sterile neutrinos assuming an integrated luminosity of 150 fb$^{-1}$. To estimate our systematic uncertainties, we marginalize over the production uncertainties using a pull term~\cite{Fogli:2002pt} in the following way. We assume that the reality is the average of the hadronic models and then take the upper and lower limits from the different models as the standard deviation of a Gaussian prior. Note that here we only consider production uncertainties discussed in \secref{fluxes_rates}. Additional uncertainties, for example related to the detector response, require a careful study and are not considered at this moment. The contours are shown in \figref{sterile_contours} along with some previous constraints and anomalies from other neutrino experiments.

\begin{figure}[t]
\centering
\includegraphics[clip,trim = 0.15cm 0 0.3cm 0 ,width=0.325\textwidth]{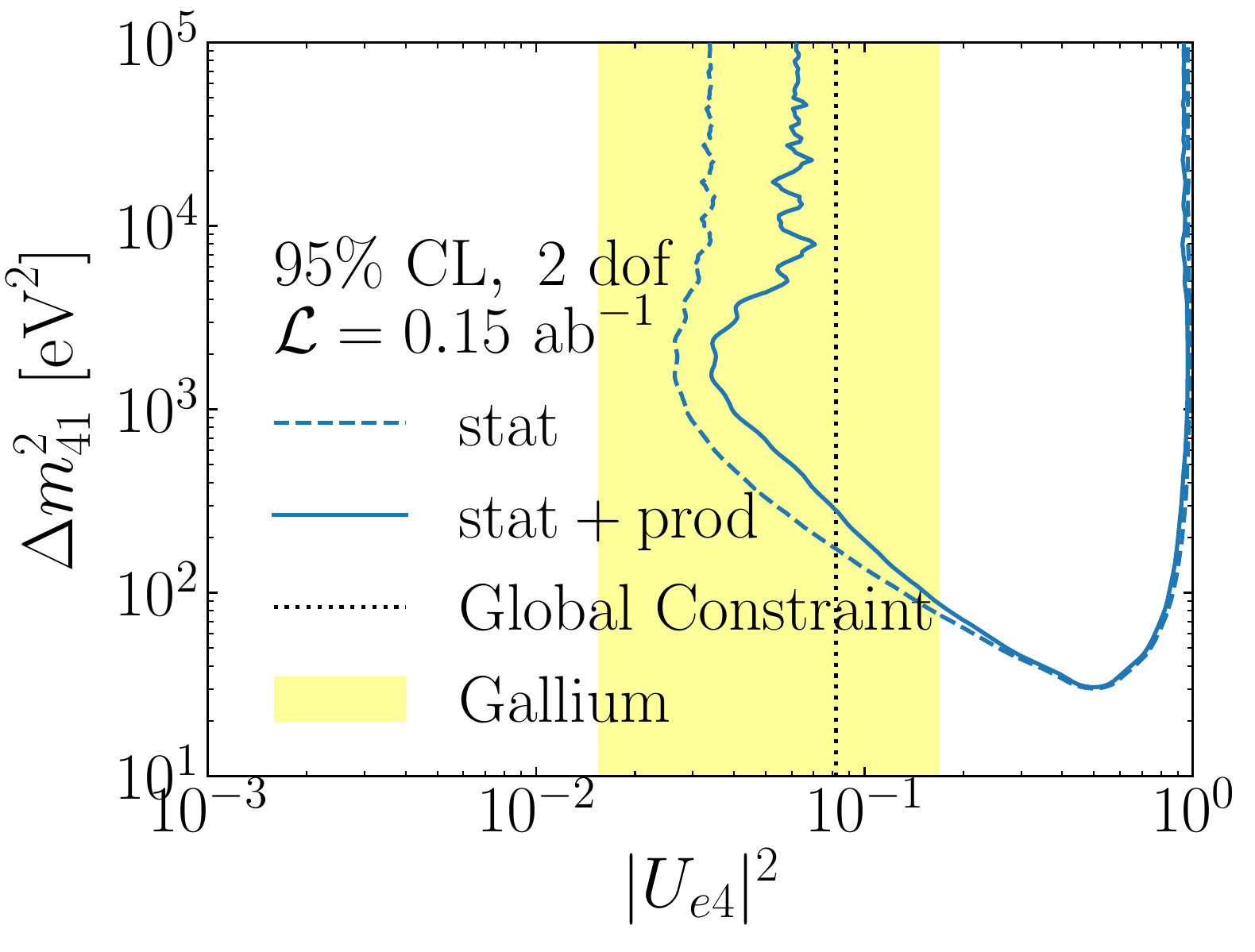}
\includegraphics[clip,trim = 0.15cm 0 0.3cm 0 ,width=0.325\textwidth]{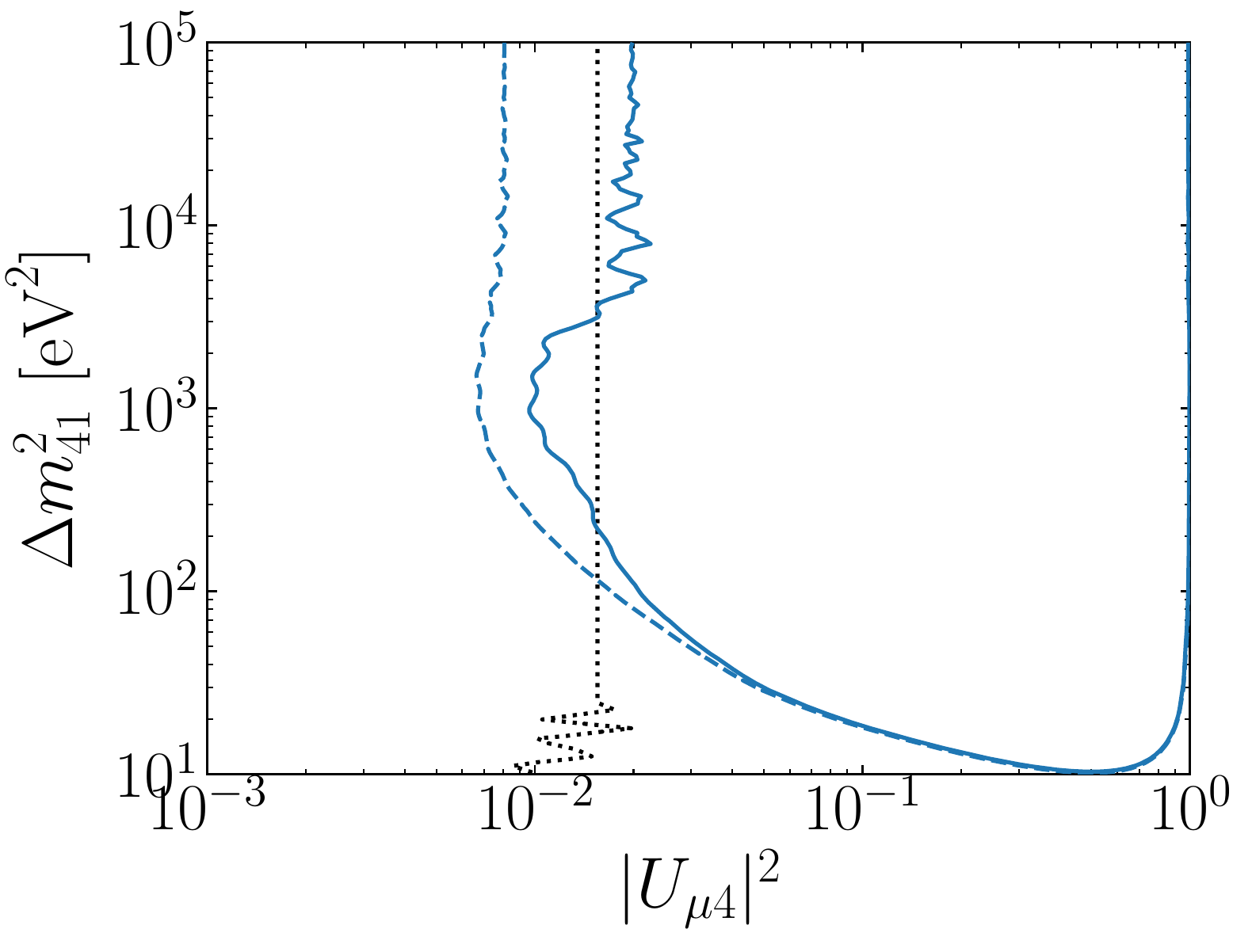}
\includegraphics[clip,trim = 0.15cm 0 0.3cm 0 ,width=0.325\textwidth]{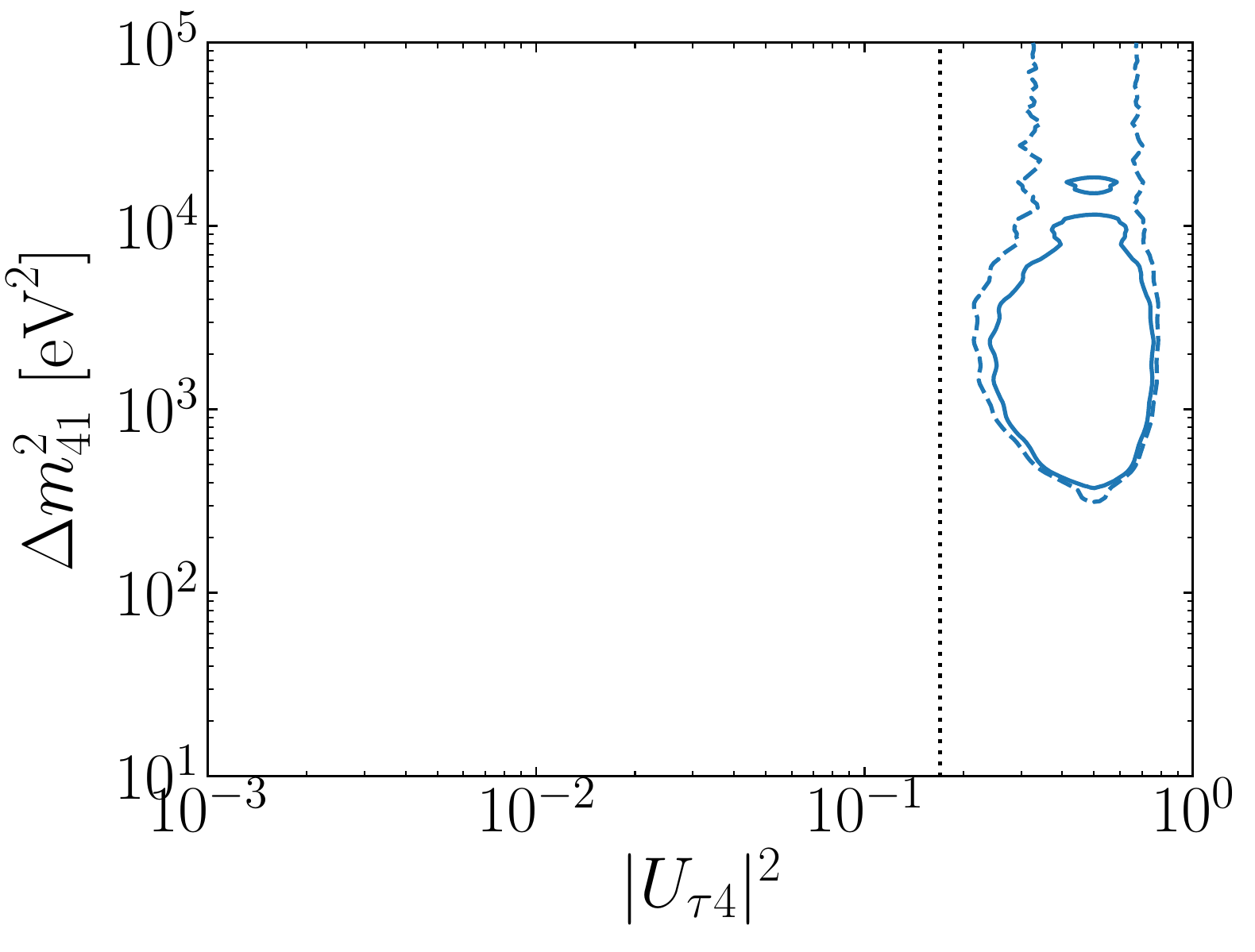}
\\
\includegraphics[clip,trim = 0.15cm 0 0.3cm 0 ,width=0.325\textwidth]{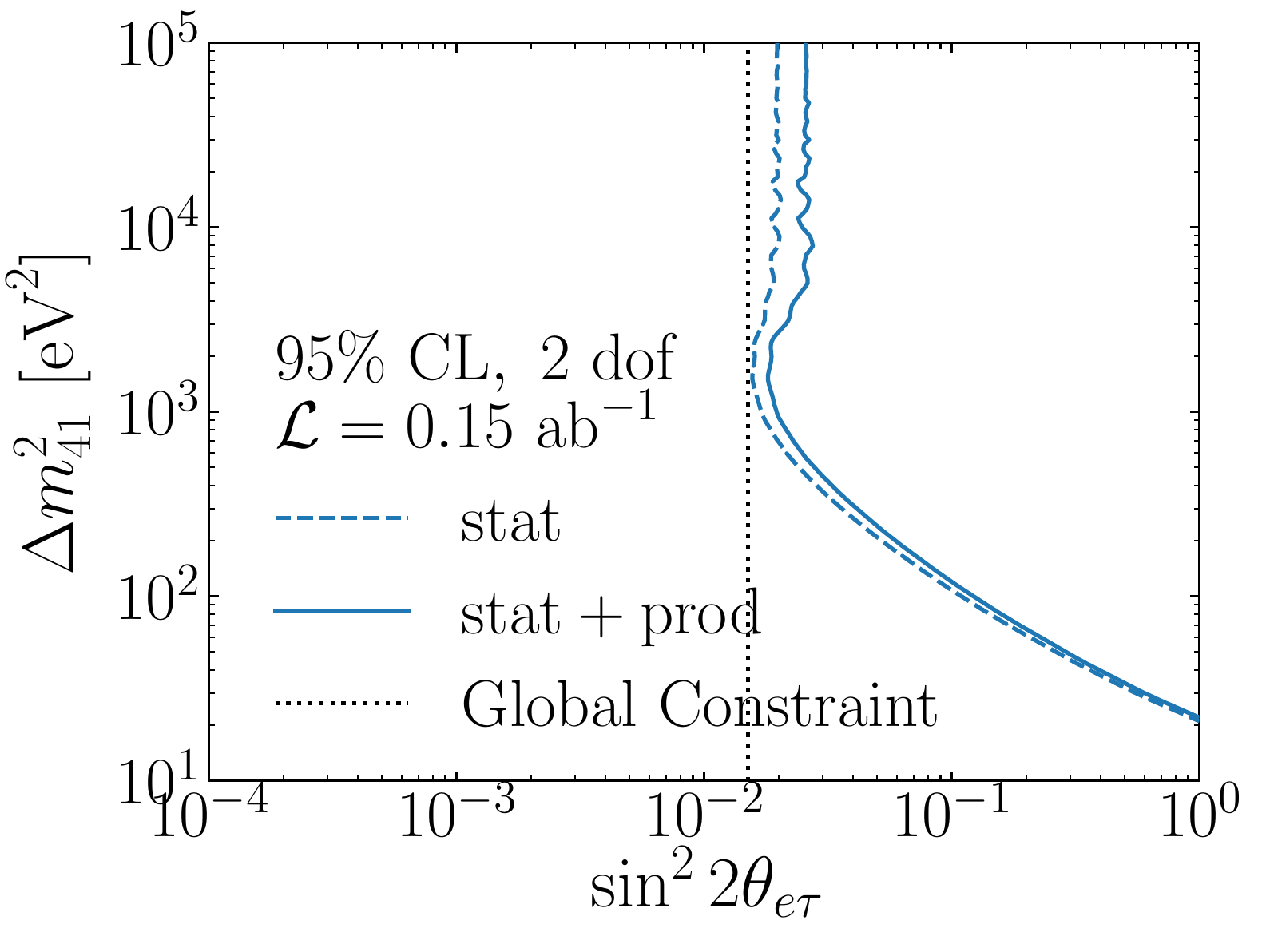}
\includegraphics[clip,trim = 0.15cm 0 0.3cm 0 ,width=0.325\textwidth]{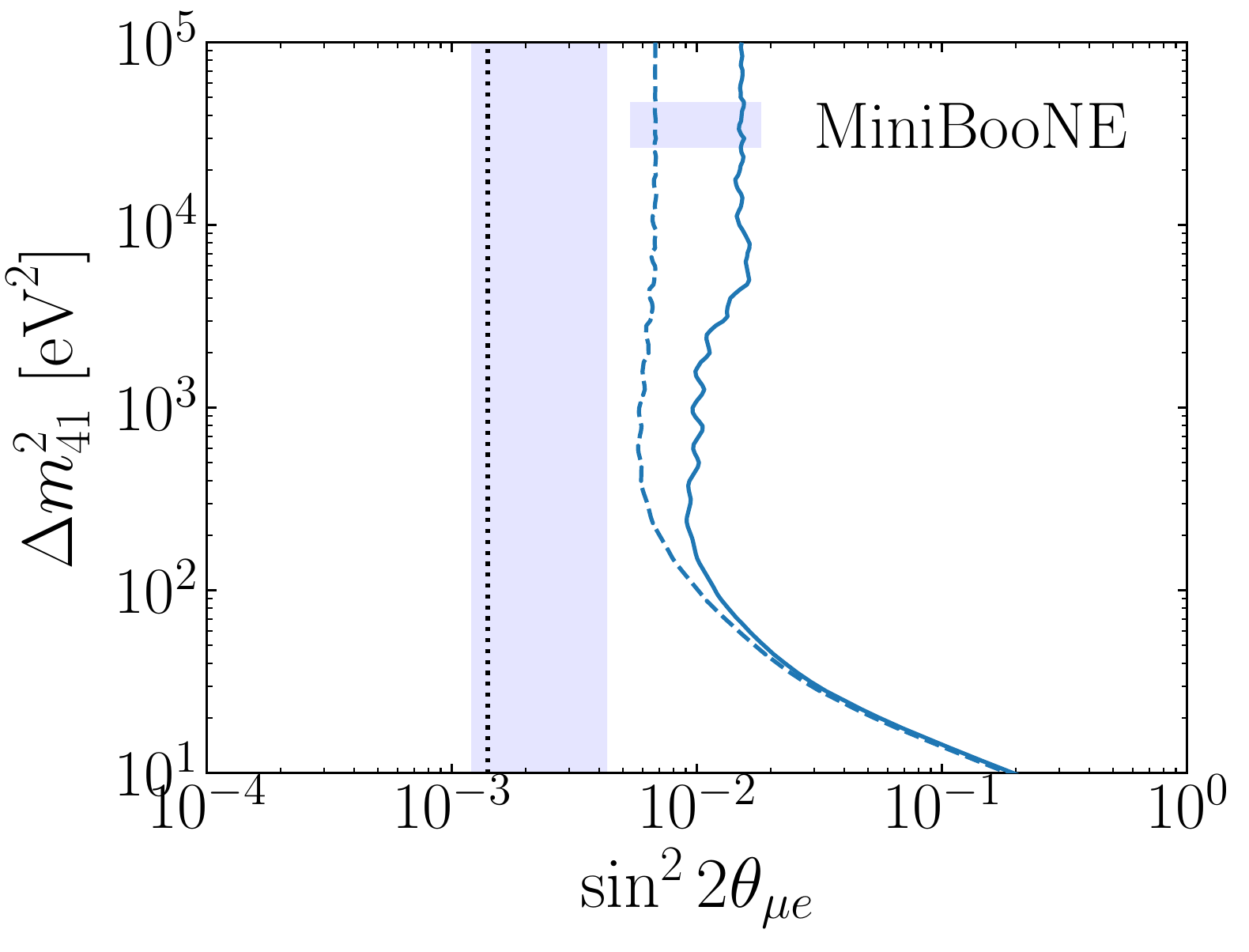}
\includegraphics[clip,trim = 0.15cm 0 0.3cm 0 ,width=0.325\textwidth]{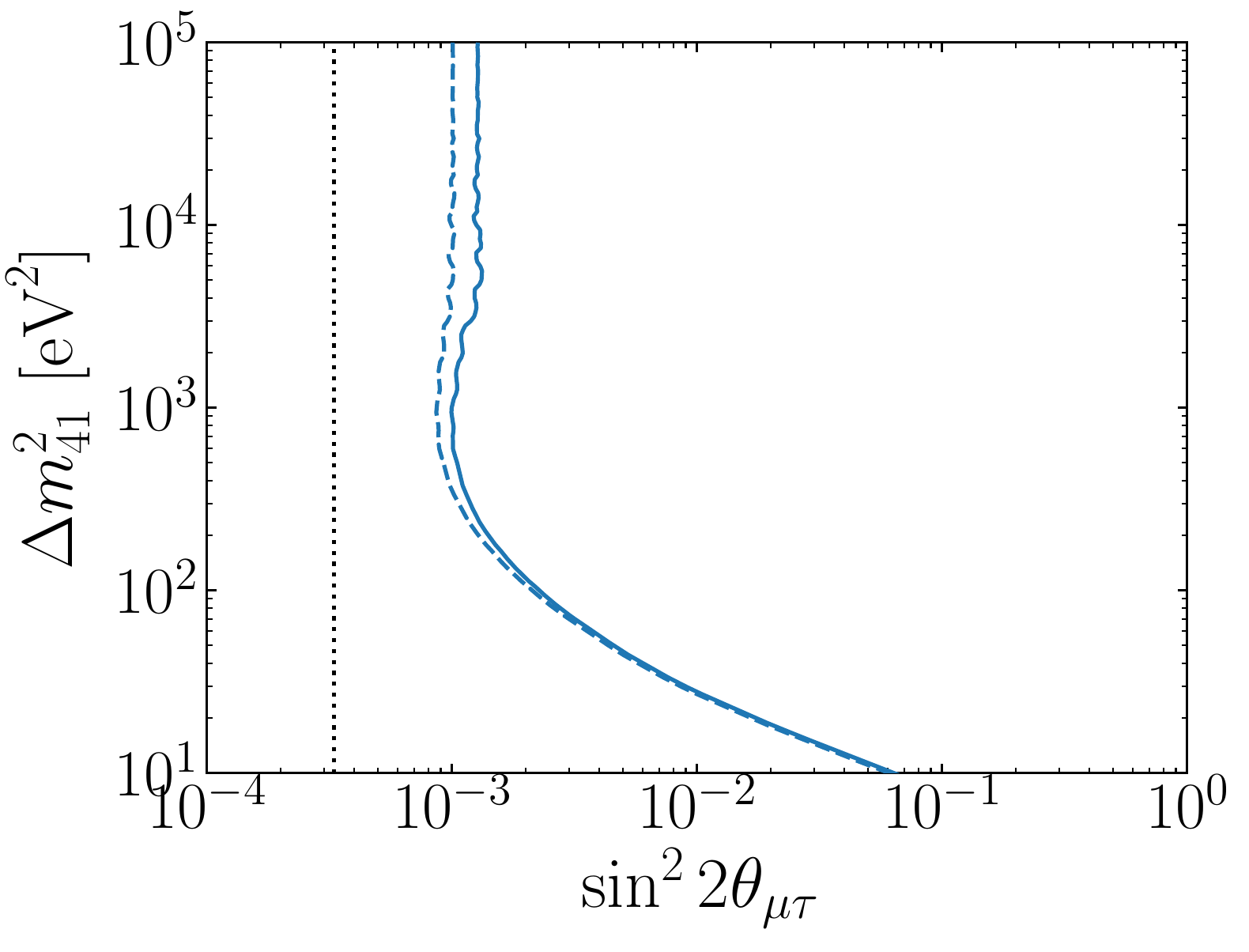}
\caption{The 95\% CL regions of sterile neutrino parameters that will be excluded by FASER assuming different normalization uncertainties. The regions to the right of the black dotted lines are already excluded by previous experiments~\cite{Astier:2001yj,Astier:2003gs,Dentler:2018sju}.
The yellow shaded region in the upper left panel is the preferred region for the Gallium anomaly in the $\nu_e$ disappearance channel~\cite{Giunti:2010zu}, and the blue shaded region in the lower center panel is the region preferred by MiniBooNE in the $\nu_\mu\to\nu_e$ appearance channel~\cite{Aguilar-Arevalo:2018gpe}.}
\label{fig:sterile_contours}
\end{figure}

We find that the effect of the production uncertainties are modest at worst and that \FASERnu will be quite competitive with global constraints in many cases. \FASERnu will be able to probe the Gallium anomaly~\cite{Giunti:2010zu} in the large $\Delta m^2_{41}$ regime directly and, in general, probe a different region of parameter space than is probed by other experiments, whose sensitivities typically peak around $\Delta m^2_{41}\sim1$ eV$^2$. The constraints continue up to large values of $\Delta m^2_{41}$, where fast oscillations result in an overall depletion in the flux.  Constraints from other experiments at these large $\Delta m^2_{41}$'s rely on normalization measurements, while FASER will have an unprecedented capability of probing large $\Delta m^2_{41}\sim1,000\,{\rm eV}^2$ directly for the first time.

\subsection{Other Probes of New Physics}
\label{sec:otherprobes}

There are numerous other potentially interesting topics in the neutrino sector that \FASERnu could be sensitive to including neutrino tridents~\cite{Czyz:1964zz,Geiregat:1990gz,Mishra:1991bv,Adams:1998yf} which are also a portal to new physics~\cite{Altmannshofer:2014pba,Magill:2017mps}, neutrino Non-Standard Interactions~\cite{Wolfenstein:1977ue,Biggio:2009nt,Dev:2019anc}, neutrino decay \cite{Chikashige:1980ui,Gelmini:1980re,Acker:1991ej} and the weak mixing angle.

In addition to neutrino searches, \FASERnu could also be used for BSM physics searches. One example would be long-lived particles (LLPs) which are both produced and decay in the emulsion detector, leading to displaced vertex signature. The precise spatial resolution of the emulsion detector allows to probe decay length in the $c\tau\gamma \sim 1~\mm$ regime, similar to the detection of displaced decaying $D$ mesons. Such LLP searches would be complementary to those at FASER with decay length of $c\tau\gamma \approx 480~\m$. Additionally, if \FASERnu is interfaced with the FASER spectrometer, one could also search for LLPs produced in \FASERnu and decaying in the FASER decay volume, probing decay lengths of $c\tau\beta\gamma \sim 1~\m$. Such searches are particularly interesting if the LLP is also produced in neutrino or dark matter interactions~\cite{Jodlowski:2019ycu}, leading to a neutral vertex in FASER$\nu$. An example of such a model, which could also explain the MiniBooNE anomaly~\cite{Aguilar-Arevalo:2018gpe}, has been discussed in Ref.~\cite{Bertuzzo:2018itn} and consists of both a heavy neutral lepton and a dark photon with a lifetime that is appropriate for FASER$\nu$. Finally, FASER$\nu$ could be sensitive to a additional neutrino production modes, for example via the decay of light mediators into $\tau$-neutrinos.

\section{Conclusions and Outlook}
\label{sec:conclusions}

At present, no neutrino produced at a particle collider has ever been detected, despite the fact that colliders are copious sources of neutrinos.  For example, in Run 3 of the 14 TeV LHC from 2021-23, roughly $10^{11}$ electron neutrinos, $10^{12}$ muon neutrinos, and $10^9$ tau neutrinos, along with comparable numbers of anti-neutrinos, will be produced in the far-forward region of the ATLAS IP.  These neutrinos will have TeV-scale energies, the highest man-made energies currently available, and detection of these neutrinos will extend the LHC's physics program in a new direction, opening up a new source of neutrinos to detailed study.  

In this paper we have shown that FASER$\nu$, a proposed small and inexpensive subdetector of FASER, will be able to detect the first collider neutrinos and usher in a new era of neutrino physics at colliders. FASER$\nu$ is a $25\,\cm \times 25\,\cm \times 1.35\,\m$ emulsion detector consisting of 1000 layers of emulsion films interleaved with 1-mm-thick tungsten plates, with a total tungsten target mass of 1.2 tons. Despite its modest size, by virtue of its location along the beam collision axis 480 m downstream from ATLAS, 1300 electron neutrinos, 20,000 muon neutrinos, and 20 tau neutrinos will interact in FASER$\nu$ in Run 3, providing an opportunity for detailed studies of $\nu_e$ and $\nu_{\tau}$ at the highest energies yet explored and $\nu_{\mu}$ neutrinos in a currently unexplored energy range between the energy ranges of accelerator experiments and IceCube.   

In this study, we have reported detailed estimates of the neutrino fluxes and interaction rates in FASER$\nu$, and we have presented the expected features and capabilities of the FASER$\nu$ emulsion detector.  Detailed considerations of event features can yield a reconstructed neutrino energy resolution of 30\% and angular resolution of $\sim 3-5~\mrad$ for CC $\nu_{\mu}$ interactions. Neutrino CC events can be identified by requiring neutral vertices with 5 or more charged tracks, including a high-energy lepton, and requiring very high energies and tracks pointing back to the ATLAS IP greatly suppresses backgrounds.  Furthermore, the \FASERnu emulsion detector module can be interfaced with the FASER spectrometer, which will provide the possibility to distinguish $\nu_{\mu}$ from $\bar{\nu}_{\mu}$ through muon charge identification, as well as better energy reconstruction and background suppression.

Assuming SM cross sections, FASER$\nu$'s physics prospects include detecting the first collider neutrino, significantly enhancing the world's sample of reconstructed $\nu_{\tau}$ events, measuring CC neutrino cross sections at uncharted energies, and probing models for forward particle production.  These observations will have interesting implications for other fields.  For example, the measurement of forward neutrino production will be a key input for high-energy neutrino measurements by large-scale Cherenkov observatories, such as IceCube, and observations of high-energy scattering rates will constrain poorly understood nuclear effects in high-energy lepton-nucleus interactions. Finally, FASER$\nu$ will probe new physics, in particular because the flavor-sensitive detector design will allow studies of channels including $\tau$, charm and beauty, which could be excellent probes of BSM effects in heavy flavor particles.

The study of collider neutrinos at the LHC will be a milestone in particle physics, opening the door to experiments at the energy frontier to make highly complementary measurements in neutrino physics, a topic typically considered to be wholly contained within the intensity frontier. As the first of its kind, \FASERnu may also pave the way for neutrino programs at future colliders. The success of FASER$\nu$ would motivate future neutrino programs at the HL-LHC, with its 20-fold increase in integrated luminosity (3 ab$^{-1}$) over Run 3, and potentially larger detectors. Such experiments will also provide timely input to the discussion of potentially much larger neutrino projects with masses of 10 to 1,000 tons at future colliders currently under discussion.

\acknowledgments

We thank CERN for the excellent performance of the LHC and the technical and administrative staff members at all FASER institutions for their contributions to the success of the FASER effort. The FASER Collaboration also gratefully acknowledge invaluable assistance from many people, including Mike Lamont and the CERN Physics Beyond Colliders study group; the LHC Tunnel Region Experiment (TREX) working group; Dominique Missiaen, Pierre Valentin, Tobias Dobers, and the CERN survey team; Francesco Cerutti, Marta Sabat\'e-Gilarte, Andrea Tsinganis, and the CERN STI group; Jonathan Gall and the CERN Civil Engineering group; the DsTau Collaboration for providing their spare emulsion films for the {\em in situ} measurements in 2018 and Yosuke Suzuki for helping with the film production; and Yoshitaka Itow, Masahiro Komatsu, Steven Mrenna, Toshiyuki Nakano, Tanguy Pierog, Hank Sobel, and Ralf Ulrich for useful discussions. This work was supported in part by grants from the Heising-Simons Foundation (Grant Nos.~2018-1135 and 2019-1179) and the Simons Foundation (Grant No.~623683).  AA is supported in part by the Albert Einstein Center for Fundamental Physics, University of Bern. The work of TA is supported by JSPS KAKENHI Grant Number JP 19H01909. PBD acknowledges the support of the U.S.~Department of Energy under Grant Contract desc0012704. JLF, FK, and JS are supported in part by U.S.~National Science Foundation Grant Nos.~PHY-1620638 and PHY-1915005.  JLF~is supported in part by Simons Investigator Award \#376204. IG~is supported in part by U.S.~Department of Energy Grant DOE-SC0010008. FK~is supported in part by DE-AC02-76SF00515 and in part by the Gordon and Betty Moore Foundation through Grant GBMF6210. ST~is supported in part by the Lancaster-Manchester-Sheffield Consortium for Fundamental Physics under STFC grant ST/P000800/1. This work is supported in part by the Swiss National Science Foundation. 

\bibliography{references}

\end{document}